\documentclass[onecolumn,useAMS,usenatbib]{mnras}

\usepackage{color}
\usepackage{amsmath}
\usepackage{graphicx}
\usepackage{amssymb}
\usepackage{psfrag}
\usepackage{threeparttable}

\newcommand{\beq}{\begin{equation}}
\newcommand{\eeq}{\end{equation}}
\newcommand{\bay}{\begin{array}}
\newcommand{\eay}{\end{array}}
\newcommand{\beqa}{\begin{align}}
\newcommand{\eeqa}{\end{align}}
\newcommand{\beqy}{\begin{eqnarray}}
\newcommand{\eeqy}{\end{eqnarray}}
\newcommand{\nn}{\nonumber}
\newcommand{\rmd}{\mathrm{d}}
\newcommand{\brac}[1]{\left({#1}\right)}
\newcommand{\pd}[2]{\frac{\partial{#1}}{\partial{#2}}}
\newcommand{\td}[2]{\frac{\rmd{#1}}{\rmd{#2}}}
\newcommand{\curl}{\nabla\times}
\renewcommand{\div}{\nabla\cdot}

\newcommand{\bB}{{\bf B}}
\newcommand{\bOm}{\boldsymbol\Omega}

\newcommand{\be}{{\bf e}}

\newcommand{\br}{{\bf r}}

\newcommand{\clL}{\mathcal{L}}
\newcommand{\clM}{\mathcal{M}}
\newcommand{\clO}{\mathcal{O}}
\newcommand{\rme}{\mathrm{e}}
\newcommand{\Mr}{{\textstyle{\frac{\pi r}{R_*}}}}

\newcommand{\skl}[1]{{\color{black}{#1}}}

\title[Non-rigid precession of magnetic stars]{Non-rigid precession of
  magnetic stars}
\author[S. K. Lander and D. I. Jones]{S. K. Lander\thanks{s.k.lander@soton.ac.uk},
         D. I. Jones\thanks{d.i.jones@soton.ac.uk}\\ \\
         Mathematical Sciences and STAG Research Centre, University of Southampton, Southampton
         SO17 1BJ, U.K.}

\begin{document}

\pagerange{\pageref{firstpage}--\pageref{lastpage}} \pubyear{0000}
\maketitle

\label{firstpage}

\begin{abstract}
  Stars are, generically, rotating and magnetised objects with a
  misalignment between their magnetic and rotation axes. Since a magnetic field induces a
  permanent distortion to its host, it provides effective rigidity
  even to a fluid star, leading to bulk stellar motion which resembles
  free precession. This bulk motion is however accompanied by induced
  interior velocity and magnetic field perturbations, which are
  oscillatory on the precession timescale. Extending
  previous work, we show that these quantities are described by a set
  of second-order perturbation equations featuring cross-terms scaling
  with the product of the magnetic and centrifugal distortions to the
  star. For the case of a background toroidal field, we reduce these
  to a set of differential equations in radial functions, and
  find a method for their solution. The resulting magnetic-field and
  velocity perturbations show complex multipolar structure and are
  strongest towards the centre of the star.
\end{abstract}


\section{Introduction}

Two of the most important pieces of stellar physics are
their magnetic fields and rotation. They have a ubiquity across many
classes of star which are otherwise governed by very different interior
and exterior physics: main-sequence stars, white dwarfs, neutron stars, or combinations
of these in binary systems. A fundamental problem is the rich variety
of ways in which rotation and magnetism interact, and the effects on
stellar properties and evolution.

In this paper we focus on one of the simplest
situations within the broad class of magneto-rotational stellar
phenomena: the dynamics of a rigidly-rotating star with a
frozen-in dynamically stable magnetic field\footnote{i.e. a field which does not
  require continuous dynamo action to maintain it, in the way that the
  Sun's does.}, symmetric about some axis. In the case where the
\emph{inclination angle} $\chi$ between the magnetic and rotational
axes is zero, i.e. the two axes coincide, the star is stationary. This
widely-studied situation has the advantage of being simple, but the disadvantage of
having limited applicability to astrophysics: observations indicate
that stars have a wide distribution of inclination angles
\citep{schmidt_nor,tauris_man,don-land,rookyard}. A non-zero
$\chi$ is essential, for example, to explain the pulsed emission seen from many
neutron stars.

The dynamics of a star with non-zero $\chi$ -- an `oblique rotator' --
is still not well understood. The classic work on the topic is by
Mestel and collaborators \citep{mestel1,mestel2}, as discussed in section \ref{fluidprec}. The
essential idea, put forward by \citet{spitzer}, is that the stellar distortion associated with the magnetic
field causes it to undergo bulk motion like that of rigid-body free
precession. \citet{mestel1} argue that internally this bulk motion has to be supported by
time-varying and non-axisymmetric velocity and magnetic
fields. Formulated in generality, this problem is intractable by
analytic methods; accordingly, Mestel and collaborators made some major simplifications
to produce solutions for their early studies. Various subsequent papers
have adopted their ideas and solutions (see, e.g., \citet{wass03},
\citet{dallosso} and \citet{lasky_glam}), but none have attempted to extend them.

The goal of the present paper is to build on the pioneering
ideas of Mestel and collaborators, but to formulate them in a more rigorous and general way. The
practical difference is that we are forced to perform perturbation
analysis to a higher order than their work, with an accompanying
increase in the complexity of the algebra. The present paper is entirely
concerned with the theory of this problem, and the ultimate
desideratum is to obtain solutions
for the perturbed velocity and magnetic fields of a fluid star with a
non-zero $\chi$. \skl{In order to keep the majority of the calculation
analytically tractable, we make two key simplifications: the
background magnetic field is assumed to be purely toroidal, and
we employ a polytropic equation of state. For numerical reasons we
have also been forced to neglect the outermost layer of the star in
our solutions.} In future work we will study the implication of our results
for observed phenomena, like the distribution and time-evolution of inclination
angles amongst stars, and the damping of precession.

The paper is structured as follows. In section \ref{fluidprec} we
summarise existing models of the internal dynamics of oblique rotators,
providing a critique of their applicability
and suggesting how to extend them. In section \ref{pert_scheme} we
formulate the oblique-rotator problem using a perturbation scheme in two small quantities, the
centrifugal and magnetic distortions, $\epsilon_\alpha$ and
$\epsilon_B$ respectively; we show that the velocity and
magnetic-field perturbations are of the same perturbative order as
cross-terms involving the interaction of the stellar rotation and background
magnetic field, and so finding these entails the solution of the
order-$\epsilon_\alpha\epsilon_B$ perturbation equations. In section
\ref{solution_zerothfirst} we systematically present solutions to all the lower-order
problems: the zeroth-order, order-$\epsilon_\alpha$ and
order-$\epsilon_B$ equations. Section \ref{solution_second} then
addresses the problem of solving the second-order equations, through a
series of stages. In section \ref{explicit_torpol} we reduce the full system of
governing vector equations to poloidal and toroidal scalar
equations. Following this, we perform a spherical-harmonic
decomposition of the perturbed magnetic field (section \ref{eqns_delB}), yielding an infinite system
of differential equations (DEs) in radial functions associated with each
spherical harmonic. Superficially these appear to be ordinary
differential equations (ODEs), but are in fact a more
complicated system of differential-algebraic equations, which cannot
be solved with ODE methods. Nonetheless we find and present a method to solve the
equations for the magnetic radial functions, when truncating at the
fourth multipole (section \ref{DAE-centrecond}). In section \ref{xi-motions} we find closed-form expressions
for the perturbed velocity field in terms of the magnetic radial
functions. After this, we present results for the
magnetic and velocity perturbations in section \ref{results}. Finally, we
discuss our findings and their applicability to different
classes of star in section \ref{conclusions}, and summarise in section \ref{summary}.

\section{Generation of precessional motion in a magnetised fluid star}
\label{fluidprec}

Precession is a rigid-body effect, and superficially one would not
expect a fluid star (or the fluid region of a neutron star) to be able
to sustain such motion. In fact, we will adopt a stellar model which
is isothermal and non-convective, so we would expect only circular
motion of fluid elements about the star's rotation axis. However, as argued by \citet{spitzer}, a
magnetic field $\bB$ can provide effective rigidity to a star, since it
provides a distortion (symmetric about the magnetic-field axis) away
from the star's spherically-symmetric hydrostatic state
\citep{chand_fermi}. If the star is now rotated about a different axis
from the magnetic one, it will undergo a secondary
rotation\footnote{referred to as `nutation' in some studies.} about the
magnetic axis, so that the bulk motion of the star resembles
precession.

In this section we begin by recapitulating the work of
\citet{mestel1}, who put this conceptual picture on a rigorous footing and
derived the form of the secondary rotation. In doing so we introduce
notation which differs in many cases from theirs. We then describe their
arguments as to why the motion is not exactly described by rigid-body
free precession, and the approaches taken to finding the non-rigid internal
dynamics that maintain the bulk precessional motion in both the original
paper of \citet{mestel1}, and a rather later follow-up study
\citep{mestel2}. We conclude the discussion with a critique of these approaches, and
describe our strategy for obtaining a more detailed solution which we
believe includes the key physics missing from their
work. \skl{Finally, we discuss the ordering of characteristic
  timescales required for non-rigid precession to occur, and compare
  the various classes of magnetic star to which our analysis may apply.}

\begin{figure}
\begin{center}
\begin{minipage}[c]{0.7\linewidth}
\psfrag{i}{$\be_x$}
\psfrag{j}{$\be_y^{(\alpha)}$}
\psfrag{k}{$\be_z^{(\alpha)}$}
\psfrag{l}{$\be_y^{(B)}$}
\psfrag{p}{$\be_z^{(B)}$}
\psfrag{t}{$\chi$}
\psfrag{r}{\bf{r}}
\psfrag{th}{$\theta$}
\psfrag{lm}{$\phi$}
\includegraphics[height=5.0cm]{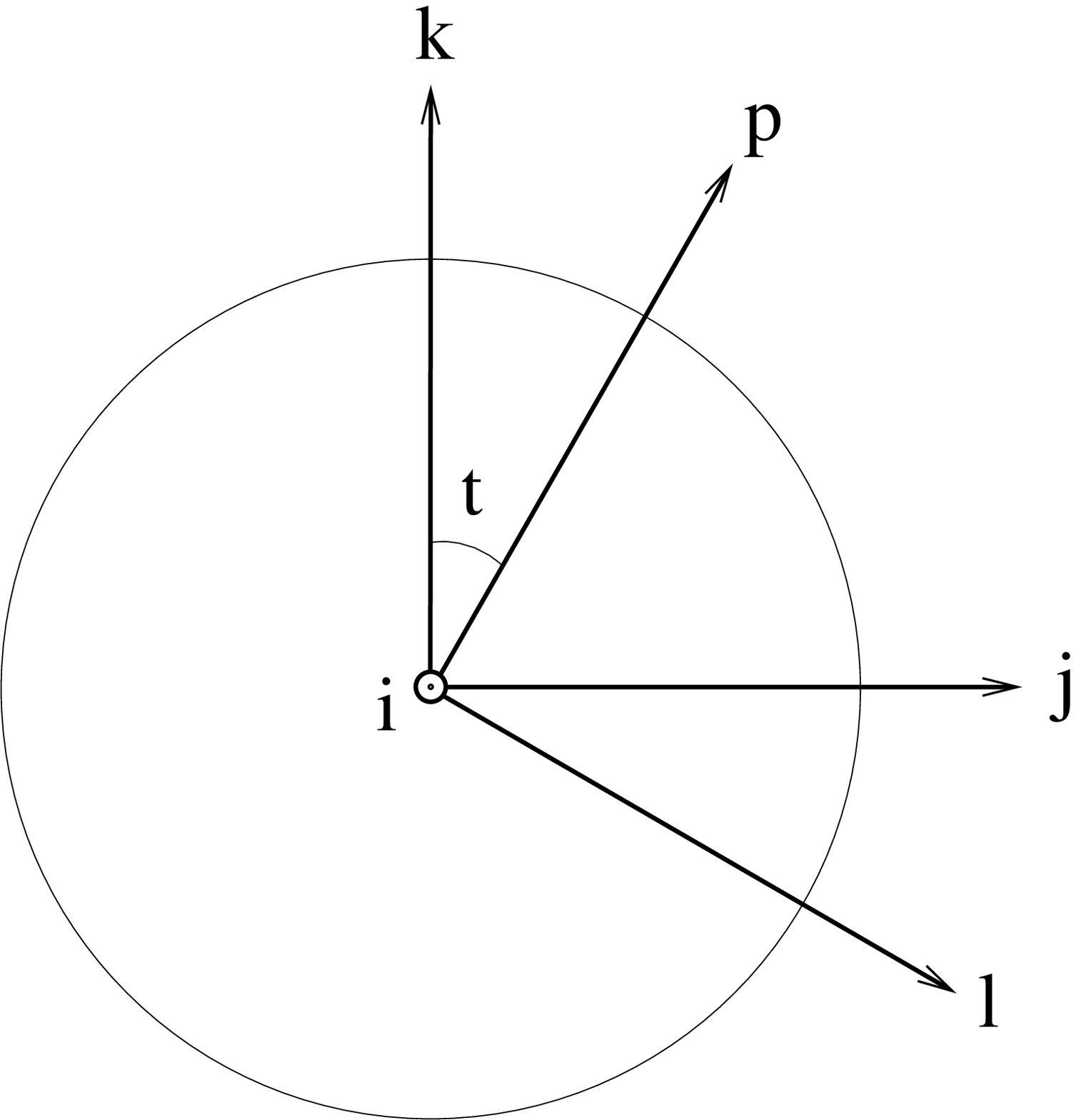}
\hfill
\includegraphics[height=5.0cm]{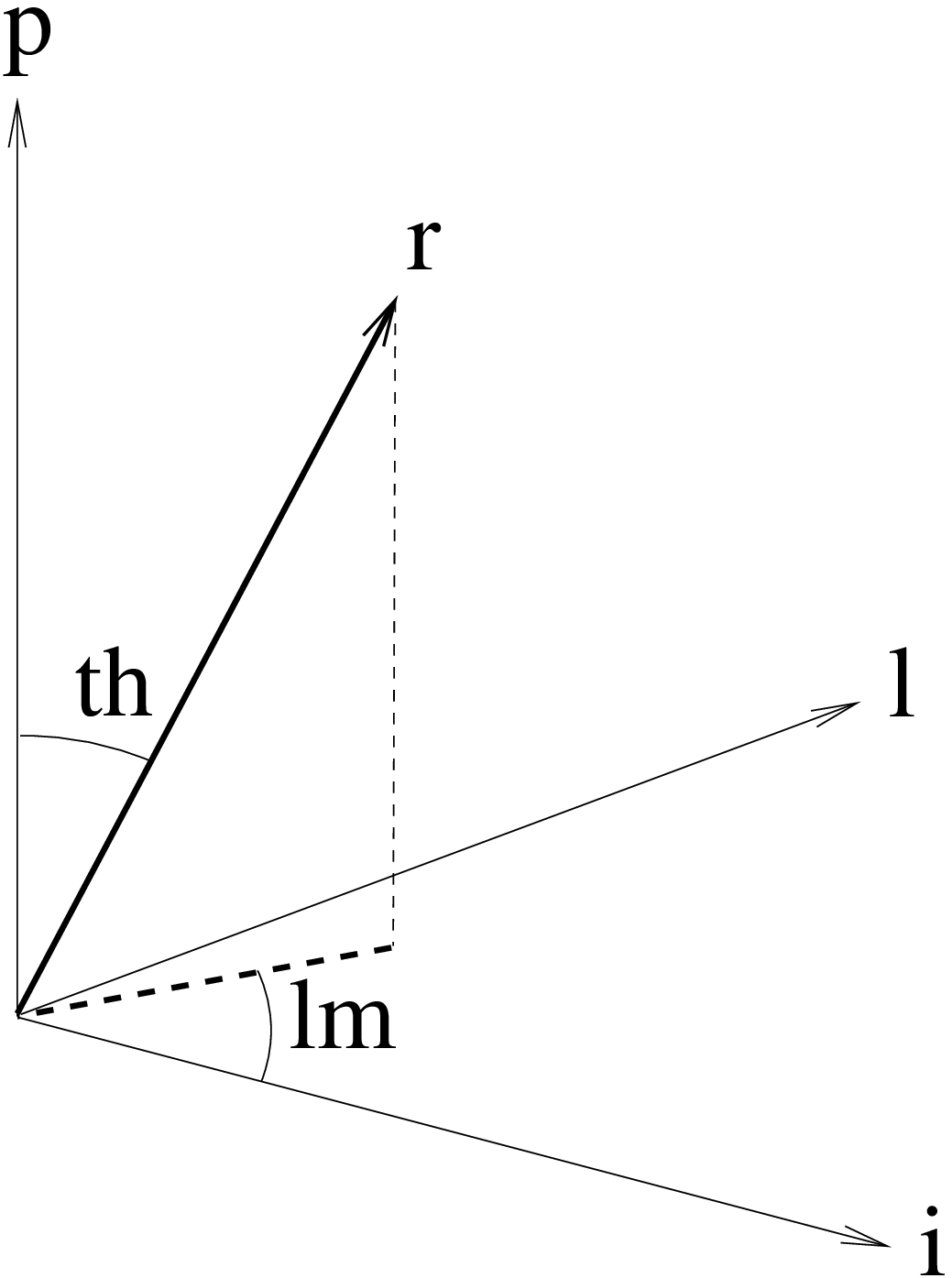}
\end{minipage}
\caption{\label{ijkl}
         Left: the magnetic and rotational triads; we assume $\be_y^{(B)},\be_z^{(B)},\be_y^{(\alpha)}$
         and $\be_z^{(\alpha)}$ are coplanar. Right: the $\be_z^{(B)}$-triad and its spherical polar coordinate system.}
\end{center}
\end{figure}

We model a star as a fluid ball rotating uniformly at angular
frequency $\alpha$ about an axis $\be_z^{(\alpha)}$, with a frozen-in magnetic field symmetric about some axis
$\be_z^{(B)}$. The axis $\be_z^{(B)}$ is misaligned by some angle $\chi$ from the
primary rotation axis $\be_z^{(\alpha)}$. At different points in our calculation we
will need to refer to both the symmetry axis of primary rotation and
that of the magnetic field, and so we form right-handed triads $({\be_x,\be_y^{(B)},\be_z^{(B)}})$ and
$(\be_x,\be_y^{(\alpha)},\be_z^{(\alpha)})$ associated with these
magnetic and rotational axes, with the $\be_x$ vector being
instantaneously common to
both (see left-hand side of figure \ref{ijkl}). In addition, we denote the spherical polar coordinate system referred to the
$\be_z^{(B)}$-triad by $(r,\theta,\phi)$; for the rest of this section we
shall work exclusively in this coordinate system (right-hand side of
figure \ref{ijkl}).

\subsection{The bulk motion of the star}
\label{bulk_motion}

Unless otherwise specified, the rest of this section follows the
reasoning of \citet{mestel1}, but sometimes explained in different
terms and with different notation. We first recall that a static,
unmagnetised star composed of homogeneous fluid would have a
spherically symmetric density field $\rho_0(r)$. Including rotation
alone adds on a small extra term\footnote{This would be axisymmetric in
  the $\be_z^{(\alpha)}$-coordinate system, but is non-axisymmetric in coordinates
  referred to the axis $\be_z^{(B)}$.}
$\delta\rho_\alpha(r,\theta,\phi)$ corresponding to a
centrifugal bulge; similarly, the density distribution of
a non-rotating magnetised fluid ball could be written as
$\rho(r,\theta)=\rho_0(r)+\delta\rho_B(r,\theta)$ to take account of magnetic
distortions $\delta\rho_B$. Hence, for a rotating, magnetised
star we may write the density of an element at the point
$(r,\theta,\phi)$, and at some instant in time, as
\beq \label{zerotimerho}
\rho(r,\theta,\phi)=\rho_0(r)+\delta\rho_B(r,\theta)+\delta\rho_\alpha(r,\theta,\phi),
\eeq
where -- for now -- we have neglected cross-terms $\sim\delta\rho_\alpha\delta\rho_B$
for being higher-order than the other density components.

The density field of a star rotating with angular velocity
$\alpha\be_z^{(\alpha)}$ has the angular momentum vector
\beq \label{angmomvec}
{\bf{J}}_\alpha =
\int \rho{\bf{r}}\times(\alpha\be_z^{(\alpha)}\times {\bf{r}})  \ \rmd V.
\eeq
However, this alone does not give an invariant angular momentum
orientated along the $\be_z^{(\alpha)}$ direction, as the $\be_y^{(\alpha)}$-component of
\eqref{angmomvec} is non-zero:
\beq \label{jcomp}
{\bf J}_\alpha\cdot\be_y^{(\alpha)}
= -\alpha\int(\be_y^{(\alpha)}\cdot\br)(\be_z^{(\alpha)}\cdot\br)\delta\rho_B \ \rmd V
\eeq
where the contributions from $\rho_0$ and $\delta\rho_\alpha$ vanish by symmetry.
To yield an invariant angular momentum we require an additional
rotation $\omega$ about the magnetic axis $\be_z^{(B)}$
with an associated angular momentum ${\bf J}_B$ such that
$({\bf J}_\alpha+{\bf J}_B)\cdot\be_y^{(\alpha)}=0$, i.e.
\beq
{\bf J}_B\cdot\be_y^{(\alpha)} =  \int\rho\br\times(\omega\be_z^{(B)}\times\br)\cdot\be_y^{(\alpha)} \ \rmd V
 = \alpha\int(\be_y^{(\alpha)}\cdot\br)(\be_z^{(\alpha)}\cdot\br)\delta\rho_B \ \rmd V.
 \label{cancel}
\eeq

Working in spherical polar coordinates with $\br=r(\sin\theta\cos\phi\ \be_x +
\sin\theta\sin\phi\ \be_y^{(B)} + \cos\theta\ \be_z^{(B)})$, and writing
 $\be_y^{(\alpha)}=\cos\chi\ \be_y^{(B)}+\sin\chi\ \be_z^{(B)}$,
$\be_z^{(\alpha)}=-\sin\chi\ \be_y^{(B)}+\cos\chi\ \be_z^{(B)}$ and
$\rmd V=r^2\sin\theta\ \rmd r\rmd\theta\rmd\phi$, we now evaluate the
integral \eqref{jcomp} in the $(\be_x,\be_y^{(B)},\be_z^{(B)})$ triad to give
\beq
{\bf J}_\alpha\cdot\be_y^{(\alpha)}
 = 2\pi\alpha\sin\chi\cos\chi\int\!\!\!\int\delta\rho_B P^0_2 (\cos\theta)
        r^4\sin\theta\ \rmd r \rmd\theta, \label{topomega}
\eeq
where $P_l^m(\cos\theta)$ denotes an associated Legendre polynomial of angular
index $l$ and azimuthal index $m$ (which, in this equation, is the ordinary
Legendre polynomial $P_2\equiv P_2^0=\frac{1}{2}(3\cos^2\theta - 1)$). We
evaluate the $\be_y^{(\alpha)}$-component of ${\bf J}_B$ in a similar
fashion to give
\beq
{\bf J}_B\cdot\be_y^{(\alpha)} = \int\rho\br\times(\omega\be_z^{(B)}\times\br)\cdot\be_y^{(\alpha)}\ \rmd V\nn\\
            = I_0 \omega\sin\chi, \label{bottomomega}
\eeq
where $I_0\equiv \frac{8\pi}{3}\int\rho_0 r^4\ \rmd r$ is the moment of
inertia of the spherically symmetric density field $\rho_0$; here the two
density perturbations are regarded as negligible parts of $\rho$ in
comparison with $\rho_0$. We now use equations \eqref{topomega},
\eqref{bottomomega} and the requirement $({\bf J}_\alpha+{\bf J}_B)\cdot\be_y^{(\alpha)}=0$ to
find the precession frequency:
\beq \label{fluidfreeprec}
\omega=-\frac{2\pi\alpha\cos\chi}{I_0}
            \int\!\!\!\int\delta\rho_B P_2(\cos\theta)r^4\sin\theta\ \rmd r\rmd\theta.
\eeq
Rewriting this expression using the $xx$ and $zz$ components of the
moment-of-inertia tensor $I_{jk}\equiv\int\rho(r^2\delta_{jk}-x_j x_k)\ \rmd V$
referred to the $\be_z^{(B)}$-triad yields
\beq \label{omega}
\omega = \alpha\cos\chi\ \frac{I_{zz}-I_{xx}}{I_0}
     \equiv \alpha\epsilon_B\cos\chi,
\eeq
the usual rigid-body result \citep{LandL}; this is not surprising
since we have not yet put any fluid (non-rigid) physics into the calculation.

\subsection{Deviation from rigid-body precession due to internal fluid dynamics}
\label{xiunderdet}

The result at the end of the previous subsection suggests that the
macroscopic dynamics of a rotating 
magnetised fluid body should resemble free precession; however the
fluid is clearly not a rigid body. This presents a question as to what degree
the magnetised fluid \emph{can} be regarded as rigid and hence how
similar the motion of a magnetised fluid is to conventional rigid-body
precession. Mestel and collaborators sought to answer this by considering the
microscopic dynamics: the effect of precession on individual fluid 
elements and the induced non-rigid velocity field.

Since we will need to distinguish
between different frames of reference here, we define for brevity the
`$\alpha$-frame' to be the one comoving with the star's primary
rotation (at frequency $\alpha$) and the `$\omega$-frame' to be the
co-precessing frame --- i.e. the rigid-body precession frame
characterised by the superimposed rotations $\alpha$ and $\omega$.

We wish to investigate the deviation of a rotating magnetised fluid
star from free precession. If the fluid moved rigidly then
each fluid element would be stationary as viewed by the co-precessing
observer in the $\omega$-frame. Since we do not expect exact rigid-body
precession here, let us define the Lagrangian displacement $\bxi$ to
be the change in position of a fluid element in the co-precessing
frame, with its time derivative 
$\dot{\bxi}$ giving the velocity of the element as viewed from the
$\omega$-frame. Following \citet{mestel1}, we will sometimes refer to
this higher-order velocity field as `the $\bxi$-motions'. On viewing the star
in the inertial frame, we will then see that the motion of a fluid
element ${\bf V}_{\mathrm{inertial}}$ is a vector sum of \emph{three} characteristic velocities: the
primary stellar rotation $\alpha$ about the rotation axis; the slower
nutation $\omega$ about the magnetic axis; and the extra velocity
field $\dot\bxi$:
\beq
{\bf V}_{\mathrm{inertial}} 
  = \alpha\be_z^{(\alpha)}\times\br + \omega\be_z^{(B)}\times\br + \dot\bxi.
\eeq

\begin{figure}
\begin{center}
\begin{minipage}[c]{0.85\linewidth}
\psfrag{om}{\textcolor[rgb]{1,0,0}{$\omega$}}
\psfrag{n3}{\textcolor[rgb]{0,0,1}{$\be_z^{(B)}$}}
\psfrag{k}{\textcolor[rgb]{0,0,1}{$\be_z^{(\alpha)}$}}
\psfrag{chi}{\textcolor[rgb]{0,0,1}{$\chi$}}
\includegraphics[width=0.5\linewidth]{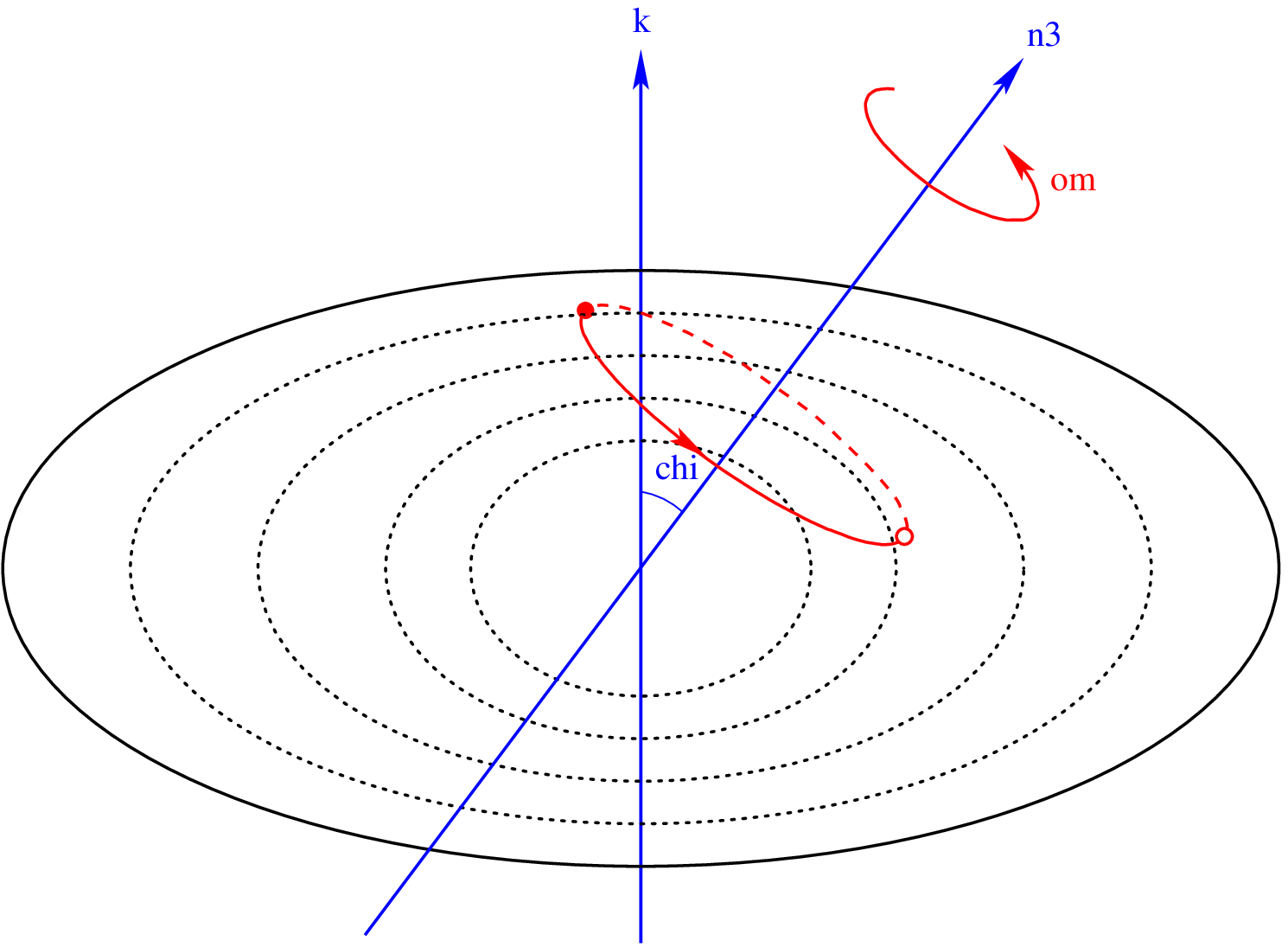}
\hfill
\includegraphics[width=0.3\linewidth]{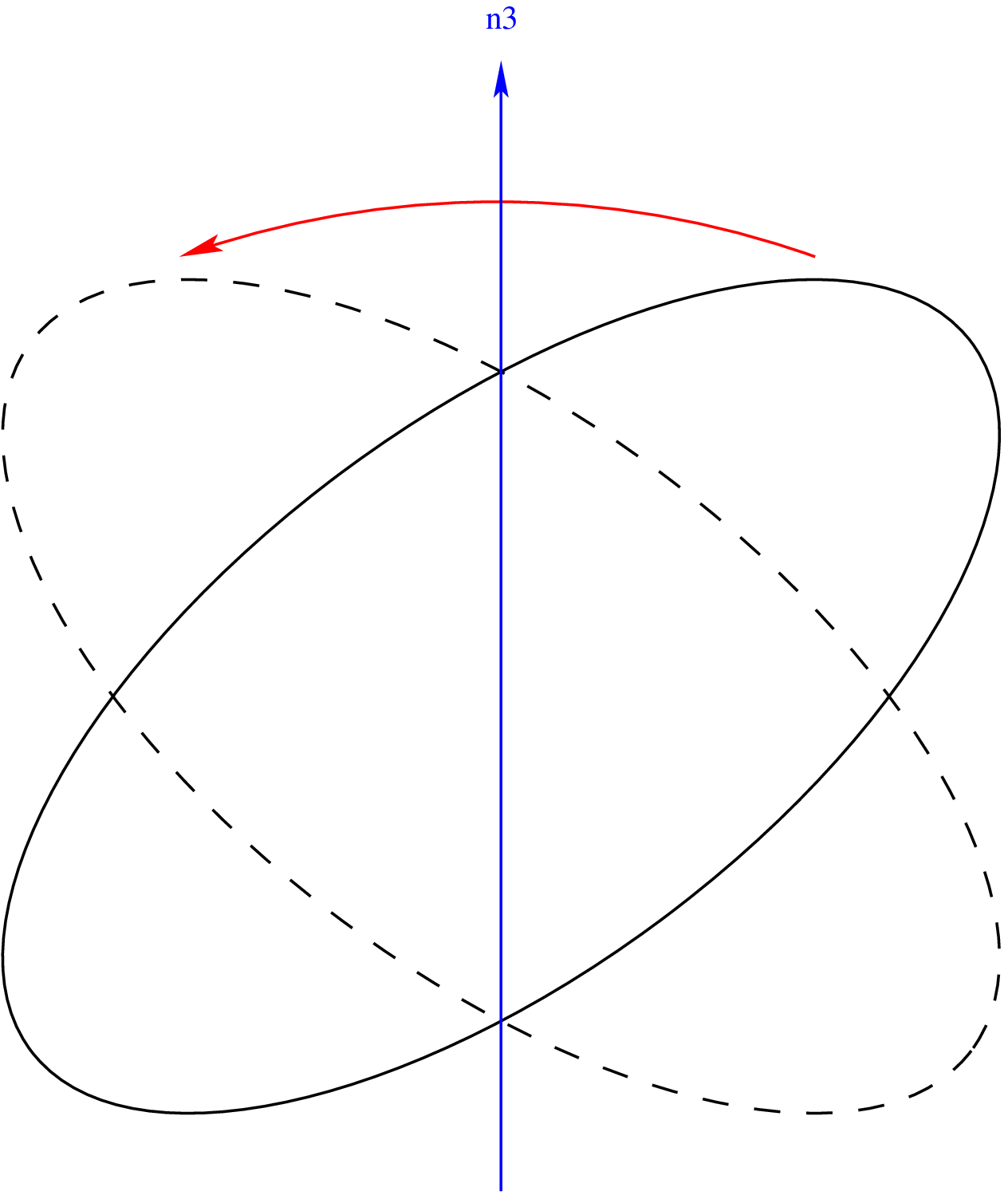}
\end{minipage}
\caption[Dynamics in the $\alpha$-frame.]
  {\label{alphaframe}
   Two views of a precessing fluid star. Left: the dynamics in the $\alpha$-frame, i.e. the frame rigidly rotating
   with rate $\alpha$, right: the dynamics in the co-precessing
   $\omega$-frame.  We show only the centrifugal contribution to the distortion
   here, so that the stellar surface (the solid
   black line) and the isopycnic surfaces (the dashed black lines) are
   spheroidal about the $\be_z^{(\alpha)}$ axis. Without a magnetic field, a fluid element would
   be stationary in the $\alpha$-frame; however the magnetic field induces a
   slow precessional motion, superimposed on the normal stellar
   rotation. This motion will cause a fluid element (the filled red
   circle) in the $\alpha$-frame to rotate about the magnetic axis
   $\be_z^{(B)}$ with period $2\pi/\omega$. A higher-order velocity field
   $\dot\bxi$ is needed to prevent the fluid element from travelling
   through regions of varying density (crossing the density
   contours as shown). An alternative viewpoint of the process can be
   obtained from the $\omega$-frame, in which the whole centrifugal bulge
   would be rotated at rate $\omega$ around the $\be_z^{(B)}$ axis
   were it not for the $\dot\bxi$ field.}
\end{center}
\end{figure}

An intuitive explanation for the existence of this extra
velocity field, presented in \citet{mestel1}, is as follows. Consider the motion of a fluid element in the
$\alpha$-frame; see the left-hand panel of figure \ref{alphaframe}. In the unmagnetised case
the element undergoes only the primary rotation $\alpha$ and so is
stationary in the corotating frame. From section \ref{bulk_motion} we
anticipate that the addition of a misaligned magnetic field will
cause the star to precess, and a fluid element in the $\alpha$-frame
will therefore undergo a slow secondary rotation (with
frequency $\omega$) about 
the magnetic axis. In doing so, however, the fluid element will be
moved through regions of differing density. Since the background
density $\rho_0$ is spherical and the magnetic distortion $\delta\rho_B$
is symmetric about its axis, the density difference will be entirely
due to the centrifugal bulge $\delta\rho_\alpha$. If we move to the
$\omega$-frame (figure \ref{alphaframe}, right-hand panel), on a macroscopic level the entire centrifugal bulge
would rotate at a rate $\omega$ about the $\be_z^{(B)}$-axis.

At this point we may argue for the existence of a field of non-rigid
$\bxi$-motions. In the $\alpha$-frame, fluid elements undergoing
rigid-body free precession (i.e. with $\dot\bxi={\bf 0}$) would be forced to
sustain large density variations over one precession
period. For this reason there will be a restoring force on each 
fluid element that acts to return it to its original density;
hence the precessional motion described above cannot be completely
rigid. Equivalently, on a macroscopic level
and in the $\omega$ frame, one would expect the global effect of the
internal $\bxi$-motions to be a restoration of the star to an
instantaneous stationary equilibrium.

At this point \citet{mestel1} used this intuitive picture to argue for the
functional form of $\delta\rho_\alpha$, the piece of the centrifugal
bulge which is time-varying in the $\omega$ frame, by simply
performing a rotation of the coordinates with respect to the rotation
axis\footnote{By a more systematic analysis later in this paper, we
  calculate $\delta\rho_\alpha$, and its angular dependence does
  indeed agree with \citet{mestel1}.}. They described a
double-perturbation formalism, in the two small parameters
$\epsilon_\alpha$ and $\epsilon_B$ -- the centrifugal and magnetic
distortions to the star, respectively. Crucially, they argued that one can complete
the calculation merely by going to first order in each of the perturbations --
i.e. using the zeroth-order, $\mathcal{O}(\epsilon_\alpha)$ and
$\mathcal{O}(\epsilon_B)$ equations, whilst neglecting higher-order
cross terms whose scaling is $\epsilon_\alpha\epsilon_B$ or higher.
However, at this level one only has a single equation to fix the three
spatial components of $\bxi$: the continuity equation, which in time-integrated form is
$\delta\rho_\alpha = -\div(\rho_0\bxi)$. To obtain a second condition,
they specialised to motions with $\div\bxi=0$, effectively
appealing to an additional buoyancy force associated with
stratification; this clearly already reduces the generality of their
results. Despite this, one extra relation is still required to close the
system. In the first of the two papers from Mestel and collaborators,
\citet{mestel1}, they chose $\xi_\phi=0$, whilst in the
second one they demanded minimisation of the kinetic
energy of $\bxi$ \citep{mestel2}.

\subsection{The need for second-order perturbations}

Although Mestel and collaborators claim that their results are likely to be qualitatively
correct, they clearly used ad-hoc assumptions to close their system of
equations. This should not have been necessary, since the original problem
of a rotating fluid star was perfectly well-defined. In particular, it
is odd that the magnetic field -- by dint of which the precession was
possible to start with -- does not enter directly at all. Instead, in our approach,
we follow the original steps of \citet{mestel1} and set up a
perturbation scheme in the small parameters $\epsilon_\alpha$ and
$\epsilon_B$, but then rigorously write out all the resulting systems
of equations. By doing so, we show that the non-rigid response of the
fluid to the precessional motion -- encoded in the velocity field
$\dot\bxi$ and a perturbed magnetic field $\delta\bB$ -- only enters
at order $\epsilon_\alpha\epsilon_B$. The resulting hierarchy of
equations -- at zeroth, first and second perturbative order -- is
unsurprisingly more complex than that considered in previous work,
but we believe these equations contain the minimal information required to get a
reliable solution.

\skl{
\subsection{Key parameters for precessing magnetic stars}

Before we begin our detailed modelling of non-rigid precession, let us
pause to define the ordering of timescales necessary for this to
occur. Firstly, by inverting equation \eqref{omega} we immediately see that
the precession period $T_\omega=2\pi/\omega$ will always be significantly longer
than the primary-rotation period $T_\alpha=2\pi/\alpha$:
\beq
T_\omega=\frac{T_\alpha}{\epsilon_B\cos\chi} \gg T_\alpha.
\eeq
Note that the timescale $T_\omega$ must also be the oscillation period of
the star's non-rigid response to precession, encoded in the perturbed
fluid velocity $\dot\bxi$ and magnetic field $\delta\bB$.
In order for the star to precess in a manner analogous to a rigid
body, it is natural to assume that the characteristic magnetic-mode
crossing timescale should be short
compared with the precession period. There is some subtlety in
defining a suitable magnetic-mode timescale, however: in non-rotating
magnetic stars the relevant timescale is the Alfv\'en-mode crossing
timescale $\tau_A$, but rotation splits each Alfv\'en mode into a pair
of co- and counter-rotating magneto-inertial modes \citep{LJP10}. If
rotation is relatively unimportant, in the sense that
$\tau_A/T_\alpha$ is small, then these modes are virtually
Alfv\'en modes; if $\tau_A/T_\alpha$ is large then one branch
effectively becomes a pure inertial mode, whereas the other branch is
virtually stationary as viewed in the rotating frame, and so its
oscillation period is far longer than $\tau_A$. This agrees with the
result of \citet{levin_dang} that rotation increases magnetic-mode timescales.

For simplicity let us assume that the magneto-inertial crossing timescale is
approximately that of a pure Alfv\'en wave $\tau_A$, in which case the criterion
for precession is
\beq
\tau_A\sim\frac{R_*}{B/\sqrt{\rho}}\sim\frac{\mathcal{M}^{1/2}}{R_*^{1/2}B}
 \leq T_\omega,
\eeq
where $R_*$ is the radius and $\mathcal{M}$ the mass of the star.
We will see in the table below that this criterion is very easily
satisfied, so the inequality will also hold even for more rapidly
rotating stars where $\tau_A$ is increased by an order of magnitude or
more.

A number of different classes of star are thought to harbour
long-lived magnetic fields and to have significant rotation
rates. These are all candidates for undergoing the non-rigid
precession mechanism upon which this paper is focussed. To orientate
the reader, table \ref{mag_star_params} gives some typical parameters for these
stars, including estimates for $\tau_A$ and $T_\omega$. These are made
by assuming that the volume-averaged magnetic field strength is equal to the
surface value $B_{\textrm{surf}}$ as inferred from observations. In reality the interior
field could be considerably stronger (especially if it is dominated by
a toroidal component), so the absolute values we report for $\tau_A$
and $T_\omega$ should be taken with caution -- however, the
ordering $\tau_A\ll T_\omega$ will hold regardless. In addition, our
estimate of $T_\omega$ for a typical Ap/Bp star is in broad
agreement with the results of the calculations of \citet{mestel2} and \citet{nittmann-wood}.

\begin{table*}
\begin{center}
\caption{\label{mag_star_params}
               \skl{Key parameters and timescales for different
                 classes of potentially precessing magnetic stars.}}
\begin{threeparttable}
\begin{tabular}{ccccccc}
\hline
class of star & mass $\mathcal{M}$/g & radius $R_*$/cm & $B_\textrm{surf}$/G &
 $T_\alpha$ & estimated $\tau_A$  & estimated $T_\omega$/yr\\
\hline
O star\tnote{1}
 & $7\times 10^{34}$ & $7\times 10^{11}$ & $10^3$ & $15$ d & $10$ yr & $5\times 10^7$\\
Ap/Bp star\tnote{2}
 & $4\times 10^{33}$ & $10^{11}$ & $10^4$ & $1$ d & $0.6$ yr & $3\times 10^5$\\
magnetic white dwarf\tnote{3}
 & $2\times 10^{33}$ & $10^{9}$ & $10^8$ & $1$ h  & $0.4$ h & $3000$\\
pulsar\tnote{4}
 & $3\times 10^{33}$ & $10^{6}$ & $10^{12}$ & $0.1$ s  & $50$ s & $2000$\\
magnetar\tnote{4}
 & $3\times 10^{33}$ & $10^{6}$ & $10^{15}$ & $10$ s  & $0.05$ s & $0.2$\\
\hline
\end{tabular}
\begin{tablenotes}
\item [1] \citet{donati02}
\item [2] \citet{land92}
\item [3] \citet{ferrario15}
\item [4] \citet{turolla}
\end{tablenotes}
\end{threeparttable}
\end{center}
\end{table*}

}

\section{Fluid precession as a second-order perturbation problem}
\label{pert_scheme}

\subsection{Full equations}

For our non-axisymmetric stellar model we need to consider a very general
form of the standard equations of motion. Firstly, we have the Euler
equation, referred to a frame that rotates \emph{non-uniformly} with
angular velocity ${\mathbf \Omega}(t)$:
\beq \label{general_Euler}
\frac{\partial{\bf v}}{\partial t} + ({\bf v} \cdot \nabla) {\bf v} + 2 {\mathbf \Omega} \times {\bf v} 
 + \frac{\rmd{\mathbf{\Omega}}}{\rmd t} \times {\bf r}  +  {\mathbf \Omega} \times ({\mathbf \Omega} \times {\bf r})
=- \nabla H - \nabla \Phi + \frac{1}{4\pi\rho} (\nabla \times {\bf B}) \times {\bf B},
\eeq
where $\Phi$ is the gravitational potential and ${\bf v}$ the fluid velocity
(at this stage we have not chosen a specific rotating frame, so this is not yet
the same thing as $\dot\bxi$).
Note the presence of the \emph{Euler force} $\td{\bOm}{t}\times\br$
(present because we are allowing for non-uniform rotation),
in addition to the more familiar Coriolis and centrifugal force
terms $2\bOm\times{\bf v}$ and $\bOm\times(\bOm\times\br)$.  As we
will deal only with barotropic fluids, with equations of state of the
form $P = P(\rho)$, we have chosen to use enthalpy $H$ in preference
to the pressure $P$; the two quantities are related by
\begin{equation}
H(\br) \equiv \int_{\hat{P}=0}^{\hat{P} = P(\br)} \frac{\rmd \hat{P}}{\rho(\hat{P})} .
\end{equation}
This choice of variable will make the perturbation equations simpler.  Together with
the Euler equation we also have the continuity, Poisson and induction
equations, the equation of state, and the solenoidal constraint on the
magnetic field:
\begin{align}
\frac{\partial \rho}{\partial t} &= -\nabla \cdot (\rho{\bf v}) ,\\
\nabla^2\Phi &= 4\pi G \rho , \\
\frac{\partial {\bf B}}{\partial t} &= \nabla \times ({\bf v} \times {\bf B}) ,\\
H &= H(\rho),\\
\div\bB &= 0.
\end{align}
Using the logic of the previous section, we argue that the
motion of our star is close to that of a freely precessing rigid body.
To some approximation, then, the motion must then consist of two
superimposed rotations, one at rate $\alpha$ about $\be_z^{(\alpha)}$, and
the other at a much slower rate
$\omega=\alpha\epsilon_B\cos\chi$ about the magnetic axis $\be_z^{(B)}$, with
$\be_z^{(B)}$ tracing out a cone of half-angle $\chi$ about $\be_z^{(\alpha)}$ at a
rate $\alpha$. However, from the point of view of an observer rotating
with the body, the centrifugal bulge of size $\epsilon_\alpha$ rotates
about $\be_z^{(B)}$, again in a cone of half-angle $\chi$, at the slow precession
frequency $\omega$; recall figure \ref{alphaframe}. It is this slow density wave which produces an
Eulerian density perturbation $\delta\rho_\alpha(t)$ (see below for the precise meaning), which in turn induces
the $\bxi$-motions that encode the non-rigid response.

\subsection{Perturbative scheme}

In order to derive sets of perturbation equations, we first need to
choose which frame to work in. There are
three natural options:\\
(i) The inertial frame, $\bOm = \bf 0$.  This is conceptually simple, but in this frame both the mass and magnetic fields are time-varying.\\
(ii) The $\alpha$-frame, ${\mathbf \Omega} = \alpha \be_z^{(\alpha)} $.  In
  this frame both the mass and magnetic fields are stationary to
  lowest order, but individual fluid elements move in large circles about $\be_z^{(B)}$.\\
(iii) The $\omega$-frame,
  ${\mathbf \Omega} = \alpha \be_z^{(\alpha)} + \omega \be_z^{(B)}$.  In this frame
  the magnetic field is stationary to lowest order, but the
  $\be_z^{(\alpha)}$ axis moves about $\be_z^{(B)}$ with a
  period $2\pi/\omega$. Consequently the mass field is also
  time-varying on the $\omega$ timescale, inducing the $\bxi$-motions. 

We will work in the $\omega$-frame.  This is the frame that is closest
to the `body frame' to which the equations of rigid bodies are
conventionally referred.  One key advantage of this is that the only
velocity component in the Euler equation is $\dot\bxi$. In the
$\omega$-frame $\bOm$ is the actual angular velocity of the body
itself, which we know from our free precession ansatz.  With respect
to the magnetic-field $\be_z^{(B)}$ triad, it is given by
\begin{equation}
{\mathbf \Omega}(t) = \omega\be_z^{(B)} + \alpha [\cos\chi\be_z^{(B)} + \sin\chi(\be_x^{(B)}\cos\omega t - \be_y^{(B)}\sin\omega t) ].
\end{equation}
From this point on we will refer all quantities to the $\be_z^{(B)}$
triad rather than that associated with the primary rotation axis
$\be_z^{(\alpha)}$; and since there is no longer any ambiguity in the
axes, we will also drop the $(B)$ superscripts on $(\be_x,\be_y,\be_z)$.

We wish to set up a perturbative scheme in which we can describe the
$\bxi$-motions.  We will follow Mestel and collaborators by expanding about a
non-rotating and unmagnetised spherical background, assuming that
terms related to the stellar rotation and magnetic field are
small. The size of these terms
can be defined by the centrifugal and magnetic ellipticities
$\epsilon_\alpha$ and $\epsilon_B$, dimensionless quantities which
scale with the ratio of their respective energies to the
gravitational binding energy of the spherical background star:
\begin{align}
\epsilon_\alpha &\sim \frac{\Omega^2 R_*^3}{G\mathcal{M}}, \\
\epsilon_B &\sim \frac{B^2 R_*^4}{G\mathcal{M}^2}.
\label{epsB_scaling}
\end{align}
Although we will assume that both $\epsilon_\alpha$ and $\epsilon_B$
are separately small, we do not make any assumption about their
\emph{relative} size, so the order $\mathcal{O}$ of each is formally
different (even though they could be numerically comparable). In
particular, we will treat second-order perturbations in the most
general case, where
\beq
\mathcal{O}(\epsilon_\alpha^2)\neq \mathcal{O}(\epsilon_\alpha\epsilon_B)
 \neq \mathcal{O}(\epsilon_B^2).
\eeq
Finally, one could envisage a different perturbative scheme with $\chi$ as the small
parameter, and perturbations being performed about a background
aligned-rotator model, but we prefer to be able to allow for an
arbitrary degree of misalignment between the rotation and magnetic axes.

In obvious notation, we can now write any given quantity (e.g. the density) as a perturbative expansion of the form
\begin{equation} 
\rho = \rho_0 + \delta \rho_{\alpha} + \delta \rho_{B} + \delta \rho_{\alpha \alpha} + \delta \rho_{\alpha B} + \delta \rho_{BB} + \dots 
\end{equation}
This enables us to expand all the terms on the right-hand side of the Euler equation.  Note that for the perturbative expansion for the magnetic field itself, which is of the form
\begin{equation}
{\bf B} = \bB_0 + \delta \bB_\alpha + \delta\bB_B + \dots ,
\end{equation}
\emph{half-integer} powers of $\epsilon_B$ will occur, e.g.
\begin{equation}
\bB_0 \sim \epsilon_B^{1/2}, \hspace{10mm}   \delta\bB_\alpha \sim \epsilon_B^{1/2} \epsilon_\alpha, \hspace{10mm}
\delta\bB_B \sim \epsilon_B^{3/2}, \dots
\end{equation}

For the left-hand side of the Euler equation \eqref{general_Euler}, we
need to think about the assumptions already made for our
precessional-like motion; these fix the \emph{leading order} scalings
of various quantities. Firstly, the secondary rotation $\omega$ scales as:
\begin{equation}
\omega \sim \epsilon_B \alpha \sim \epsilon_\alpha^{1/2} \epsilon_B .
\end{equation}
The angular velocity itself can be written as
\begin{equation}
{\bf \Omega}(t) = \omega \be_z + {\bf \Omega}_\alpha ,
\end{equation}
where
\begin{equation}
\label{eq:Omega_alpha}
{\bf \Omega}_\alpha \equiv \alpha [\cos\chi \be_z + \sin\chi({\be_x}  \cos\omega t - {\be_y} \sin\omega t) ] ,
\end{equation}
so that the two pieces of $\bf \Omega$ have scalings:
\begin{align}
\omega \be_z &\sim \omega \sim \epsilon_\alpha^{1/2} \epsilon_B , \\
{\bf \Omega}_\alpha &\sim \alpha \sim \epsilon_\alpha^{1/2} .
\end{align}
We assume that, to leading order, the displacements $\bxi$ are sourced by the motion of the centrifugal bulge, so that
\begin{equation}
\xi \sim \epsilon_\alpha R_* \sim \epsilon_\alpha .
\end{equation}
To leading order, all time derivatives are due to quantities varying on the timescale of $\omega^{-1}$, so that, symbolically,
\begin{equation}
\frac{\partial}{\partial t} \rightarrow \omega \sim \epsilon_\alpha^{1/2} \epsilon_B .
\end{equation}
In particular, the leading-order piece of the fluid velocity must then scale as
\begin{equation}
\dot\xi \sim \omega \xi \sim \epsilon_\alpha^{3/2} \epsilon_B .
\end{equation}
We can then use these results to write down the scalings of the
leading-order parts of the first three terms on the left-hand side of
equation \eqref{general_Euler}:
\begin{align}
\frac{\partial {\dot\bxi}}{\partial t} &\sim \omega^2 \xi \sim \epsilon_\alpha^2 \epsilon_B^2 ,  \\
({\dot\bxi} \cdot \nabla) {\dot\bxi} &\sim (\dot\xi)^2 \sim   \epsilon_\alpha^3 \epsilon_B^2 ,  \\
{\mathbf \Omega} \times {\dot\bxi} &\sim \alpha (\epsilon_\alpha \epsilon_B \alpha)  \sim   \epsilon_\alpha^{2} \epsilon_B .
\end{align}
We will find below that all these leading-order pieces are of sufficiently high order that they will not be needed in our analysis.

Given that we have prescribed the exact form of ${\bf \Omega}(t)$ we
can compute the fourth and fifth terms on the left-hand side of
equation \eqref{general_Euler} exactly, but for now let us just note
their scalings with $\epsilon_\alpha$ and
$\epsilon_B$.   For the fourth term on the left-hand side of equation \eqref{general_Euler}, we have a leading-order piece that scales as
\begin{equation}
\frac{\rmd{\mathbf{\Omega}}}{\rmd t} \times {\bf r} \sim \alpha\omega  R_* \sim \epsilon_\alpha \epsilon_B ,
\end{equation}
which will be relevant.  For the fifth term on the left-hand side of the Euler equation, there will be a leading order piece of the form
\begin{equation}
 {\mathbf \Omega}_\alpha \times ({\mathbf \Omega}_\alpha \times {\bf r}) \sim \alpha^2 \sim \epsilon_\alpha ,
 \end{equation}
which will be needed,  pieces given by
\begin{equation}
(\omega {\mathbf e_z}) \times ({\mathbf \Omega_\alpha} \times {\bf r}) 
+   {\mathbf \Omega_\alpha} \times [(\omega {\mathbf e_z}) \times {\bf r}] \sim \omega \alpha \sim  \epsilon_\alpha \epsilon_B ,
\end{equation}
which will also be needed, and a piece
\begin{equation}
(\omega {\mathbf e_z}) \times [(\omega {\mathbf e_z}) \times {\bf r}] 
\sim \omega^2 \sim  \epsilon_\alpha \epsilon_B^2 .
\end{equation}
which will turn out to be of too high an order to be important in this paper.

Finally, to find the equation-of-state relations needed to close each
system of perturbation equations we perform a Taylor expansion of
$H=H(\rho)$ about the point $\rho=\rho_0$:
\begin{align}
H(\rho) &= H(\rho_0)+(\rho-\rho_0)\td{H}{\rho}+\frac{(\rho-\rho_0)^2}{2!}\td{^2H}{\rho^2}+\dots\nn\\
             &= H(\rho_0)+\delta\rho_\alpha\td{H}{\rho}+\delta\rho_B\td{H}{\rho}
                    +\delta\rho_{\alpha B}\td{H}{\rho} + \delta\rho_\alpha\delta\rho_B\td{^2H}{\rho^2}+\dots
\end{align}
Now comparing this with
$H=H_0+\delta H_\alpha+\delta H_B+\delta H_{\alpha B}+\dots$, we can
read off the relations for different perturbative orders.

We can now insert all of these perturbative expansions into the full
set of equations, to give sets of perturbation equations to be solved
simultaneously.  It is convenient to label these sets according to the
order in $\epsilon_\alpha$ and $\epsilon_B$ to which terms are
retained in the Euler equation.

\subsection{Zeroth-order equations}

To zeroth order in $\epsilon_\alpha$ and $\epsilon_B$, the Euler equation simply gives the hydrostatic force balance of a non-rotating unmagnetised star:
\begin{equation}
\label{eq:zeroth_Euler}
0 = -\nabla H_0 - \nabla \Phi_0 .
\end{equation}
To the same order, we have Poisson's equation
\begin{equation}
\nabla^2 \Phi_0 = 4\pi G \rho_0 ,
\end{equation}
and the equation of state
\begin{equation}
H_0 = H_0(\rho_0) .
\end{equation}
The solution is static and spherical, so that only the radial component of equation  (\ref{eq:zeroth_Euler}) is non-trivial, leaving us with three equations in the three unknowns $H_0, \Phi_0, \rho_0$, each of which will depend only upon the radial coordinate $r$.

\subsection{Order-$\epsilon_B$ equations}
\label{order-B-eqns}

To order $\epsilon_B$, the Euler equation gives the perturbation to
the spherical star caused by a magnetic field $\bB_0$:
\begin{equation}
0 = -\nabla \delta H_B - \nabla \delta  \Phi_B + \frac{1}{4 \pi \rho_0} (\nabla \times \bB_0) \times \bB_0  .
\end{equation}
To the same order, we have Poisson's equation
\begin{equation}
\nabla^2 \delta \Phi_B = 4\pi G \delta \rho_B ,
\end{equation}
the equation of state
\begin{equation}
\delta H_B = \td{H}{\rho} \delta\rho_B ,
\end{equation}
and the solenoidal constraint
\begin{equation}
\nabla \cdot \bB_0 = 0 .
\end{equation}
We have six equations in the six unknowns  $ (B_0)_r, (B_0)_\theta, (B_0)_\phi, \delta H_B, \delta \Phi_B, \delta \rho_B$. 
By assumption, the leading order piece of the magnetic field, $\bB_0$, is static and axisymmetric, so all these unknown quantities will be functions only of $r$ and $\theta$.

\subsection{Order-$\epsilon_\alpha$ equations}

To order $\epsilon_\alpha$, the Euler equation gives the perturbation to the spherical star caused by the rotation $\mathbf \Omega_\alpha$:
\begin{equation}
\label{eq:alpha_Euler}
 {\mathbf \Omega}_\alpha \times ({\mathbf \Omega}_\alpha \times {\bf r}) 
 =  -\nabla \delta H_\alpha - \nabla \delta  \Phi_\alpha  .
\end{equation}
To the same order, we have Poisson's equation
\begin{equation}
\nabla^2 \delta \Phi_\alpha = 4\pi G \delta \rho_\alpha ,
\end{equation}
and the equation of state
\begin{equation}
\delta H_\alpha = \td{H}{\rho} \delta \rho_\alpha .
\end{equation}
As can be expected given the form of equation \eqref{eq:Omega_alpha}
above, the centrifugal force term on the left hand side of the
above perturbed Euler equation is time varying, and has no particular
symmetry, so we can expect the unknown quantities $(\delta H_\alpha,
\delta \rho_\alpha, \delta \Phi_\alpha)$ to depend upon all four
coordinates $(t, r, \theta, \phi)$.  The centrifugal term on the
left-hand side of equation (\ref{eq:alpha_Euler}) can easily be
written as the gradient of a scalar, so that the equation reduces to a
single scalar equation, leaving three equations in the  three
unknowns.

\subsection{Order-$\epsilon_\alpha \epsilon_B$ equations}

To order $\epsilon_\alpha \epsilon_B$, the Euler equation gives the perturbation to the spherical star caused by the interaction between the rotation and the magnetic field.  Explicitly, we find:
\begin{equation} \label{Euler_aB}
\frac{\rmd{\mathbf \Omega_\alpha}}{\rmd t} \times {\bf r} 
+
\omega \be_z \times ({\mathbf \Omega_\alpha} \times {\bf r})
+
{\mathbf \Omega_\alpha}  \times (\omega \be_z  \times {\bf r})
 =
 -\nabla \delta H_{\alpha B}  - \nabla \delta  \Phi_{\alpha B}
 - \frac{\delta\rho_\alpha}{4\pi \rho_0^2} (\nabla \times \bB_0) \times \bB_0 
 +  \frac{1}{4\pi \rho_0} [ (\nabla \times \delta \bB_\alpha) \times \bB_0  + (\nabla \times \bB_0) \times \delta\bB_\alpha ]   .
\end{equation}
To the same order, we have Poisson's equation
\begin{equation}
\nabla^2 \delta \Phi_{\alpha B} = 4\pi G \delta \rho_{\alpha B} ,
\end{equation}
and the equation of state
\begin{equation}
\delta H_{\alpha B} = \td{H}{\rho} \delta\rho_{\alpha B} + \td{^2 H}{\rho^2} \delta \rho_{\alpha}\delta \rho_B  .
\end{equation}

To close these equations, we can make use of the  continuity and the
induction equations.  The  leading-order non-zero part of the
continuity equation contains terms which scale as $\epsilon_\alpha^{3/2}
\epsilon_B$: 
\begin{equation} \label{cont_aB}
\frac{\partial \delta \rho_\alpha}{\partial t} + \nabla \cdot (\rho_0 \dot\bxi) = 0 .
\end{equation}
The leading-order non-zero part of the induction equation contains terms that scale as $\epsilon_\alpha^{3/2} \epsilon_B^{3/2}$:
\begin{equation} \label{induct_aB}
\frac{\partial\delta\bB_\alpha}{\partial t} - \nabla \times (\dot\bxi \times \bB_0) = 0 .
\end{equation}
Together, these form a set of nine equations in the nine unknowns $(\delta \rho_{\alpha B}, \delta H_{\alpha B}, \delta \Phi_{\alpha B},\dot\bxi, \delta\bB_\alpha$).  One could add the further equation
\begin{equation}
\nabla \cdot \delta \bB_\alpha = 0 ,
\end{equation}
but this is redundant, as the induction equation guarantees that this
constraint is preserved by the evolution, so provided the initial data
is divergence free, the solution at later times will be too. Since
this is the only perturbation of $\bB$ we will need to consider, we
will drop its $\alpha$ subscript in future.

\subsection{Equations of order $\epsilon_\alpha^2$,
  $\epsilon_B^2$, and higher}

As discussed earlier, there are three different sets of second-order
perturbation equations: one set with $\mathcal{O}(\epsilon_\alpha^2)$ terms,
another with $\mathcal{O}(\epsilon_B^2)$ terms, and a third
with $\mathcal{O}(\epsilon_\alpha\epsilon_B)$ terms. We saw from the previous
subsection that the latter set encodes the non-rigid motions in which
we are interested. The reason for neglecting the other two is
\emph{not} that they are of higher order -- in fact,
$\mathcal{O}(\epsilon_\alpha\epsilon_B)$ quantities will never
numerically be the largest of the three -- but simply that they
contain more mundane information.

The order-$\epsilon_\alpha^2$
equations merely describe a correction $\delta\rho_{\alpha\alpha}$ to the result of calculating the
centrifugal bulge from the order-$\epsilon_\alpha$ equations. This
correction scales with $\alpha^4$ and therefore does not become
important except for very rapidly-rotating stellar
models. Analogously, the order-$\epsilon_B^2$ equations encode a
correction $\delta\rho_{BB}$, scaling with $B^4$, to the magnetic
distortion as given by the order-$\epsilon_B$
calculation. Here a stronger statement may be made: there is
\emph{no known star} for which the magnetic energy is anywhere near
large enough to warrant including this higher-order correction
\citep{reis09}.

Although we will not consider any higher-order quantities than those
discussed, we note that the density perturbation $\delta\rho_{\alpha B}$
should be time-dependent, and therefore through a higher-order
analogue of the continuity equation \eqref{cont_aB} will induce a velocity
perturbation of higher order than $\dot\bxi$. Similarly,
$\delta\rho_{\alpha\alpha}$ induces a velocity field, but merely a
rapid-rotation correction to $\dot\bxi$. Finally, $\delta\rho_{BB}$ is
stationary, so the associated velocity correction is identically zero.

\section{Solution of the zeroth- and first-order equations}
\label{solution_zerothfirst}

Before tackling the second-order
$\mathcal{O}(\epsilon_\alpha\epsilon_B)$ system of equations we wish
to solve, we first need results from the three lower-order systems of
equations: the zeroth-order, $\mathcal{O}(\epsilon_B)$
and $\mathcal{O}(\epsilon_\alpha)$ sets. The first two are readily
found in the literature, but the latter calculation is non-standard
and so we report it in full.

We will specialise to the case of a $\gamma=2$ polytropic equation of
state. This has the advantage that virtually the entire calculation
may be carried out analytically, and is the simplest case for which
the fluid motion is compressible. Although this polytrope better
mimics the pressure-density distribution of neutron stars than for other
classes of star, we will argue later that many features of our
solutions are likely to be applicable to a generic magnetic
star. A detailed discussion of the applicability, and limitations, of
our model may be found in section \ref{conclusions}.

\subsection{The zeroth-order equations}

The density profile of a $\gamma=2$ polytrope in hydrostatic
equilibrium is a classic result \citep{chandra39}, and is
given by
\beq \label{rho_0}
\rho_0=\rho_0(r)=\rho_c\frac{R_*\sin\brac{\Mr}}{\pi r},
\eeq
where $\rho_c$ is the central density.

\subsection{The order-$\epsilon_B$ equations}

The study of magnetic-field distributions in axisymmetric stars also
has a considerable pedigree. For purely poloidal and poloidal-toroidal
fields one needs to solve for both the magnitude and direction of the
field, encapsulated in a magnetic streamfunction \citep{chand_prend,monaghan}. Purely toroidal fields are the simplest of all, however, as
the direction is known (azimuthal, i.e. along $\be_\phi$) and one need
only determine the functional form of the magnitude. One such solution
for the toroidal field is:
\beq \label{B0}
\bB_0=B_\phi\be_\phi = \Lambda\rho_0 r\sin\theta\be_\phi,
\eeq
where $\Lambda$ is a constant governing the strength of the field
\citep{roxburgh}.
This corresponds to a Lorentz force per unit mass
$(\curl\bB_0)\times\bB_0/4\pi\rho_0$ which is the gradient of the 
following scalar \citep{LJ09}:
\beq \label{M_toroidal}
M = -\frac{\Lambda^2}{4\pi} r^2\sin^2\theta\rho_0
    = -\frac{\Lambda^2 R_*\rho_c}{4\pi^2} r\sin(\Mr)\sin^2\theta.
\eeq
Later we will also need to relate the ellipticity $\epsilon_B$ to the
coefficient $\Lambda$. It seems, however, that a closed-form
expression for $\epsilon_B$ is only possible for an incompressible
star. For our $\gamma=2$ polytrope we will need to be content with
using the scaling of $\epsilon_B$ from equation \eqref{epsB_scaling},
together with the fact that the mass and average magnetic field
$\langle B\rangle$ scale as follows:
\beq
\clM\sim\rho_c R_*^3 \ \ ,\ \ \langle B\rangle\sim\Lambda\rho_c R_*.
\eeq
The scaling of the ellipticity with stellar parameters is then:
\beq
\epsilon_B \sim \frac{\langle B\rangle{}^2 R_*^4}{G\clM^2}
                  \sim \frac{\Lambda^2}{G}.
\eeq
$\epsilon_B$ and $\Lambda^2/G$ are related by a dimensionless constant
of proportionality $k_B$, which must be determined numerically. Using the
code from \citet{LJ09}, we find that, for a $\gamma=2$ polytrope,
\beq \label{epsB}
\epsilon_B = k_B \frac{\Lambda^2}{G} \approx -0.019 \frac{\Lambda^2}{G},
\eeq
although the constant depends only weakly on the equation of state;
it is again $-0.019$ for a polytrope with $\gamma=5/3$, and is $-0.015$
for $\gamma=4/3$.

This completes the description of order-$\epsilon_B$ quantities needed
as input for the second-order perturbation equations later. In
particular, although
the quantities $\delta\rho_B,\delta H_B$ or $\delta\Phi_B$ appear
directly in the order-$\epsilon_B$ equations, we will not need
explicit forms for these.

\subsection{The order-$\epsilon_\alpha$ equations}

In the $\omega$-frame, the centrifugal distortion $\delta\rho_\alpha$ is neither stationary nor
axisymmetric. It is sourced by the centrifugal force
$\bOm_\alpha\times(\bOm_\alpha\times\br)$, whose form we 
know from our free-precession ansatz on $\bOm_\alpha$. Our first task is
to write this force as the gradient of a scalar function.

\subsubsection{The centrifugal force}

We will work in spherical polar coordinates, and begin by converting
the Cartesian expression for $\bOm_\alpha$ from equation \eqref{eq:Omega_alpha} into
spherical-polar form:
\beq
\bOm_\alpha = \alpha\Big\{
                              [\cos\chi\cos\theta+\sin\chi\sin\theta\cos(\omega t+\phi)]\be_r
                            +  [-\cos\chi\sin\theta+\sin\chi\cos\theta\cos(\omega t+\phi)]\be_\theta
                            - \sin\chi\sin(\omega t+\phi)\be_\phi
                                \Big\}.
\eeq
We use this to calculate the $r$, $\theta$ and $\phi$ components of
the centrifugal force. To express this force as the gradient of some
scalar $F_\alpha$, we compare the three components of the force with
the general expression for $\nabla F_\alpha$. After suitable
integrations we can find $F_\alpha$ explicitly and thus
write the centrifugal force as
\beq \label{fcent}
\bOm_\alpha\times(\bOm_\alpha\times\br)
 = \nabla\Big\{
         \frac{\alpha^2 r^2}{4} \Big[
                        -2\sin^2\theta
                        + \sin^2\chi(3\sin^2\theta-2)
                        + \sin(2\chi)\sin(2\theta)\cos\lambda
                        + \sin^2\chi\sin^2\theta\cos(2\lambda) \Big] \Big\},
\eeq
which reduces to the standard result for a stationary rotating star in
the limit $\chi\to 0$, as expected. In equation \eqref{fcent} we
introduced a time-shifted azimuthal coordinate
\beq
\lambda\equiv\phi+\omega t
\eeq
to simplify the expression.
Note that we may freely replace the original azimuthal coordinate
$\phi$ with the new one $\lambda$, since it only acts to redefine the
origin of time in our equations, and since derivatives are unaffected:
\beq
\pd{f(\theta,\lambda)}{\phi} = \pd{f}{\lambda}\pd{\lambda}{\phi}
 = \pd{f}{\lambda}\pd{ }{\phi}(\omega t+\phi) = \pd{f}{\lambda}
 \implies \pd{ }{\phi} = \pd{ }{\lambda}.
\eeq
The stationary background field $\bB_0$ discussed in the previous
subsection may, of course, be written equivalently in terms of
$\lambda$, i.e. $\bB_0=B_\phi\be_\phi=B_\lambda\be_\lambda$.

\subsubsection{The $\clO(\epsilon_\alpha)$ system as a single Helmholtz equation}

Having found an expression for the perturbed centrifugal force, we now
turn to the set of $\mathcal{O}(\epsilon_\alpha)$ equations. We wish to
reduce the three equations in three variables to a single equation in
one variable. Although it is equivalent, in principle, to work with a final
equation in either $\delta\rho_\alpha$, $\delta H_\alpha$ or
$\delta\Phi_\alpha$, we choose the latter -- as only in this case is it
straightforward to impose the necessary boundary conditions at the
stellar surface. For a polytropic equation of state the enthalpy may
be written as
\beq
H = \frac{\gamma P}{(\gamma-1)\rho}=\frac{\gamma k\rho^{\gamma-1}}{\gamma-1}.
\eeq
Substituting this into the $\clO(\epsilon_\alpha)$ Euler equation, we have
\beq
\bOm_\alpha\times(\bOm_\alpha\times\br)
 = -k\gamma\delta\rho_\alpha\nabla(\rho^{\gamma-2})
     -k\gamma\rho^{\gamma-2}\nabla\delta\rho_\alpha
      - \nabla\delta\Phi_\alpha.
\eeq
This is clearly particularly simple for our chosen polytropic index,
$\gamma=2$. Together with the result of \eqref{fcent} the equation becomes
\beq
\nabla\Big\{
         \frac{\alpha^2 r^2}{4} \Big[
                        -2\sin^2\theta
                        + \sin^2\chi(3\sin^2\theta-2)
                        + \sin(2\chi)\sin(2\theta)\cos\lambda
                        + \sin^2\chi\sin^2\theta\cos(2\lambda) \Big] \Big\}
= -2k\nabla\delta\rho_\alpha-\nabla\delta\Phi_\alpha,
\eeq
whose first integral is
\beq
C+\frac{\alpha^2 r^2}{4} \Big[ -2\sin^2\theta + \sin^2\chi(3\sin^2\theta-2)
                                            + \sin(2\chi)\sin(2\theta)\cos\lambda
                                            + \sin^2\chi\sin^2\theta\cos(2\lambda) \Big]
 = -2k\delta\rho_\alpha-\delta\Phi_\alpha,
\eeq
where $C$ is an integration constant. Finally, we use the Poisson
equation in the above to replace $\delta\rho_\alpha$, to give:
\beq \label{helm_inhom}
\brac{\nabla^2+\frac{2\pi G}{k}}\delta\Phi_\alpha
 = -\frac{2\pi G}{k}\Big\{ C + \frac{\alpha^2 r^2}{4} \Big[ -2\sin^2\theta + \sin^2\chi(3\sin^2\theta-2)
                                            + \sin(2\chi)\sin(2\theta)\cos\lambda
                                            + \sin^2\chi\sin^2\theta\cos(2\lambda) \Big] \Big\}.
\eeq
This is an inhomogeneous Helmholtz equation, which we solve next,
beginning with the homogeneous solution.

\subsubsection{Homogeneous solution}

The general solution of the homogeneous Helmholtz equation
\beq
\brac{\nabla^2+\frac{2\pi G}{k}}\delta\Phi_\alpha^\textrm{homog} = 0
\eeq
is a standard result \citep{arfken}, which in spherical polar
coordinates may be written as the infinite sum
\beq
\delta\Phi_\alpha^\textrm{homog}
 = \sum\limits_{l=0}^{\infty} \sum\limits_{m=-l}^{l}
      b_l^m j_l\brac{\scriptstyle{\sqrt{\frac{2\pi G}{k}}r}} Y_l^m(\theta,\lambda),
\eeq
where $b_l^m$ are constants, $j_l$ are spherical Bessel functions and
$Y_l^m$ spherical harmonics.

\subsubsection{Particular solution}

We begin by decomposing the various angular pieces of the centrifugal
potential \eqref{fcent} into spherical harmonics:
\begin{align}
1 &= 2\sqrt{\pi} Y_0^0,\\
\sin^2\theta
 &= \frac{2}{3}\brac{1 - P_2^0} = -\frac{4\sqrt{\pi}}{3}\brac{Y_0^0-\frac{1}{\sqrt{5}}Y_2^0},\\
\sin(2\theta)\cos\lambda
 &= \sin(2\theta)\brac{\frac{\rme^{i\lambda}}{2}+\frac{\rme^{-i\lambda}}{2}}
    = -\frac{1}{3}P_2^1 \rme^{i\lambda} + 2P_2^{-1}\rme^{-i\lambda}
    = -2\sqrt{\frac{2\pi}{15}}\brac{Y_2^1 - Y_2^{-1}},\label{Y21}\\
\sin^2\theta\cos(2\lambda)
 &= \sin^2\theta\brac{\frac{\rme^{2i\lambda}}{2}+\frac{\rme^{-2i\lambda}}{2}}
    = \frac{1}{6}P_2^2 \rme^{2i\lambda} + 4P_2^{-2}\rme^{-2i\lambda}
    = 4\sqrt{\frac{\pi}{30}}\brac{Y_2^2 + Y_2^{-2}}. \label{Y22}
\end{align}
Using the above relations and equation \eqref{fcent}, the right-hand
side of the inhomogeneous Helmholtz equation \eqref{helm_inhom} may,
therefore, be written as a sum over $l=0,2$ and $m=0,\pm 2$:
\beq \label{RHS_inhom}
-\frac{2\pi^{3/2} G\alpha^2 r^2}{k}
    \Bigg[  2\brac{\frac{C}{\alpha^2 r^2}-\frac{1}{3}}Y_0^0
               +\frac{1}{\sqrt{5}}\brac{\frac{2}{3}-\sin^2\chi}Y_2^0
               -\frac{\sin(2\chi)}{\sqrt{30}}(Y_2^1-Y_2^{-1})
               +\frac{\sin^2\chi}{\sqrt{30}}(Y_2^2-Y_2^{-2})
         \Bigg]
\eeq
and so we expect the same angular structure for the particular
solution $\delta\Phi_\alpha^\textrm{PS}$:
\beq
\delta\Phi_\alpha^\textrm{PS}
 = \delta\Phi_0^0(r)Y_0^0 + \delta\Phi_2^0(r)Y_2^0 
     + \delta\Phi_2^1(r)Y_2^1 + \delta\Phi_2^{-1}(r)Y_2^{-1} 
     + \delta\Phi_2^2(r)Y_2^2 + \delta\Phi_2^{-2}(r)Y_2^{-2}.
\eeq
Now, using
\beq
\nabla^2(f(r)Y_l^m)
 = f(r)\nabla^2 Y_l^m +
         Y_l^m\brac{\td{^2}{r^2}+\frac{2}{r}\td{ }{r}}f(r)
\eeq
and
\beq
\nabla^2 Y_l^m = -\frac{l(l+1)}{r^2}Y_l^m
\eeq
we rewrite the left-hand side of \eqref{helm_inhom}:
\beq \label{LHS_inhom}
\brac{\nabla^2+\frac{2\pi G}{k}}\delta\Phi_\alpha^\textrm{PS}
 = \sum\limits_{l=0}^2\sum_{m=-l}^l
        \brac{\td{^2}{r^2}+\frac{2}{r}\td{ }{r}+\frac{2\pi G}{k}-\frac{l(l+1)}{r^2}}\delta\Phi_l^m(r)Y_l^m.
\eeq
Next, we solve for the radial functions $\delta\Phi_l^m$ by equating
the different $Y_l^m$ terms from the left- and right-hand
sides of the inhomogeneous Helmholtz equation, i.e. the terms from
\eqref{RHS_inhom} with their counterparts from equation
\eqref{LHS_inhom}. Before continuing, we recall that our final aim is
to calculate time-dependent quantities in a precessing, magnetised
fluid star: the perturbations to the velocity and magnetic field
which oscillate over a precession timescale. However, two of the six
terms in $\delta\Phi_\alpha$ are stationary:
\beq
\pd{\delta\Phi_0^0}{t} = \pd{\delta\Phi_2^0}{t} = 0,
\eeq
so we need not solve for these terms, and will drop them in the rest
of the calculation. Next we turn to the time-dependent terms. Equating components of
$Y_2^1$ we have
\beq
\brac{\td{^2}{r^2}+\frac{2}{r}\td{ }{r}+\frac{2\pi G}{k}-\frac{6}{r^2}}\delta\Phi_2^1
  = \frac{2\sin(2\chi)\pi^{3/2} G\alpha^2 r^2}{\sqrt{30} k}.
\eeq
To solve this, we make the ansatz that $\delta\Phi_2^1$ is a quadratic
function of $r$, finding that its constant and linear pieces are zero,
so that
\beq
\delta\Phi_2^1 = \sqrt{\frac{\pi}{30}} \sin(2\chi)\alpha^2 r^2.
\eeq
A similar procedure applied to the other components gives
\beq
\delta\Phi_2^{-1} = -\delta\Phi_2^1 = -\sqrt{\frac{\pi}{30}} \sin(2\chi)\alpha^2 r^2
\ \ ,\ \ 
\delta\Phi_2^2 = \delta\Phi_2^{-2} = -\sqrt{\frac{\pi}{30}} \sin^2\chi\alpha^2 r^2.
\eeq

\subsubsection{Full interior solution}

The full solution for the perturbed gravitational potential inside the
star $\delta\Phi_\alpha^\textrm{int}$ -- with yet-to-be-determined coefficients $b_l^m$ -- is given by
the sum of the particular solution and the homogeneous solution. For
the latter, we need only consider the terms in the infinite sum with
the same multipolar order as the time-dependent pieces in the
particular solution (again, since $\delta\Phi_\alpha$ is
time-dependent by ansatz):
\begin{align}
\delta\Phi_\alpha^\textrm{int}
 &= \brac{b_2^1j_2\brac{\scriptstyle{\sqrt{\frac{2\pi G}{k}}}r}
               + \sqrt{\frac{\pi}{30}}\sin(2\chi)\alpha^2 r^2 }Y_2^1 
   + \brac{b_2^{-1}j_2\brac{\scriptstyle{\sqrt{\frac{2\pi G}{k}}}r}
               - \sqrt{\frac{\pi}{30}}\sin(2\chi)\alpha^2 r^2 }Y_2^{-1}\nn\\
 &\ \ \ 
   + \brac{b_2^2j_2\brac{\scriptstyle{\sqrt{\frac{2\pi G}{k}}}r}
               - \sqrt{\frac{\pi}{30}}\sin^2\chi \alpha^2 r^2 }Y_2^2 
   + \brac{b_2^{-2}j_2\brac{\scriptstyle{\sqrt{\frac{2\pi G}{k}}}r}
               - \sqrt{\frac{\pi}{30}}\sin^2\chi \alpha^2 r^2 }Y_2^{-2}.  
\end{align}

\subsubsection{Exterior solution and boundary conditions}

In order to fix the constants $b_l^m$ for the interior gravitational
potential, we need to match the interior and exterior gravitational
potentials appropriately at the stellar surface.
The exterior gravitational potential $\delta\Phi_\alpha^\textrm{ext}$ is governed
by Laplace's equation:
\beq
\nabla^2\delta\Phi_\alpha^\textrm{ext}=0.
\eeq
Discarding unphysical terms which diverge at infinity, and keeping
only those multipoles which feature in the interior solution, we are
left with
\beq
\delta\Phi_\alpha^\textrm{ext}
 = \frac{1}{r^3}\brac{ c_2^1 Y_2^1 + c_2^{-1} Y_2^{-1} + c_2^2 Y_2^2 + c_2^{-2} Y_2^{-2} }.
\eeq
We need to match the four multipoles above to the corresponding ones
in the interior solution at the stellar surface $R_*$, such that
\begin{align}
\delta\Phi_\alpha^\textrm{int}|_{r=R*} &=\delta\Phi_\alpha^\textrm{ext}|_{r=R*},\\
\nabla\delta\Phi_\alpha^\textrm{int}|_{r=R*} &=\nabla\delta\Phi_\alpha^\textrm{ext}|_{r=R*}.
\end{align}
Since the angular pieces of both interior and exterior solutions are
the same $Y_l^m$ functions, imposing the boundary conditions
reduces to matching the radial functions of interior and exterior
solutions, and the radial derivatives of these. To perform this
matching, we now use the fact that $\epsilon_\alpha\ll 1$ and
$|\epsilon_B|\ll 1$, so that it is legitimate to assume the stellar
surface for the time-independent background is spherical. For a
$\gamma=2$ polytrope in hydrostatic equilibrium, the stellar surface
is located at a radius \citep{chandra39} of
\beq
R_* = \sqrt{\frac{\pi k}{2G}}.
\eeq
Matching the interior and exterior solutions at this radius, we find
that the four constants in the interior gravitational potential
solution are:
\begin{align}
b_2^1  &= -b_2^{-1} 
           = -\sqrt{\frac{5}{6}}\frac{\pi^{3/2}k\alpha^2}{2G}\sin(2\chi),\\
b_2^2  &= b_2^{-2}
           = \sqrt{\frac{5}{6}}\frac{\pi^{3/2}k\alpha^2}{2G}\sin^2\chi.
\end{align}

\subsubsection{The first-order perturbation in the density}

Finally, we may use Poisson's equation and our expression for
$\delta\Phi_\alpha^\textrm{int}$ to find the time-dependent piece of
the centrifugal bulge, and therefore of the mass distribution of our
precessing star:
\beq \label{delta_rho_alpha}
\delta\rho_\alpha
 = -\frac{5\pi\alpha^2}{16 G}
       j_2\brac{\Mr}
       \brac{ \sin(2\chi)\sin(2\theta)\cos\lambda + \sin^2\chi\sin^2\theta\cos(2\lambda) },
\eeq
where we have used the relation $R_*=\sqrt{\pi k/2G}$ to simplify the
result and replace all instances of the polytropic constant $k$. The angular
dependence of this result is in agreement with that of
\citet{mestel1}, but unlike their expression we have calculated the
explicit radial dependence (for the specific case of a $\gamma=2$ polytrope).

\section{Solution of the second-order equations}
\label{solution_second}

We are interested in the time-dependent velocity and magnetic fields
that are generated in a precessing magnetised fluid star. Looking at
the system of second-order equations
\eqref{Euler_aB}-\eqref{induct_aB}, however, we see that
higher-order fluid terms (e.g. $\delta H_{\alpha B}$) are also
present. These quantities represent the tiny distortion to
the star induced by the perturbed Lorentz force, which we are not
interested in solving for. We can remove them and simplify the system
of equations by taking the curl of the Euler 
equation \eqref{Euler_aB}; the equation of state then becomes
redundant, as does the Poisson 
equation. The centrifugal-force terms from the Euler equation also
vanish, since these may be written as the gradient of a scalar:
\beq
(\omega\be_z)\times(\bOm_\alpha\times\br)
 + \bOm_\alpha\times((\omega\be_z)\times\br)
 = \nabla\Big[\frac{\alpha\omega r^2}{2}
                          (\sin\chi\sin(2\theta)\cos\lambda-2\cos\chi\sin^2\theta)\Big].
\eeq
The remaining set of perturbation equations reads:
\begin{align} \label{curled_second-order}
\curl\Big\{ \td{\bOm_\alpha}{t} \times {\bf r} \Big\}
 &=  \curl\Big\{
- \frac{\delta\rho_\alpha}{4\pi{\rho_0}^2} (\nabla \times \bB_0) \times \bB_0
 +  \frac{1}{4\pi\rho_0} [ (\nabla \times \delta\bB) \times \bB_0  + (\nabla \times\bB_0) \times\delta\bB ]
                \Big\},\\
\pd{\delta\rho_\alpha}{t} &= -\div(\rho_0\dot\bxi),\\
\pd{\delta\bB}{t} &= \nabla \times ({\dot\bxi} \times \bB_0)  ,\\
\div\delta\bB &= 0.
\end{align}
Now, the curled Euler equation has only the three components of
$\delta\bB$ as unknown variables. Because each term of the equation is
divergence-free, however (since the divergence of a curl is zero), it
may be expressed as the sum of a poloidal and a toroidal piece.
Thus this curled Euler equation only involves two degrees of freedom, not
three; it must be reducible to one scalar equality governing the various
poloidal components, and another for the toroidal components. Together with the
solenoidal nature of $\delta\bB$, we get a well-defined system of
three equations in three unknowns (the components of $\delta\bB$):
\begin{align} \label{curled_Euler_pol}
\Big[\curl\Big\{ \td{\bOm_\alpha}{t} \times {\bf r} 
                           +\frac{\delta\rho_\alpha}{4\pi\rho_0^2} (\nabla \times\bB_0) \times \bB_0\Big\}\Big]_{\rm{poloidal}}
 &=  \Big[\curl\Big\{
            \frac{1}{4\pi\rho_0} [ (\nabla \times \delta\bB) \times \bB_0  + (\nabla \times\bB_0) \times\delta\bB ]
                \Big\}\Big]_{\rm{poloidal}},\\
\Big[\curl\Big\{ \td{\bOm_\alpha}{t} \times {\bf r} 
                           +\frac{\delta\rho_\alpha}{4\pi\rho_0^2} (\nabla \times\bB_0) \times \bB_0\Big\}\Big]_{\rm{toroidal}}
 &=  \Big[\curl\Big\{
            \frac{1}{4\pi\rho_0} [ (\nabla \times \delta\bB) \times \bB_0  + (\nabla \times\bB_0) \times\delta\bB ]
                \Big\}\Big]_{\rm{toroidal}},\label{curled_Euler_tor}\\
\div\delta\bB &= 0.
\end{align}
In the full set of Euler equations, the divergence-free nature of
$\delta\bB$ did not count as an equation, merely a constraint, giving
an initial condition $\div\delta\bB(t=0)=0$. The induction equation
would then ensure $\div\delta\bB=0$ for all time. In the
above system we have no induction equation, so $\div\delta\bB=0$ is
elevated to the role of an equation in its own right. This is also
the case in, for example, hydromagnetic equilibrium equations;
cf. section \ref{order-B-eqns}.

Now assume we have solved the above for $\delta\bB$. We have two equations left over -- the continuity and induction
equations:
\begin{align} \label{continuity-xi}
\pd{\delta\rho_\alpha}{t} &= -\div(\rho_0\dot\bxi) ,\\
\pd{\delta\bB}{t} &= \nabla \times ({\dot\bxi} \times \bB_0) .
 \label{induction-xi}
\end{align}
These represent three equations (not four, since the induction
equation again only represents two equations -- a poloidal and a
toroidal degree of freedom) in three unknowns: the components of the
velocity $\dot\bxi$. At this stage, having already found the perturbed
magnetic field $\delta\bB$, one then has enough information to use
the continuity and induction equations to obtain a solution
for $\dot\bxi$; see section \ref{xi-motions}.

The above gives us a strategy to solve the second-order perturbation
equations, but to proceed we need to find explicit forms of the
schematically-written equations \eqref{curled_Euler_pol} and
\eqref{curled_Euler_tor}. To this end, we first describe a general
decomposition of a solenoidal field.

\subsection{Preliminaries on solenoidal fields}
\label{solenoidal-fields}

If a vector field ${\bf F}$ is divergence-free, it has one fewer degree
of freedom than its number of dimensions -- in the case of a
three-dimensional star, it has two degrees of freedom. These are the poloidal and toroidal
components of the vector field. These two components may be written in
terms of the gradients of scalar functions\footnote{The poloidal-field
  scalar function is conventionally denoted $\Phi$; we use $\Upsilon$
  instead, to avoid confusion with the gravitational potential.} $\Psi$ and $\Upsilon$, in what
is sometimes known as the Mie representation of the vector field (see,
e.g., \citet{backus}):
\beq \label{Mie_rep}
{\bf F} = \underbrace{\curl(\Psi\br)}_{\text{toroidal}}
              + \underbrace{\curl(\nabla\Upsilon\times\br)}_{\text{poloidal}}.
\eeq
Note that unlike the special case of axisymmetric fields, we cannot generally
orientate our axes so that the toroidal unit vector is aligned with
the azimuthal direction. Now, if we have an equation involving a number of different poloidal and
toroidal vector fields, we can separate the equation into one equality
governing the toroidal fields and another for the poloidal
fields. Furthermore, we can reduce each of these equations to a
relation involving just the poloidal/toroidal scalar functions,
e.g. if ${\bf P}_1, {\bf P}_2, {\bf P}_3$ are three poloidal fields then
\begin{align}
{\bf P}_1 + {\bf P}_2 = {\bf P}_3
  &\implies \curl(\nabla\Upsilon_1\times\br)
                 +  \curl(\nabla\Upsilon_2\times\br) 
                 =  \curl(\nabla\Upsilon_3\times\br)\nn\\
  &\implies  \curl(\nabla[\Upsilon_1+\Upsilon_2-\Upsilon_3]\times\br)=0\nn\\
  &\implies  \Upsilon_1+\Upsilon_2-\Upsilon_3 = \mathcal{A},
\end{align}
using the distributive properties of the cross product and curl. $\mathcal{A}$ represents the gauge freedom that one may add on
any terms to the solution which satisfy
\beq
\curl(\nabla\mathcal{A}\times\br)=0.
\eeq
One may equate the scalar functions of toroidal fields in an analogous
manner; in this case one may add to the solution any term $\mathcal{B}$ which
satisfies
\beq
\nabla\mathcal{B}\times\br = 0.
\eeq

\subsection{An explicit toroidal-poloidal split of the curled Euler
  equation}
\label{explicit_torpol}

Our first task is to find explicit expressions for the poloidal and
toroidal terms in the curled Euler equations,
\eqref{curled_Euler_pol} and \eqref{curled_Euler_tor}, by comparing the terms with the Mie
representation \eqref{Mie_rep}.

Firstly, note that the time derivative of the perturbed angular
velocity may be written as the gradient of a scalar:
\beq \label{dOm_dt}
\td{\bOm_\alpha}{t} = \nabla\brac{-\alpha\omega\sin\chi r\sin\theta\sin\lambda}
     = \nabla\brac{-\alpha^2\sin\chi\cos\chi\epsilon_B r\sin\theta\sin\lambda}.
\eeq
The curl of the Euler force $\curl(\textrm{d}\bOm_\alpha/\textrm{d}t \times\br)$ is therefore 
a purely poloidal vector.

The next term from \eqref{curled_Euler_pol} and \eqref{curled_Euler_tor} involves $\delta\rho_\alpha$ and the background Lorentz
force. We may simplify this using the result that the Lorentz force
divided by the density is the gradient of a scalar $M$:
\beq
\curl\Big\{ \frac{\delta\rho_\alpha}{4\pi{\rho_0^2}}
        (\curl\bB_0) \times \bB_0 \Big\}
 = \curl\brac{\frac{\delta\rho_\alpha}{\rho_0}\nabla M}
 = \nabla\brac{\frac{\delta\rho_\alpha}{\rho_0}} \times \nabla M.
\eeq
Next expand the two gradients in components:
\begin{align}
\nabla\brac{\frac{\delta\rho_\alpha}{\rho_0}} \times \nabla M
 &= \brac{ \pd{(\delta\rho_\alpha/\rho_0)}{r}\be_r
               + \frac{1}{r}\pd{(\delta\rho_\alpha/\rho_0)}{\theta}\be_\theta
               + \frac{1}{r\sin\theta}\pd{(\delta\rho_\alpha/\rho_0)}{\lambda}\be_\lambda}
         \times 
            \brac{ \pd{M}{r}\be_r + \frac{1}{r}\pd{M}{\theta}\be_\theta }\nn\\
 &= -\frac{1}{r^2\sin\theta}\pd{M}{\theta}\pd{(\delta\rho_\alpha/\rho_0)}{\lambda}\be_r
       + \frac{1}{r\sin\theta}\pd{M}{r}\pd{(\delta\rho_\alpha/\rho_0)}{\lambda}\be_\theta
       + \frac{1}{r} \brac{ \pd{M}{\theta}\pd{(\delta\rho_\alpha/\rho_0)}{r} 
                                     -\pd{M}{r}\pd{(\delta\rho_\alpha/\rho_0)}{\theta} }\be_\lambda.
\label{gradrho_gradM}
\end{align}

We now have to extract the poloidal and toroidal contributions to this
vector.  To do so, first note that a toroidal field cannot have any
component in the $\be_r$-direction, and therefore the $\be_r$
component of any solenoidal vector ${\bf S}$ must come from the vector's
poloidal piece. Next, we use vector identities to identify the
relation between the poloidal-field scalar $\Upsilon$ and the
$r$-component of ${\bf S}$:
\begin{align}
\curl(\nabla\Upsilon\times\br)
 &= (\div\br)\nabla\Upsilon-\br\nabla^2\Upsilon+(\br\cdot\nabla)\nabla\Upsilon
       -(\nabla\Upsilon\cdot\nabla)\br\nn\\
 &= 2\nabla\Upsilon-\br\nabla^2\Upsilon+r\pd{ }{r}(\nabla\Upsilon),
\end{align}
and therefore
\beq
S_r=\big[ 2\nabla\Upsilon-\br\nabla^2\Upsilon+r\pd{ }{r}(\nabla\Upsilon)\big]_r
     = -\frac{1}{r\sin\theta}\pd{ }{\theta}\brac{\sin\theta\pd{\Upsilon}{\theta}}
         - \frac{1}{r\sin^2\theta}\pd{^2\Upsilon}{\lambda^2}.
\eeq
Decomposing $\Upsilon$ and $S_r$ into sums of spherical
harmonics, $\Upsilon=\sum_{l,m} \Upsilon_l^m Y_l^m\ ,\ S_r=\sum_{l,m} G_l^m Y_l^m$, we can
then identify the spherical-harmonic components of
$\Upsilon$ from those of $S_r$ by using: 
\beq
S_r  = -\sum\limits_{l=0}^\infty \sum\limits_{m=-l}^l
             \Upsilon_l^m(r) r\nabla^2 Y_l^m
       = \sum\limits_{l=0}^\infty \sum\limits_{m=-l}^l
                \frac{l(l+1) \Upsilon_l^m(r)}{r} Y_l^m,
\eeq
i.e.
\beq \label{S_r}
\Upsilon_l^m = \frac{r}{l(l+1)} G_l^m\ \  (\forall\ l,m).
\eeq
To apply these results to the vector
$\nabla(\delta\rho_\alpha/\rho_0)\times\nabla M$, we need an explicit
expression for its $r$-component. Using the expressions from
\eqref{rho_0},\eqref{M_toroidal},\eqref{delta_rho_alpha}, we have
\beq \label{poloidal_rcomp}
\Big[\nabla\brac{\frac{\delta\rho_\alpha}{\rho_0}}\times\nabla M \Big]_r
  = -\frac{1}{r^2\sin\theta}\pd{M}{\theta}\pd{(\delta\rho_\alpha/\rho_0)}{\lambda}
  = \frac{5\Lambda^2\alpha^2}{32G}  j_2\brac{\Mr}
         \sin(2\theta)
         \big( \sin(2\chi)\cos\theta\sin\lambda 
                  + \sin^2\chi\sin\theta\sin(2\lambda) 
             \big).
\eeq
In order to use \eqref{S_r} to find the radial functions $\Upsilon_l^m$ of the
poloidal-field scalar, we must convert the angular terms in the above
expression to spherical harmonics:
\begin{align}
\sin(2\theta)\cos\theta\sin\lambda
 = 2(\sin\theta-\sin^3\theta)\sin\lambda
   &= -\frac{1}{5i}(\textstyle{\frac{2}{3}}P_3^1+P_1^1)(\rme^{i\lambda}-\rme^{-i\lambda})\nn\\
   &= -\frac{1}{5i}( P_1^1\rme^{i\lambda}+2P_1^{-1}\rme^{-i\lambda}
                                +\textstyle{\frac{2}{3}}P_3^1\rme^{i\lambda}+8P_3^{-1}\rme^{-i\lambda})\nn\\
   &= -\frac{2\sqrt{\pi}}{5i}\brac{
                  \textstyle{\sqrt{\frac{2}{3}}}(Y_1^1+Y_1^{-1}) 
                  +\textstyle{\frac{4}{\sqrt{21}}}(Y_3^1+Y_3^{-1}) }. \label{Y1131}\\
\sin(2\theta)\sin\theta\sin(2\lambda)
   &= -\frac{1}{15i}P_3^2(\rme^{2i\lambda}-\rme^{-2i\lambda})\nn\\
   &= -\frac{1}{15i}(P_3^2\rme^{2i\lambda}-120P_3^{-2}\rme^{-2i\lambda})\nn\\
   &= \frac{4\sqrt{\pi}}{15i}\sqrt{\frac{30}{7}}( Y_3^2-Y_3^{-2} ). \label{Y32}
\end{align}
Substituting these relations into \eqref{poloidal_rcomp}, we can use
\eqref{S_r} to show that
\begin{align}
\Upsilon_1^1=\Upsilon_1^{-1}   &= \frac{i\pi^{1/2}\Lambda^2\alpha^2\sin(2\chi)}{16\sqrt{6}G}
                                       r j_2\brac{\frac{\pi r}{R_*}}, \\
\Upsilon_3^1=\Upsilon_3^{-1 }  &= \frac{i\pi^{1/2}\Lambda^2\alpha^2\sin(2\chi)}{48\sqrt{21}G}
                                       r j_2\brac{\frac{\pi r}{R_*}}, \\
\Upsilon_3^2=-\Upsilon_3^{-2} &= -\frac{i\pi^{1/2}\sqrt{10}\Lambda^2\alpha^2\sin^2\chi}{96\sqrt{21}G}
                                       r j_2\brac{\frac{\pi r}{R_*}}.
\end{align}
Finally, then, the scalar function $\Upsilon$ is given by
\begin{align} \label{phi_deltalor}
\Upsilon &= \Upsilon_1^1 Y_1^1 + \Upsilon_1^{-1} Y_1^{-1} +  \Upsilon_3^1 Y_3^1 + \Upsilon_3^{-1} Y_3^{-1}
            +  \Upsilon_3^2 Y_3^2 + \Upsilon_3^{-2} Y_3^{-2} \nn\\
       &= \frac{5\alpha^2\Lambda^2}{192G} r j_2(\Mr)
              \brac{ \sin(2\chi)\sin\theta(1+\cos^2\theta)\sin\lambda
                         + \sin^2\chi\sin^2\theta\cos\theta\sin(2\lambda) }
              + \mathcal{A},
\end{align}
where $\mathcal{A}$ represents the gauge freedom in the solution; see
section \ref{solenoidal-fields}.
From this we may now calculate the complete poloidal vector
$\curl(\nabla\Upsilon\times\br)$. The toroidal vector is then given by the
difference ${\bf T}$ between the full expression \eqref{gradrho_gradM} and the
poloidal vector:
\beq
\nabla\Psi\times\br
    = {\bf T} \equiv \nabla\brac{\frac{\delta\rho_\alpha}{\rho_0}}\times\nabla M
               - \curl(\nabla\Upsilon\times\br),
\eeq
where here -- and for the rest of the derivation -- we suppress the
lengthy explicit expressions for
$\nabla(\delta\rho_\alpha/\rho_0)\times\nabla M$ and its poloidal component.
The $\theta$ and $\lambda$ components of this equation give us
expressions for $\Psi$:
\begin{align}
\frac{1}{\sin\theta}\pd{\Psi}{\lambda} = T_\theta
 &\implies \Psi=\int \sin\theta T_\theta\ \rmd\lambda + f(r,\theta),\\
-\pd{\Psi}{\theta} = T_\lambda
 &\implies \Psi=-\int T_\lambda\ \rmd\theta + g(r,\lambda).
\end{align}
From this we identify a function $h(r,\theta,\lambda)$ common to both
expressions for $\Psi$, and deduce that the full expression for $\Psi$
is 
\beq
\Psi=h(r,\theta,\lambda)+f(r,\theta)+g(r,\lambda).
\eeq
After considerable rearrangement and simplification,
we find that $\Psi$ is given by
\begin{align}
\Psi &= \frac{5\alpha^2\Lambda^2}{384G}
                \Big[ (\Mr j_1-j_2)(2-3\sin^2\theta) + 6\brac{1+\Mr\cot(\Mr)}j_2\sin^2\theta \Big]
               (\sin(2\chi)\sin(2\theta)\cos\lambda+\sin^2\chi\sin^2\theta\cos(2\lambda))\nn\\
  &\ \ \  + \frac{5\alpha^2\Lambda^2\sin^2\chi}{3072G}(j_2-\Mr j_1) + \mathcal{B}.
\end{align}
We now exploit the `gauge freedom' of $\Psi$ by choosing the function
$\mathcal{B}$ as follows:
\beq
\mathcal{B} = -\frac{5\alpha^2\Lambda^2\sin^2\chi}{3072G}(j_2-\Mr j_1).
\eeq
This clearly satisfies the requirement that
$\nabla\mathcal{B}\times\br=0$ (from section \ref{solenoidal-fields}),
since $\nabla\mathcal{B}$ must be parallel to $\br$. Therefore we may add
this onto the original expression for $\Psi$, cancelling the final
term:
\beq \label{psi_deltalor}
\Psi+\mathcal{B} = \frac{5\alpha^2\Lambda^2}{384G}
                \Big[ (\Mr j_1-j_2)(2-3\sin^2\theta) + 6\brac{1+\Mr\cot(\Mr)}j_2\sin^2\theta \Big]
               (\sin(2\chi)\sin(2\theta)\cos\lambda+\sin^2\chi\sin^2\theta\cos(2\lambda))+\mathcal{C}.
\eeq
Note that one is still free to add additional terms $\mathcal{C}$
satisfying the condition $\nabla\mathcal{C}\times\br=0$.
To summarise, we have found scalar functions -- $\Upsilon$ from \eqref{phi_deltalor}
and $\Psi+\mathcal{B}$ from \eqref{psi_deltalor} -- representing the two degrees of freedom of
$\curl[\delta\rho_\alpha(\curl\bB_0)\times\bB_0/4\pi\rho_0^2]$, one of
the three pieces of the perturbed Lorentz force. We can
now replace the schematic poloidal and toroidal pieces of this quantity in equations
\eqref{curled_Euler_pol} and \eqref{curled_Euler_tor} with explicit
results in terms of scalar functions.

The final terms in the curled Euler equations are the two other
pieces of the perturbed Lorentz force. Again, we know that the curl of
these force terms must be solenoidal and therefore expressible in
terms of \emph{another} pair of poloidal and toroidal scalars:
\beq \label{curldL}
\curl\Big\{\frac{(\curl\delta\bB)\times\bB_0+(\curl\bB_0)\times\delta\bB}{4\pi\rho_0}\Big\}
 = \curl(\Psi_{\delta\mathcal{L}}\br) + \curl(\nabla\Upsilon_{\delta\mathcal{L}}\times\br).
\eeq
These scalars involve the unknown $\delta\bB$, but they can
nonetheless now be calculated explicitly, since we know all the other
terms in equations \eqref{curled_Euler_pol} and \eqref{curled_Euler_tor}. Equating the terms
under the curl, using the results \eqref{dOm_dt}, \eqref{phi_deltalor}
and \eqref{psi_deltalor}:
\begin{align}
\Upsilon_{\delta\mathcal{L}}
 &= -\alpha\omega\sin\chi r\sin\theta\sin\lambda+\Upsilon\nn\\
 &= \frac{\alpha^2\Lambda^2\sin\chi}{G} r\sin\theta\sin\lambda
             \Big\{ \textstyle{\frac{5}{192}}j_2(\Mr) \big[2\cos\chi(1+\cos^2\theta)+\sin\chi\sin(2\theta)\cos\lambda\big]
                         -k_B \cos\chi \Big\} + \mathcal{A} ,\label{Upsilon_dL}\\
\Psi_{\delta\mathcal{L}}
 &= \Psi +\mathcal{B}\nn\\
 &= \frac{5\alpha^2\Lambda^2}{384G}
                \Big[ (\Mr j_1-j_2)(2-3\sin^2\theta) + 6\brac{1+\Mr\cot(\Mr)}j_2\sin^2\theta \Big]
               (\sin(2\chi)\sin(2\theta)\cos\lambda+\sin^2\chi\sin^2\theta\cos(2\lambda))  + \mathcal{C},
\label{Psi_dL}
\end{align}
where we have used the scaling of $\epsilon_B$ taken from
\eqref{epsB} to simplify the first equation.

\subsection{Equations for $\delta\bB$}
\label{eqns_delB}

With our expressions for the two scalar functions $\Upsilon_{\delta\clL}$ and
$\Psi_{\delta\clL}$, we now have an explicit expression for the curl of the
perturbed Lorentz force -- see equation \eqref{curldL}. Inverting the
curl operator from this equation gives us an expression for the two
terms involving the unknown $\delta\bB$, in terms of the known
quantities $\Upsilon_{\delta\clL}$ and $\Psi_{\delta\clL}$. In doing
so, however, we have to allow for the possibility of this force
containing an additional, and unknown, irrotational term $\nabla\Xi$:
\beq \label{dL+Xi}
\frac{(\curl\delta\bB)\times\bB_0+(\curl\bB_0)\times\delta\bB}{4\pi\rho_0}
 = \Psi_{\delta\clL}\br + \nabla\Upsilon_{\delta\clL}\times\br +\nabla\Xi.
\eeq
Note that the gauge-freedom functions $\mathcal{A}$ and $\mathcal{C}$
may be absorbed into $\nabla\Xi$, since the addition of these terms to
$\Psi_{\delta\clL}$ and $\Upsilon_{\delta\clL}$ produces the following
contribution to the right-hand side of equation \eqref{dL+Xi}:
\begin{align}
\mathcal{C}\br + \nabla\mathcal{A}\times\br,
\end{align}
and these terms are curl-free by the definitions of $\mathcal{A}$ and
$\mathcal{C}$. If we compare \eqref{dL+Xi} with the original
second-order Euler equation \eqref{Euler_aB}, we see that $\Xi$ must include
all the information about the second-order fluid perturbations
(e.g. $\delta H_{\alpha B}$), which was thrown away by taking the curl
of the Euler equation.

Equation \eqref{dL+Xi} gives three scalar equations in four unknowns -- the three
components of $\delta\bB$ and the new scalar $\Xi$. The set of
equations is then closed with the divergence-free condition on
$\delta\bB$. We are, at last, in a position to obtain a set of
differential equations featuring $\delta\bB$ explicitly.

Our equations for the perturbed field involve vector operations on
$\delta\bB$ and $\bB_0$. Operations of this form are generally
simplest performed in a vector spherical harmonic basis, and so we
write $\delta\bB$ as an infinite sum over these basis vectors,
and $\bB_0$ in its (known) vector-spherical-harmonic form:
\begin{align} \label{deltaB_VSH}
\delta\bB &= \sum\limits_{l=1}^\infty\sum\limits_{m=-l,m\neq 0}^{l}
                         \brac{ U_l^mY_l^m\be_r + V_l^m\nabla Y_l^m
                                    + W_l^m\be_r\times\nabla Y_l^m },\\
\bB_0   &= -2\sqrt{\frac{\pi}{3}}\Lambda r^2\rho_0 \be_r\times\nabla Y_1^0.
\label{B0_VSH}
\end{align}
Recall that we are looking for time-dependent solutions, $\sim
e^{im\lambda}=e^{i(m\phi+\omega t)}$, so we explicitly exclude $m=0$
terms from the sum for $\delta\bB$ (though our reasoning is applicable
to this case too). In what follows, we will use
$\sum_{l,m}$ as shorthand for the summation used for
$\delta\bB$ above.
First, we look at the divergence-free condition for $\delta\bB$ with
the above decomposition:
\begin{align}
0=\div\delta\bB
  &=\sum\limits_{l,m} \left\{
        \nabla(U_l^mY_l^m)\cdot\be_r + U_l^m Y_l^m\div\be_r
        + \nabla V_l^m\cdot\nabla Y_l^m+V_l^m\nabla^2 Y_l^m +\div[W_l^m\curl(Y_l^m\be_r)]\right\}\nn\\
  &=\sum\limits_{l,m} \left[
        \nabla(U_l^mY_l^m)\cdot\be_r + U_l^m Y_l^m\div\be_r
          + V_l^m\nabla^2 Y_l^m \right] \nn\\
  &=\sum\limits_{l,m}
        \brac{U_l^m{}' + \frac{2}{r}U_l^m - \frac{l(l+1)}{r^2}V_l^m}Y_l^m,
\label{divdeltaB}
\end{align}
where $U_l^m{}'=(U_l^m)'=\rmd U_l^m/\rmd r$. Now, equation \eqref{divdeltaB} implies that
\beq \label{elimV}
V_l^m = \frac{r^2 U_l^m {}' + 2rU_l^m}{l(l+1)}.
\eeq
Note that this result shows how a solenoidal vector field loses one degree of freedom. 
Next we rewrite the curl of $\delta\bB$, using standard vector identities:
\begin{align}
\curl\delta\bB
  &=\sum\limits_{l,m} \left[
        \nabla(U_l^m Y_l^m)\times\be_r + \nabla V_l^m\times\nabla Y_l^m
         + \nabla W_l^m\times(\be_r\times\nabla Y_l^m) + W_l^m\curl(\be_r\times\nabla Y_l^m)\right]\nn\\
  &= \sum\limits_{l,m} \left\{
        U_l^m\nabla Y_l^m\times\be_r + V_l^m{}'\be_r\times\nabla Y_l^m
         - \nabla Y_l^m(\nabla W_l^m\cdot\be_r)\right.\nn\\
  & \left.\hspace{1cm} + W_l^m\left[\be_r\nabla^2 Y_l^m-\nabla Y_l^m(\div\be_r)
                                 + (\nabla Y_l^m\cdot\nabla)\be_r-(\be_r\cdot\nabla)\nabla Y_l^m\right]\right\}\nn\\
  &= \sum\limits_{l,m} \left[ -\frac{l(l+1)W_l^m}{r^2} Y_l^m\be_r - W_l^m{}'\nabla Y_l^m
         + (V_l^m{}'-U_l^m)\be_r\times\nabla Y_l^m \right].
\label{curldeltaB}
\end{align}
Comparing this result from earlier ones in this subsection, we see
that this is an embodiment of the known result that the curl of a poloidal
  field is toroidal, and vice versa.
We use the above expression to calculate one of the unknown pieces of
the perturbed Lorentz force:
\beq \label{deltaLor1}
(\curl\delta\bB)\times\bB_0
 = -\Lambda\rho_0 \sum\limits_{l,m}
       \bigg\{
            \big[ W_l^m{}'\sin\theta Y_l^m{}_{,\theta}
                       +(V_l^m{}'-U_l^m)Y_l^m{}_{,\lambda} \big]\be_r
+          l(l+1)W_l^mY_l^m\nabla(\cos\theta) \bigg\},                 
\eeq
where we have used the following vector identities
\begin{align}
\be_r\times(\be_r\times\nabla Y_1^0) &= -\nabla Y_1^0= -\frac{\sqrt{3/\pi}}{2}\nabla(\cos\theta),\\
\nabla Y_l^m\times(\be_r\times\nabla Y_1^0)
   &= \be_r \nabla Y_l^m\cdot\nabla Y_1^0
      = -\frac{\sqrt{3/\pi}}{2r^2}\sin\theta Y_l^m{}_{,\theta}\be_r,\\
(\be_r\times\nabla Y_l^m)\times(\be_r\times\nabla Y_1^0)
   &= \nabla Y_l^m\times\nabla Y_1^0
      = \frac{\sqrt{3/\pi}}{2r^2} Y_l^m{}_{,\lambda}\be_r.
\end{align}
Very similar relations may be used to calculate the other unknown
piece of the perturbed Lorentz force:
\begin{align}
(\curl\bB_0)\times\delta\bB
 = -\Lambda\rho_0 \sum\limits_{l,m}
   & \left[ (V_l^m Y_l^m{}_{,\lambda} + W_l^m\sin\theta Y_l^m{}_{,\theta})\brac{\frac{2}{r}+\frac{\rho'_0}{\rho_0}}\be_r
               +2W_l^m\cos\theta\nabla Y_l^m\right.\nn\\
   & \left.\ \  + \brac{2r+\frac{r^2\rho'_0}{\rho_0}}U_l^mY_l^m\be_r\times\nabla(\cos\theta)
                - 2V_l^m\cos\theta\be_r\times\nabla Y_l^m \right].
\label{deltaLor2}
\end{align}
We are now in a position to reduce our original system of equations --
\eqref{dL+Xi} plus $\div\delta\bB=0$ -- to three equations (per $l,m$)
in the unknown radial functions $U_l^m$, $W_l^m$ and $\Xi_l^m$, where
the latter is defined through $\Xi=\sum_{l,m} \Xi_l^m Y_l^m$.
To find this simpler set of equations, we take the expressions for the
perturbed Lorentz force from equations \eqref{deltaLor1} and
\eqref{deltaLor2}, and use the divergence-free condition
\eqref{divdeltaB} to eliminate all $V_l^m$ in favour of $U_l^m$. We
also rewrite the right-hand side of \eqref{dL+Xi} slightly so that it
is manifestly in vector spherical harmonic form. The resulting set of
equations is:
\begin{align}
0= \sum\limits_{l,m}\Bigg\{
      &  \frac{4\pi}{\Lambda}\big[ (r\tilde\Psi_l^m+\Xi_l^m{}')Y_l^m\be_r + \Xi_l^m\nabla Y_l^m
                                                 - r\tilde\Upsilon_l^m\be_r\times\nabla Y_l^m \big]
           + \left[ W_l^m{}'+\brac{\frac{2}{r}+\frac{\rho'_0}{\rho_0}}W_l^m \right]\sin\theta Y_l^m{}_{,\theta}\be_r
            \nn\\
      & + \left[ r^2 U_l^m{}'' + \brac{6+\frac{r\rho'_0}{\rho_0}}rU_l^m{}' 
                      +\brac{6-l(l+1)+\frac{2r\rho'_0}{\rho_0}}U_l^m \right]\frac{Y_l^m{}_{,\lambda}}{l(l+1)}\be_r
          +\brac{2r+\frac{r^2\rho'_0}{\rho_0}}U_l^mY_l^m\be_r\times\nabla(\cos\theta)
            \nn\\
      & - \frac{2}{l(l+1)}(r^2 U_l^m{}' + 2rU_l^m)\cos\theta\be_r\times\nabla Y_l^m
         +2W_l^m\cos\theta\nabla Y_l^m + l(l+1)W_l^m Y_l^m\nabla(\cos\theta)  \Bigg\}.
\label{U_W_Xi_set}
\end{align}
In the above $\tilde\Psi_l^m$ and $\tilde\Upsilon_l^m$ are the radial
functions in spherical-harmonic decompositions of $\Psi_{\delta\mathcal{L}}$
and $\Upsilon_{\delta\mathcal{L}}$:
\beq
\Psi_{\delta\mathcal{L}}=\sum_{l,m} \tilde\Psi_l^m(r)Y_l^m\ ,\ 
\Upsilon_{\delta\mathcal{L}}=\sum_{l,m} \tilde\Upsilon_l^m(r)Y_l^m,
\eeq
defined with tildes to distinguish them from the
$\Psi_l^m,\Upsilon_l^m$ defined in the previous subsection. The
explicit form of the functions $\tilde\Psi_l^m,\tilde\Upsilon_l^m$ is
given in appendix \ref{Ylm_decomp_Ups_Psi}.

Note that equation \eqref{U_W_Xi_set} is \emph{not} expressed in a
vector-spherical-harmonic basis, as it contains product terms like
$Y_l^m\nabla(\cos\theta)$. Writing a term like this in a vector-spherical-harmonic basis appears
to require an infinite sum in itself, which would have to be performed
for each $l,m$ in the original (infinite) sum. Instead of facing this
unappealing double infinite sum, we look at the spherical polar
coordinate components of equation \eqref{U_W_Xi_set}. Our aim will be
to convert these into three ODEs, per $l,m$, in the three unknown
radial functions. As we will see later, the angular dependence can be
removed using the spherical harmonic orthogonality relation, but to do
so we need to cast all angular dependence of the components of
\eqref{U_W_Xi_set} in the form of spherical harmonics. To this end, we
will need three key identities:
\begin{align} \label{sinYlm}
\sin\theta Y_l^m{}_{,\theta} &= lQ_{l+1}Y_{l+1}^m - (l+1)Q_lY_{l-1}^m \ \ \ ,\\
\cos\theta Y_l^m &= Q_{l+1}Y_{l+1}^m + Q_l Y_{l-1}^m \ \ \ ,
\label{cosYlm}\\
Y_l^m{}_{,\lambda} &= im Y_l^m\ \ \ ,
\end{align}
where we have defined
\beq
Q_l\equiv \sqrt{\frac{(l-m)(l+m)}{(2l-1)(2l+1)}}.
\eeq
Using these relations, it is clear that all angular terms in the
$r$-component of equation \eqref{U_W_Xi_set} may be rewritten
accordingly. The $\theta$- and $\lambda$-components of the equation, however, involve 
combinations like $\cos\theta Y_l^m{}_{,\theta}$, which cannot be
readily simplified (we encounter a similar problem as for the vector-spherical-harmonic
basis, where one needs to invoke an infinite sum). To avoid this
problem, we leave the $r$-component of \eqref{U_W_Xi_set} unchanged,
but multiply its $\theta$ and $\lambda$ components
by $\sin\theta$, giving us the set:
\begin{align}
0=\sum\limits_{l,m}\bigg\{
 & \frac{4\pi}{\Lambda} (r\tilde\Psi_l^m+\Xi_l^m{}') Y_l^m
   + \frac{im}{l(l+1)}\left[  r^2 U_l^m{}'' + \brac{6+\frac{r\rho'_0}{\rho_0}}rU_l^m{}'
                 + \brac{6-l(l+1)+\frac{2r\rho'_0}{\rho_0}}U_l^m \right] Y_l^m \nn\\
 &  +\left[ W_l^m{}' + \brac{2+\frac{r\rho'_0}{\rho_0}}\frac{W_l^m}{r} \right]\sin\theta Y_l^m{}_{,\theta} \bigg\},\\
0=\sum\limits_{l,m}\bigg\{
 & \frac{4\pi}{\Lambda} \brac{\frac{\Xi_l^m}{r}\sin\theta Y_l^m{}_{,\theta} + im\tilde\Upsilon_l^m Y_l^m}
     + \frac{2im}{l(l+1)} (r U_l^m{}' +2U_l^m) \cos\theta Y_l^m
     + [2\cos\theta\sin\theta Y_l^m{}_{,\theta}-l(l+1)\sin^2\theta Y_l^m]\frac{W_l^m}{r} \bigg\}, \\
0=\sum\limits_{l,m}\bigg\{
 & \frac{4\pi}{\Lambda} \brac{ \frac{im\Xi_l^m}{r} Y_l^m - \tilde\Upsilon_l^m\sin\theta Y_l^m{}_{,\theta}}
     - \brac{2+\frac{r\rho'_0}{\rho_0}}U_l^m\sin^2\theta Y_l^m
     -\frac{2}{l(l+1)}(rU_l^m{}'+2U_l^m)\cos\theta\sin\theta Y_l^m{}_{,\theta}\nn\\
 &  + \frac{2im W_l^m}{r}\cos\theta Y_l^m \bigg\}.
\end{align}
The latter two equations in the above set may now also be rewritten,
by the use of the following auxiliary relations (which may be derived
from \eqref{sinYlm} and \eqref{cosYlm}):
\begin{align} \label{cossinYlm}
\cos\theta\sin\theta Y_l^m{}_{,\theta}
 &= l Q_{l+1}Q_{l+2}Y_{l+2}^m + [lQ_{l+1}^2-(l+1)Q_l^2]Y_l^m - (l+1)Q_{l-1}Q_lY_{l-2}^m, \\
\sin^2\theta Y_l^m = Y_l^m-\cos^2\theta Y_l^m
 &= -Q_{l+1}Q_{l+2}Y_{l+2}^m+ (1-Q_l^2-Q_{l+1}^2)Y_l^m - Q_{l-1}Q_lY_{l-2}^m.
\label{sinsinYlm}
\end{align}
With these relations, we are now able to eliminate all instances of trigonometric functions
and derivatives of $Y_l^m$ in our set of equations:
\begin{align}
\label{infsumYlm1}
0= \sum\limits_{l,m} \Bigg\{
 &  \frac{4\pi}{\Lambda} (r\tilde\Psi_l^m+\Xi_l^m{}') Y_l^m
        + \frac{im}{l(l+1)}\left[  r^2 U_l^m{}'' + \brac{6+\frac{r\rho'_0}{\rho_0}}rU_l^m{}'
                 + \brac{6-l(l+1)+\frac{2r\rho'_0}{\rho_0}}U_l^m \right] Y_l^m \nn\\
 &  + \left[ W_l^m{}' + \brac{2+\frac{r\rho'_0}{\rho_0}}\frac{W_l^m}{r} \right]
                                    \Big[ lQ_{l+1}Y_{l+1}^m - (l+1)Q_lY_{l-1}^m \Big] \Bigg\},\\
\label{infsumYlm2}
0= \sum\limits_{l,m} \Bigg\{
 &  \frac{4\pi}{\Lambda} \left\{ \frac{\Xi_l^m}{r}[lQ_{l+1}Y_{l+1}^m-(l+1)Q_l Y_{l-1}^m]
                    + im\tilde\Upsilon_l^m Y_l^m  \right\}
     + \frac{2im}{l(l+1)} (r U_l^m{}'+2U_l^m)(Q_{l+1}Y_{l+1}^m+Q_l Y_{l-1}^m) \nn\\
 &  + \frac{W_l^m}{r}\Big\{ l(l+3)Q_{l+1}Q_{l+2}Y_{l+2}^m
                                     + \big[ l(l+3)Q_{l+1}^2 + (l-2)(l+1)Q_l^2 - l(l+1) \big]Y_l^m
                                     + (l-2)(l+1)Q_{l-1}Q_l Y_{l-2}^m\Big\} \Bigg\},\\
\label{infsumYlm3}
0= \sum\limits_{l,m} \Bigg\{
 &  \frac{4\pi}{\Lambda} \left\{ \frac{im\Xi_l^m}{r} Y_l^m 
                    - \tilde\Upsilon_l^m[lQ_{l+1}Y_{l+1}^m-(l+1)Q_l Y_{l-1}^m] \right\}
     + U_l^m\bigg\{ \brac{\frac{2(l-1)}{(l+1)}+\frac{r\rho'_0}{\rho_0} }Q_{l+1}Q_{l+2} Y_{l+2}^m\nn\\
 &            - \left[ 2\brac{1-\frac{(l+2)Q_l^2}{l}-\frac{(l-1)Q_{l+1}^2}{(l+1)}}
                             + (1-Q_l^2-Q_{l+1}^2)\frac{r\rho'_0}{\rho_0} \right]Y_l^m
               + \brac{\frac{2(l+2)}{l}+\frac{r\rho'_0}{\rho_0}}Q_{l-1}Q_l Y_{l-2}^m\bigg\}\nn\\
 & - \frac{2}{l(l+1)} r U_l^m{}'\Big\{ 
          lQ_{l+1}Q_{l+2}Y_{l+2}^m + [lQ_{l+1}^2-(l+1)Q_l^2]Y_l^m -(l+1)Q_{l-1}Q_l Y_{l-2}^m \Big\}
    + \frac{2im W_l^m}{r} (Q_{l+1}Y_{l+1}^m+Q_lY_{l-1}^m) \Bigg\}.
\end{align}
Note that in the above, spherical-harmonic terms with $l<|m|$ are
automatically excluded from the sum by the form of their $Q_l$
prefactors (which are zero for $l=\pm m$).
Next we will eliminate all the angular terms in the above, by using the
spherical-harmonic orthogonality relation:
\beq
\int Y_{\hat{l}}^{\hat{m}} Y_l^m{}^*\ \sin\theta\rmd\theta\rmd\lambda
 = \delta_{\hat{l}l} \delta_{\hat{m}m}.
\eeq
Let us relabel the indices $l,m$ in equations \eqref{infsumYlm1}, \eqref{infsumYlm2} and
\eqref{infsumYlm3} to $\hat{l},\hat{m}$, to match the notation of the above
relation. Multiplying these three equations by
$Y_l^m{}^*$, integrating and using the above relation, then picks out
individual elements in the sums. This reduces the three infinite
sums to three equalities (without summation) for each value of $m\neq
0$ and $l\geq 1$. These equalities are now just DEs in the radial
coordinate, as the angular dependence has dropped out:
\begin{align}
0 =& \frac{4\pi}{\Lambda} (r\tilde\Psi_l+\Xi'_l)
          + \frac{im}{l(l+1)}\left[  r^2 U''_l + \brac{6+\frac{r\rho'_0}{\rho_0}}rU'_l
                 + \brac{6-l(l+1)+\frac{2r\rho'_0}{\rho_0}}U_l \right]\nn\\
   & + (l-1)Q_l \left[ W'_{l-1}+\brac{2+\frac{r\rho'_0}{\rho_0}}\frac{W_{l-1}}{r} \right]
      - (l+2)Q_{l+1} \left[ W'_{l+1}+\brac{2+\frac{r\rho'_0}{\rho_0}}\frac{W_{l+1}}{r} \right], \label{eqn1+Xi}\\
0 =& \frac{4\pi}{\Lambda} \left[ (l-1)Q_l\frac{\Xi_{l-1}}{r}-(l+2)Q_{l+1}\frac{\Xi_{l+1}}{r} + im\tilde\Upsilon_l\right]
       + \frac{2imQ_l}{(l-1)l} (rU'_{l-1}+2U_{l-1}) 
       + \frac{2imQ_{l+1}}{(l+1)(l+2)}(rU'_{l+1}+2U_{l+1})\nn\\
   & + (l-2)(l+1)Q_{l-1}Q_l \frac{W_{l-2}}{r}
       + \big[l(l+3)Q_{l+1}^2+(l-2)(l+1)Q_l^2-l(l+1)\big] \frac{W_l}{r}
       + l(l+3)Q_{l+1}Q_{l+2}\frac{W_{l+2}}{r},\label{eqn2+Xi}\\
0=& \frac{4\pi}{\Lambda}\left[ \frac{im\Xi_l}{r} 
              - (l-1)Q_l \tilde\Upsilon_{l-1}+(l+2)Q_{l+1}\tilde\Upsilon_{l+1} \right]
         - \frac{2}{(l-1)}Q_{l-1}Q_l r U'_{l-2}
         + Q_{l-1}Q_l\brac{ \frac{2(l-3)}{(l-1)}+\frac{r\rho'_0}{\rho_0} } U_{l-2}\nn\\
   &  - \frac{2}{l(l+1)} [lQ_{l+1}^2-(l+1)Q_l^2] r U'_l
       - \bigg[ 2\brac{1-\frac{(l+2)Q_l^2}{l}-\frac{(l-1)Q_{l+1}^2}{(l+1)}}
                     + (1-Q_l^2-Q_{l+1}^2)\frac{r\rho'_0}{\rho_0} \bigg] U_l\nn\\
   &  + \frac{2Q_{l+1}Q_{l+2}}{(l+2)}r U'_{l+2}
       + Q_{l+1}Q_{l+2}\brac{ \frac{2(l+4)}{(l+2)}+\frac{r\rho'_0}{\rho_0} }U_{l+2}
       + 2im Q_l \frac{W_{l-1}}{r} + 2imQ_{l+1}\frac{W_{l+1}}{r}.\label{eqn3+Xi}
\end{align}
The above equations form an infinite set for different values of $l$
and $m$. Individual equations couple together $U$ and $W$ functions
with different angular indices $l$, but terms with different azimuthal
index $m$ decouple, allowing us to suppress $m$. Since the angular structure of the source
terms $\Upsilon,\Psi$ in equations \eqref{eqn1+Xi}-\eqref{eqn3+Xi} only includes terms with azimuthal index $|m|=1$
and $2$ (see appendix \ref{Ylm_decomp_Ups_Psi}), we expect the solution to reflect this, i.e. $U_l^m=W_l^m=0$ for
$|m|\geq 3$ (and as before, we exclude the stationary $m=0$ terms). In
addition, in the case of an orthogonal rotator ($\chi=\pi/2$), only the $|m|=2$ terms survive.

For brevity we have shown equations for general $l$, but by doing so
some of the above equations as written feature
$U$ and $W$ functions with $l<|m|$, and in particular with negative
$l$ (e.g. the $U_{l-2}$ terms when $l=m=1$). The origin of these
terms, however, is in $l<|m|$ spherical harmonics from
\eqref{infsumYlm1}-\eqref{infsumYlm3} -- and these are identically
zero (see note after these equations). Accordingly, equations
\eqref{eqn1+Xi}-\eqref{eqn3+Xi} and those which follow them in this
section are presented on the understanding 
that any instances of $U$ or $W$ functions with $l<|m|$ are undefined
and should be excluded. This is then also consistent with the original sum
\eqref{deltaB_VSH} in which the $U$ and $W$ functions first appeared.

To find the perturbed magnetic field $\delta\bB$ we need only the functions
$U_l$ and $W_l$, and are not interested in solving for the other
unknown set of functions $\Xi_l$. We 
next use \eqref{eqn3+Xi} to eliminate all instances of $\Xi_l$ terms in
equations \eqref{eqn1+Xi} and \eqref{eqn2+Xi}. After simplification, the result is the
following two coupled DEs for each value of $l$:
\begin{align}
0 =& \frac{4\pi}{\Lambda} \left[ mr\tilde\Psi_l - i(l-1)Q_l(r\tilde\Upsilon'_{l-1}+\tilde\Upsilon_{l-1}) 
                                                    + i(l+2)Q_{l+1}(r\tilde\Upsilon'_{l+1}+\tilde\Upsilon_{l+1}) \right]\nn\\
      & + Q_{l-1}Q_l\bigg\{ -\frac{2}{(l-1)}r^2\bar{U}''_{l-2}
                + \left[ \frac{2(l-5)}{(l-1)}+\frac{r\rho'_0}{\rho_0}\right]r\bar{U}'_{l-2}
                + \left[ \frac{2(l-3)}{(l-1)}+\frac{r\rho'_0}{\rho_0}+r\brac{\frac{r\rho'_0}{\rho_0}}'\right]\bar{U}_{l-2}\bigg\}\nn\\
      & + \frac{(m^2+2(l+1)Q_l^2-2lQ_{l+1}^2)}{l(l+1)}r^2\bar{U}''_l
         + \left[\frac{6m^2}{l(l+1)}-2+\frac{2(l+4)Q_l^2}{l}+\frac{2(l-3)Q_{l+1}^2}{(l+1)}
                       +\brac{\!\frac{m^2}{l(l+1)}\!-\!1\!+\!Q_l^2\!+\!Q_{l+1}^2\!}\!\frac{r\rho'_0}{\rho_0} \right]r\bar{U}'_l\nn\\
      & + \left[ \frac{6m^2}{l(l+1)}-m^2 -2+\frac{2(l+2)Q_l^2}{l}+\frac{2(l-1)Q_{l+1}^2}{(l+1)} 
                       + \brac{\frac{2m^2}{l(l+1)}-1+Q_l^2+Q_{l+1}^2}\frac{r\rho'_0}{\rho_0}
                       - (1-Q_l^2-Q_{l+1}^2)r\brac{\frac{r\rho'_0}{\rho_0}}' \right]\bar{U}_l \nn\\
      & + Q_{l+1}Q_{l+2}\bigg\{ \frac{2}{(l+2)}r^2\bar{U}''_{l+2}
                 + \left[ \frac{2(l+6)}{(l+2)}+\frac{r\rho'_0}{\rho_0} \right]r\bar{U}'_{l+2}
                 + \left[ \frac{2(l+4)}{(l+2)}+\frac{r\rho'_0}{\rho_0}+r\brac{\frac{r\rho'_0}{\rho_0}}' \right]\bar{U}_{l+2}  \bigg\}\nn\\
      & + (l-3)mQ_l rX'_{l-1} + mQ_l\left[ (3l-5)+(l-1)\frac{r\rho'_0}{\rho_0} \right]X_{l-1}
          - (l+4)mQ_{l+1} rX'_{l+1} - mQ_{l+1}\left[ (3l+8)+(l+2)\frac{r\rho'_0}{\rho_0} \right]X_{l+1},
 \label{DAE1} \\
0 =& \frac{4\pi i}{\Lambda}\left\{ -(l-1)(l-2)Q_{l-1}Q_l\tilde\Upsilon_{l-2}
                      + [ m^2+(l-1)(l+1)Q_l^2+l(l+2)Q_{l+1}^2 ]\tilde\Upsilon_l - (l+2)(l+3)Q_{l+1}Q_{l+2}\tilde\Upsilon_{l+2}  \right\}\nn\\
      &  - (l-1)Q_{l-2}Q_{l-1}Q_l\left[ \frac{2}{(l-2)}r\bar{U}'_{l-3}-\brac{\!\frac{2(l-4)}{(l-2)}+\frac{r\rho'_0}{\rho_0}\!}\!\bar{U}_{l-3} \right]
          + \frac{Q_l}{l}\bigg\{ 2\left[ \frac{m^2}{(l\!-\!1)}+lQ_{l-1}^2-(l\!-\!1)Q_l^2+(l\!+\!2)Q_{l+1}^2 \right]r\bar{U}'_{l-1} \nn\\ 
      &  + \bigg[ \frac{4m^2}{(l\!-\!1)} -2l\big[l-1-(l\!+\!1)Q_{l-1}^2\big] + 2(l\!-\!2)\big[(l\!-\!1)Q_l^2 \!-\! (l\!+\!2)Q_{l+1}^2\big]
                       - l\big[ (l\!-\!1)(1\!-\!Q_{l-1}^2\!-\!Q_l^2) + (l\!+\!2)Q_{l+1}^2 \big]\frac{r\rho'_0}{\rho_0} \bigg]\bar{U}_{l-1} \bigg\}\nn\\
      &  + \frac{Q_{l+1}}{(l+1)}\bigg\{ 2\left[ \frac{m^2}{(l+2)}+(l-1)Q_l^2-(l+2)Q_{l+1}^2+(l+1)Q_{l+2}^2 \right]r\bar{U}'_{l+1}
           + \bigg[ \frac{4m^2}{(l+2)} +2(l+1)\big[ l+2-lQ_{l+2}^2 \big]\nn\\
      &               +2(l+3)[(l-1)Q_l^2 - (l+2)Q_{l+1}^2]
                       + (l+1)\big[ (l+2)(1-Q_{l+1}^2-Q_{l+2}^2) + (l-1)Q_l^2 \big]\frac{r\rho'_0}{\rho_0} \bigg]\bar{U}_{l+1} \bigg\}\nn\\
      &  - (l+2)Q_{l+1}Q_{l+2}Q_{l+3}\left[ \frac{2}{(l+3)}r\bar{U}'_{l+3}+\brac{\frac{2(l+5)}{(l+3)}+\frac{r\rho'_0}{\rho_0}}\bar{U}_{l+3} \right]
           + mQ_{l-1}Q_l[ (l-2)(l+1)-2(l-1) ]X_{l-2}\nn\\
      &  + m\left\{ -l(l+1) + [ (l-2)(l+1)-2(l-1) ]Q_l^2 + [ l(l+3)+2(l+2) ]Q_{l+1}^2 \right\}X_l
           + mQ_{l+1}Q_{l+2}[ l(l+3)+2(l+2) ]X_{l+2}.
 \label{DAE2}
\end{align}
where we have defined
\beq \label{Ubar-X}
X_l\equiv\frac{W_l}{r}, \bar{U}_l\equiv iU_l.
\eeq
The former definition gives a toroidal function with the same
dimension as the poloidal functions $U_l$; the latter produces a set
of DEs where all coefficients are real (note that the $\tilde\Upsilon_l$
functions are all imaginary, so $i\tilde\Upsilon_l$ is real).

\subsubsection{Analysis of equations}
\label{analysis_of_eqs}

In principle, both equations \eqref{DAE1} and \eqref{DAE2} must be
solved for all $l>|m|$, but upon closer inspection we see that half of
the equations are trivial. To see this, note that only the $l=2$ and $4$ terms of
$\tilde\Psi_l^m$ and the $l=1$ and $3$ terms of
$\tilde\Upsilon_l^m$ are non-zero ($\tilde\Upsilon^{\pm 2}_1$ is
also zero, as expected since $l<|m|$). In addition, the equations only couple
$U_l$ to $U_{l+2},U_{l+4},\dots$, with the same being true for
$X_l$. The result is that \eqref{DAE1} for $l=2,4,6,\dots$ couples to
\eqref{DAE2} for $l=1,3,5,\dots$, and to all the source
terms. These are clearly the relevant equations to solve for our
problem. On the other hand, equation \eqref{DAE1} for odd $l$ couples to \eqref{DAE2} for
even $l$, but since there are no terms sourcing the variables involved in
these equations we may take
\beq \label{Xeven-Uodd}
X_{\textrm{even }l}=U_{\textrm{odd }l}=0.
\eeq
Next, we argue that the terms with negative $m$ are equal to
plus-or-minus the corresponding terms with positive $m$ (as usual for
problems involving spherical-harmonic decompositions).

Although equations \eqref{DAE1} and \eqref{DAE2} are rather lengthy, it is
enough to note that they take the schematic forms:
\begin{align}
0 &= \textrm{terms of the form }\{ m\tilde\Psi_l^m,\tilde\Upsilon_l^m,U_l^m,mX_l^m\},\\
0 &= \textrm{terms of the form }\{ \tilde\Upsilon_l^m,m^2\tilde\Upsilon_l^m,U_l^m,mX_l^m\},
\end{align}
respectively (where we have restored the suppressed $m$ superscripts
on the variables). From the forms of the source terms, \eqref{Ups_lm}
and \eqref{Psi_lm}, we see that
\beq
\tilde\Upsilon_l^1=\tilde\Upsilon_l^{-1},\tilde\Upsilon_l^2=-\tilde\Upsilon_l^{-2},
\tilde\Psi_l^1=-\tilde\Psi_l^{-1},\tilde\Psi_l^2=\tilde\Psi_l^{-2},
\eeq
and therefore we expect
\beq \label{UX-parity}
U_l^1=U_l^{-1},U_l^2=-U_l^{-2},
X_l^1=-X_l^{-1},X_l^2=X_l^{-2}.
\eeq
We have now accumulated a number of results -- specifically those of
equations \eqref{elimV}, \eqref{Ubar-X}, \eqref{Xeven-Uodd} and \eqref{UX-parity} -- which allow us to simplify the original
expression \eqref{deltaB_VSH} for $\delta\bB$ and rewrite it in terms of the new variables
used in our system of DEs \eqref{DAE1}, \eqref{DAE2}:
\begin{align}
\delta\bB = \sum\limits_{l=1}^\infty \bigg\{
 & -i\bar{U}_{2l}^1(Y_{2l}^1+Y_{2l}^{-1})\be_r - i\frac{(r^2\bar{U}_{2l}^1{}'+2r\bar{U}_{2l}^1)}{l(l+1)}\nabla(Y_{2l}^1+Y_{2l}^{-1})
     + rX_{2l-1}^1\be_r\times\nabla(Y_{2l-1}^1-Y_{2l-1}^{-1}) \nn\\
 & -i\bar{U}_{2l}^2(Y_{2l}^2-Y_{2l}^{-2})\be_r - i\frac{(r^2\bar{U}_{2l}^2{}'+2r\bar{U}_{2l}^2)}{l(l+1)}\nabla(Y_{2l}^2-Y_{2l}^{-2})
     + rX_{2l+1}^2\be_r\times\nabla(Y_{2l+1}^2+Y_{2l+1}^{-2}) \bigg\}.
\label{deltaB-positivem}
\end{align}
Since the quantities $\bar{U}_l^m,X_l^m,(Y_l^1-Y_l^{-1}),(Y_l^2+Y_l^{-2}) $ are all
real, whilst $(Y_l^1+Y_l^{-1})$ and $(Y_l^2-Y_l^{-2})$ are imaginary, it is clear that $\delta\bB$
is also real, as expected.

\subsection{Differential algebraic equations; conditions at the centre}
\label{DAE-centrecond}

The first of the above pair of equations, \eqref{DAE1}, is second-order in
$\bar{U}_l$, but the second \eqref{DAE2} is only first-order. If we were to define
a new variable
\beq
Z_l\equiv \bar{U}'_l
\eeq
and substitute this back into the pair of equations above, the first
equation would reduce to first order, and the second would be
algebraic, with no derivatives. Thus, instead of a conventional
system of ODEs, we have a system of differential algebraic equations
(DAEs). This is an unwelcome result, as DAEs are in many senses more
difficult to solve than ODEs. Extra care is needed in choosing
suitable boundary conditions for the differentiated variables, to
ensure that they are consistent -- i.e. that they satisfy any
algebraic relations stemming from the DAE system. Furthermore,
although initial-value problems for DAEs have been relatively
well-explored, we have a boundary-value problem, since we will have
conditions to impose both at the centre and outer
boundary. See, e.g., \citet{petzold} for more discussion of the difficulties of solving DAEs.

Despite these problems we have been able to find a solution method,
albeit one with limited applicability. For this method, we chose to
work with equation \eqref{DAE1} in its original form, but coupled to
the $r$-derivative of equation \eqref{DAE2}:
\begin{align}
0 =& \frac{4\pi i}{\Lambda}\left\{ -(l-1)(l-2)Q_{l-1}Q_l\tilde\Upsilon'_{l-2}
                      + [ m^2+(l-1)(l+1)Q_l^2+l(l+2)Q_{l+1}^2 ]\tilde\Upsilon'_l - (l+2)(l+3)Q_{l+1}Q_{l+2}\tilde\Upsilon'_{l+2}  \right\}\nn\\
      &  - (l-1)Q_{l-2}Q_{l-1}Q_l\left[  \frac{2}{(l-2)}r\bar{U}''_{l-3}
                     -\brac{\frac{2(l-5)}{(l-2)}+\frac{r\rho'_0}{\rho_0}}\bar{U}'_{l-3}-\brac{\frac{r\rho'_0}{\rho_0}}' \bar{U}_{l-3} \right]
           + \frac{Q_l}{l}\bigg\{ 2\bigg[ \frac{m^2}{(l-1)}+lQ_{l-1}^2-(l-1)Q_l^2\nn\\
      &                                      +(l+2)Q_{l+1}^2 \bigg]r\bar{U}''_{l-1}
                + \bigg[ \frac{6m^2}{(l-1)} -2l(l-1) +2l(l+2)Q_{l-1}^2 + 2(l-3)(l-1)Q_l^2
                      - 2(l-3)(l+2)Q_{l+1}^2 \nn\\
      & - l\big[ (l-1)(1-Q_{l-1}^2-Q_l^2) + (l+2)Q_{l+1}^2 \big]\frac{r\rho'_0}{\rho_0} \bigg]\bar{U}'_{l-1} 
                - l\big[ (l-1)(1-Q_{l-1}^2-Q_l^2) + (l+2)Q_{l+1}^2 \big]\brac{\frac{r\rho'_0}{\rho_0}}'\bar{U}_{l-1}  \bigg\}\nn\\    
      &  + \frac{Q_{l+1}}{(l+1)}\bigg\{ 2\left[ \frac{m^2}{(l+2)}+(l-1)Q_l^2-(l+2)Q_{l+1}^2+(l+1)Q_{l+2}^2 \right]r\bar{U}''_{l+1}
             + \bigg[ \frac{6m^2}{(l+2)} +2(l+1)(l+2)+2(l-1)(l+4)Q_l^2\nn\\
      &               -2(l+2)(l+4)Q_{l+1}^2-(l-1)(l+1)Q_{l+2}^2
                       + (l+1)\big[ (l+2)(1-Q_{l+1}^2-Q_{l+2}^2) + (l-1)Q_l^2 \big]\frac{r\rho'_0}{\rho_0} \bigg]\bar{U}'_{l+1} \nn\\
      &     + (l+1)\big[ (l+2)(1-Q_{l+1}^2-Q_{l+2}^2) + (l-1)Q_l^2 \big]\brac{\frac{r\rho'_0}{\rho_0}}'\bar{U}_{l+1} \bigg\}
           - (l+2)Q_{l+1}Q_{l+2}Q_{l+3}\bigg[ \frac{2}{(l+3)}r\bar{U}''_{l+3}\nn\\
      &           +\brac{\frac{2(l+6)}{(l+3)}+\frac{r\rho'_0}{\rho_0}}\bar{U}'_{l+3}+\brac{\frac{r\rho'_0}{\rho_0}}'\bar{U}_{l+3} \bigg]
           + mQ_{l-1}Q_l[ (l-2)(l+1)-2(l-1) ]X'_{l-2}\nn\\
      &  + m\left\{ -l(l+1) + [ (l-2)(l+1)-2(l-1) ]Q_l^2 + [ l(l+3)+2(l+2) ]Q_{l+1}^2 \right\}X'_l
           + mQ_{l+1}Q_{l+2}[ l(l+3)+2(l+2) ]X'_{l+2}.
 \label{diff_DAE2}
\end{align}
Now we really do have a conventional system of coupled ODEs, where
both equations (per $l,m$) are second
order in $U_l$ and first-order in $X_l$ -- and may be solved with
conventional numerical methods. However, the set of solutions to this new
system of equations is larger than those which solve our original
problem. In particular, if we integrate equation
\eqref{diff_DAE2} we recover the original equation plus some arbitrary
integration constant. Plugging a solution to the differentiated
equation back into the original system thus generically results in an
substantial $r$-independent error; see appendix \ref{numerical-details}.

In order to fix the integration constant, we need information from the
original equation \eqref{DAE2}. We note that since all derivatives are
premultiplied by $r$, if we evaluate \eqref{DAE2} at the centre of the
star these are zero, assuming sufficient regularity of the $\bar{U}$
and $X$ functions. In addition, the source terms
$\tilde\Psi_l^m,\tilde\Upsilon_l^m$ are zero. We are thus left with a
set of algebraic equations in $U_l(0)$ and $X_l(0)$ (one per value of
$l$):
\begin{align}
0 =& \frac{(l-4)(l-1)}{(l-2)}Q_{l-2}Q_{l-1}Q_l\bar{U}_{l-3}(0)
         + \frac{Q_l}{l}\left[ \frac{2m^2}{(l-1)}-l(l-1)+l(l+1)Q_{l-1}^2+(l-2)(l-1)Q_l^2
                                      -(l-2)(l+2)Q_{l+1}^2\right]\bar{U}_{l-1}(0) \nn\\
 & + \frac{Q_{l+1}}{(l+1)}\left[ \frac{2m^2}{(l+2)}+(l+1)(l+2)+(l-1)(l+3)Q_l^2
                                      -(l+2)(l+3)Q_{l+1}^2 -l(l+1)Q_{l+2}^2\right]\bar{U}_{l+1}(0)\nn\\
 & - \frac{(l+2)(l+5)}{(l+3)}Q_{l+1}Q_{l+2}Q_{l+3}\bar{U}_{l+3}(0)
     + mQ_{l-1}Q_l\left[\frac{(l-2)(l+1)}{2}-l+1\right]X_{l-2}(0)\nn\\
 & + m\left\{-\frac{l(l+1)}{2}+\left[ \frac{(l-2)(l+1)}{2}-l+1 \right]Q_l^2
               +\left[ \frac{l(l+3)}{2}+l+2 \right]Q_{l+1}^2 \right\}X_l(0)
      + mQ_{l+1}Q_{l+2}\left[\frac{l(l+3)}{2}+l+2\right]X_{l+2}(0).
\label{alg_cent}
\end{align}
In numerical solutions of systems of differential equations like
\eqref{DAE1},\eqref{diff_DAE2}, one truncates the infinite set of equations
at some finite value of the angular index $l_\textrm{max}$. For a particular
$l_\textrm{max}$, equation \eqref{alg_cent} may be `solved' -- it does
not give numerical values for all the variables at the centre, of
course, since this requires information from the full equations
including the source terms, but it does allow one to express these
variables in terms of each other. For example, for $l_\textrm{max}=4$
and $m=1$, one can show that equation \eqref{alg_cent} implies that
\begin{align}
\bar{U}_4^1(0) &= -\frac{55}{6\sqrt{6}}\bar{U}_2^1(0),\label{U41_cent}\\
X_3^1(0) &= -\frac{7}{12}\sqrt{\frac{35}{2}}\bar{U}_2^1(0). \label{X31_cent}
\end{align}
Note that for this example, and in fact for all $l_\textrm{max}$, $X_1(0)$ drops out of
equation \eqref{alg_cent} completely, as its prefactor is always zero.

Unfortunately, applying the same method to higher values of
$l_\textrm{max}$ does not give solutions to the original system of
DAEs. We believe this is because the relations obtained at the centre
from \eqref{alg_cent} become more complex -- relating, for example, all
quantities to a combination of $\bar{U}_2(0)$ \emph{and}
$\bar{U}_4(0)$, instead of $\bar{U}_2(0)$ alone. As a result, it seems
that the centre conditions do not give sufficient information to fix
the integration constant for higher $l_\textrm{max}$.

It would naturally have been more
satisfactory to go to higher values of $l_\textrm{max}$ to check the
convergence of the solution in the limit $l\to\infty$. However, the low-$l,m$
nature of the source terms gives us reason to believe that the
full $l\to\infty$ sum will be dominated by lower multipoles, and so we anticipate
that our results will be representative of the full, untruncated,
solution. We will later find some hints of how the higher-$l$
components behave, in section \ref{possible_multipoles}.

\subsection{Exterior solution}

To understand what boundary conditions are appropriate, let us first
recall the physical meaning of the quantities in our equations.
From \eqref{deltaB_VSH} and \eqref{elimV}, we see that the $U_l$
functions are associated with the poloidal component of $\delta\bB$,
and the $W_l$ functions with the toroidal
component. We may, therefore, regard the two unknown quantities in the
above coupled DEs as being $\delta\bB_{\textrm{pol}}$ and
$\delta\bB_{\textrm{tor}}$. Note that by eliminating the $\Xi_l$
functions, we have produced a set of equations with no dependence on
second-order fluid perturbations like $\delta H_{\alpha B}$; see the
discussion after equation \eqref{dL+Xi}. This means we only need to
impose boundary conditions on the perturbed magnetic field, rather
than on the fluid too.

We assume that the exterior of the star is vacuum; it may be more
realistic to model a charge-carrying magnetosphere, but the aim of
this study is to isolate the physics of a quasi-rigidly precessing
magnetised fluid star. Assuming a vacuum exterior means there are no
particles to carry an electric current $\delta{\bf j}$. Using
Amp\`ere's law, this gives us:
\beq
\curl\delta\bB = 4\pi\delta{\bf j}=0\ \ \textrm{ ($r>R_*$)}.
\eeq
From \eqref{curldeltaB}, we require that each of the three components
of the vector-spherical-harmonic decomposition of $\curl\delta\bB$ be zero outside the star, i.e.
\beq
-\frac{l(l+1)W_l^m}{r^2} = -W_l^m{}' 
 = V_l^m{}' - U_l^m  = 0\ \ \textrm{ ($r>R_*$)}.
\eeq
As for the equations for the interior, the exterior equations decouple for different
azimuthal index, and so we may suppress the $m$ indices. Using the
above result, and eliminating $V_l^m$ in favour of $U_l^m$ using \eqref{elimV}, the
equations for the exterior are:
\begin{align} \label{Uexteqn}
0=& r^2U''_l + 4rU'_l+[2-l(l+1)]U_l \ \ \textrm{ ($r>R_*$)},\\
0=& W_l\ \ \textrm{ ($r>R_*$)}.
 \label{Xexteqn}
\end{align}
The latter equation immediately tells us that there is no exterior
toroidal field. The former relation is a Cauchy-Euler equation, which
we may solve by first making the
ansatz that $U_l=r^p$; plugging this in to equation \eqref{Uexteqn}
gives us an indicial equation, which can be factorised to give the
values of $p$:
\beq
0 = p^2 + 3p + [2-l(l+1)]p = (p+l+2)(p-l+1) \implies p=-l-2,p=l-1.
\eeq
Thus the general exterior solution $U^{\textrm{ext}}_l$ for $U_l$ is
\beq
U^{\textrm{ext}}_l(r) = u^\textrm{ext}_l r^{-l-2}+\hat{u}^\textrm{ext}_l r^{l-1}.
\eeq
The latter term in this expression diverges at infinite radius,
however, and so is unphysical. Therefore we set $\hat{u}^\textrm{ext}_l=0$,
leaving us with
\beq \label{Uext_orig}
U^{\textrm{ext}}_l(r) = u^\textrm{ext}_l r^{-l-2}.
\eeq
Note that this gives us the expected physical result: the dipole
component ($l=1$) of a poloidal field should decay as $1/r^3$, the
quadrupole as $1/r^4$, and so on. Differentiating the above, we get a
condition on $U_l$ which does not involve the unknown constants
$u_l^\textrm{ext}$:
\beq \label{Uext_diff}
(U^\textrm{ext}_l)' = -(l+2)u_l^\textrm{ext}r^{-(l+3)} = -\frac{(l+2)}{r}U^\textrm{ext}_l
\implies rU'_l + (l+2)U_l = 0\ \ \textrm{ ($r>R_*$)}.
\eeq

\subsection{Internal solution behaviour as outer boundary $\to R_*$}
\label{surface_issues}

Within our model, the most natural place to impose boundary conditions
matching the interior and exterior solutions would be at the stellar
surface $R_*$. A closer look at equations \eqref{DAE1} and
\eqref{DAE2}, however, indicates problems with doing so. In
particular, the equations 
feature terms of the form $\rho'_0/\rho_0$ and the radial derivative
thereof, both of which diverge as $r\to R_*$ for a $\gamma=2$
polytrope\footnote{These quantities also diverge for $\gamma=4/3$ -- often used to
model the pressure-density relation for main-sequence stars -- and for
$\gamma=5/3$, a typical white-dwarf model. This already hints at a
more general problem with the model.}.

The divergent terms originate from the perturbed Lorentz force; see equation
\eqref{Euler_aB}. Of these three terms, two have a denominator of
$\rho_0$ and one has a denominator of $\rho_0^2$; we therefore need corresponding
factors in the numerators to avoid the Lorentz force diverging when
$\rho_0\to 0$. The
second of the three Lorentz-force terms from \eqref{Euler_aB} involves
a cross product with $\bB_0$, which in turn scales with $\rho_0$, and
so this term is indeed well-behaved. The first and third terms will diverge,
however, unless $\curl\bB_0\to 0$ at least as fast as $\rho_0$ --
which it does not; it contains one term which instead scales with $\rho'_0$ (this can
be seen by direct evaluation of the curl of equation \eqref{B0}, or by
applying relation \eqref{curldeltaB} to $\bB_0$ in the form given in
equation \eqref{B0_VSH}), and $\rho'_0$ is non-zero at $R_*$. As
originally noted by \citet{mestel2}, this divergence is a problem before even
reaching the surface, since the perturbed Lorentz force -- an
$\mathcal{O}(\epsilon_\alpha\epsilon_B)$ quantity -- will eventually become
numerically larger than the $\mathcal{O}(\epsilon_\alpha)$ centrifugal
force from equation \eqref{eq:Omega_alpha} at some radius $r<R_*$. At
this point our perturbative ordering breaks down, and our scheme and
equations become invalid.

To see how this problem would manifest itself in practice, we solved the system of equations
\eqref{DAE1} and \eqref{diff_DAE2} up to an outer radius
$R_\textrm{out}<R_*$, and investigated the behaviour of the solutions
in the limit $R_\textrm{out}\to R_*$. As $R_\textrm{out}$ was
increased, the amplitude of the radial functions $\bar{U},X$ was seen
to diverge. At the inner and outer boundaries, however, the solutions
were well-behaved.

The simplest resolution to this problem is to place an outer boundary
$R_\textrm{out}$ somewhere beneath the
stellar surface. For the stellar model studied in \citet{mestel2},
this was chosen to be $0.99R_*$. Our numerical results from varying
the cut-off radius, on the other hand, 
showed that the solutions were relatively insensitive to the exact
cut-off value provided $R_\textrm{out}\lesssim 0.93R_*$; accordingly, we will
place the outer boundary for the calculation
at a value $R_\textrm{out}=0.9R_*$. For main-sequence stars and white dwarfs this is an
arbitrary choice, and a discussion of its validity is given at the end of the paper,
in section \ref{conclusions}. For neutron stars, however, we can give
a more quantitative argument for placing the outer boundary at
$0.9R_*$: it is (approximately) the radius at which the star's fluid core gives way to a solid,
elastic crust, and so provides a natural cut-off for our fluid
calculation. Next we discuss how our solutions might be affected by
elastic forces in a neutron-star crust, if we were to extend our
modelling to include this region.
 
\subsection{Conditions in the crust of a neutron star}
\label{cc-boundary}

To extend our analysis to include the effects of a neutron-star crust,
we would need to include elastic-force terms in our perturbation
equations. This is not straightforward, however, as the resulting
terms have a separate perturbative scaling from those in our
problem. In addition, one has to define what the relaxed state of
the crust is -- i.e. the state in which the stresses are zero (and
therefore the crust is described by fluid terms alone). In a real
neutron star, the crust's elastic stress pattern is likely to be
complicated and to depend on the star's rotational and seismic history.

For the purposes of this paper, we just wish to make a qualitative
assessment of how elasticity might
modify our solution; for this there are two limiting cases. In one
limit, the crust is so strong that it can maintain its spheroidal
centrifugal distortion even as the star's instantaneous rotation
vector is substantially misaligned from the primary rotation vector
(recall figure \ref{alphaframe});
the crust would therefore undergo rigid-body free precession with no
additional $\bxi$-motions. In the other limit, the crust is so weak
that elastic forces have negligible contribution at all perturbative
orders we consider. We will argue next that a neutron-star crust falls
between these two limits. A separate issue to consider is whether the
stresses could be large enough to exceed the elastic yield limit of
the crust. \skl{Finally, we discuss the coupling between the crust and
core in a magnetised precessing neutron star.}

\subsubsection{Importance of elastic force terms}

We begin by assessing the size of elastic force terms.
For the sake of simplicity, let us define the crust as being relaxed
in the presence of its magnetic distortion; this is not crucial for
the argument, but makes it clearer. In this case, the combined
background and $\mathcal{O}(\epsilon_B)$ equations will look the same
as in the fluid case, since no elastic terms will appear. If we then
subtract force terms of these two orders from the full, unperturbed
Euler equation \eqref{general_Euler}, we are left with an equation containing
$\mathcal{O}(\epsilon_\alpha)$ terms, $\mathcal{O}(\epsilon_\alpha\epsilon_B)$
terms, and the elastic-force terms per unit mass $\delta{\bf
  F}_\textrm{el}$, whose leading-order pieces scale as
\beq
\delta{\bf F}_\textrm{el}\sim \frac{\mu}{\rho}\nabla^2\bxi
 \sim \frac{\mu}{\rho}\frac{\xi}{R_*^2}\sim \frac{\mu\epsilon_\alpha}{\rho R_*},
\eeq
where $\mu$ is the shear modulus. To find the perturbative order of
this quantity we nondimensionalise it by dividing by
$P R^2/\mathcal{M}$:
\beq
\frac{\delta{\bf F}_\textrm{el}}{P R_*^2/\mathcal{M}}
 \sim \frac{\mu}{P\rho}\frac{\mathcal{M}}{R_*^3}
 \sim \frac{\mu}{P}\epsilon_\alpha.
\eeq
Since this quantity scales with a new dimensionless parameter, $\mu/P$, it prevents us from
separating out the $\mathcal{O}(\epsilon_\alpha)$ and
$\mathcal{O}(\epsilon_\alpha\epsilon_B)$ equations as before. 
The ratio $\mu/P\lesssim 0.01$ for neutron star crusts -- so one can
think of the elastic forces as representing a numerically small correction
to the $\mathcal{O}(\epsilon_\alpha)$ equations, and therefore a small
correction to our existing $\delta\rho_\alpha$ solution, equation
\eqref{delta_rho_alpha}. On the other hand, they represent a potentially dominant
piece of the $\mathcal{O}(\epsilon_\alpha\epsilon_B)$ equations, and
therefore could drastically change the solution for $\delta\bB$.

The conclusion we draw from this is that the crust is weak enough that
the centrifugal bulge will always remain approximately symmetrical
about the primary-rotation axis, but strong enough that our solutions
for $\dot\bxi$ and $\delta\bB$ could be radically altered in the
crustal region. Since we are obliged to place an outer boundary some
distance within the star anyway, to ensure the validity of our
perturbative scheme, we find it natural simply to exclude the crust
from our modelling altogether, and place an outer boundary at the
crust-core interface, $r=0.9R_*$.

\subsubsection{Exceeding the crustal breaking strain}

Like any elastic medium, a neutron star crust has a yield strain
$\sigma_\textrm{max}$, beyond which value it ceases to respond
elastically to additional strain and instead breaks\footnote{`Break'
  in this context is likely to be a plastic deformation rather than a
  brittle fracture \citep{jones03}.}. Recent molecular-dynamics simulations of
neutron-star crustal matter \citep{horowitz} indicate that this happens at a very large
value, $\sigma_\textrm{max}\sim 0.1$. We need to compare this with the
strain tensor induced by the $\bxi$-motions, $\nabla\bxi$:
\beq
\nabla\bxi \sim \frac{\epsilon_\alpha R_*}{R_*} \sim \epsilon_\alpha
 \sim 0.2\brac{\frac{\alpha}{\textrm{1 kHz}}}^2,
\eeq
where the right-hand side is an estimate of $\epsilon_\alpha$ from
\citet{jonesand_02}. Comparing this with $\sigma_\textrm{max}$, we
conclude that the crust will not break unless the rotation frequency
exceeds $\sim 700$ Hz, about half the Keplerian frequency for a
neutron star. Rotation rates more rapid than this value start to
violate $\epsilon_\alpha\ll 1$, one of our original ans\"atze, anyway
-- so this case is outside the confines of our model. We may therefore assume the crust
never breaks, without additional loss of generality.

\skl{
\subsubsection{Crust-core coupling}

This paper focusses on fluid stars whose only `rigidity' comes from
their magnetic fields. By contrast the solid crust of a neutron
star can, of course, undergo conventional rigid-body free precession without
the aid of any magnetic field -- provided it has some permanent
distortion misaligned from the rotation axis. If we temporarily ignore
the magnetic rigidity of the core, this fluid region will be unable to
precess, and the resulting dynamics of the neutron star will be a
precessing crust atop a rigidly rotating core.

In fact, the star's magnetic field would generally thread the crust
and core, and couple together the two regions on an Alfv\'en timescale
$\tau_A$. In the case where $\tau_A\lesssim \sqrt{T_\alpha T_\omega}$, \citet{levin_dang} showed that the core's
motion is strongly coupled to that of the crust, so that the
whole star would precess as one\footnote{More accurately, only the \emph{charged}
  component of the core (protons and electrons) would couple to the
  crust, not the neutron superfluid component.}. Thus, in general a neutron
star's precession should combine two effects: the non-rigid precession
on which this paper is focussed (related to the magnetic distortion of
the fluid), and a coupling to the rigid-body
precession of the crust (related only to the crust's permanent
distortion). We close by noting that for our particular stellar model,
the latter effect may not actually occur: purely toroidal magnetic field lines
are always non-radial and so no field line will cross from
the crust to the core.
}

\subsection{Boundary conditions for our model}

Our system of equations, \eqref{DAE1} and \eqref{diff_DAE2}, is second-order in $\bar{U}_l$ and first-order in
$X_l$. This means that for each $\bar{U}_l$ we need to impose two boundary
conditions, and for each $X_l$ one boundary condition.

We begin with boundary conditions at the centre, where we are obliged
to use certain relations to fix the `integration constant' in our problem, and thus to ensure our
results are solutions of the original problem as encoded in equations
\eqref{DAE1} and \eqref{DAE2} (see subsection
\ref{DAE-centrecond}). For $l_\textrm{max}=4$ and $m=1$, the centre conditions
\eqref{U41_cent} and \eqref{X31_cent} fix $\bar{U}^1_4$ and $X^1_3$ in terms
of $\bar{U}_2^1$, but the value of $\bar{U}_2^1$ itself at the centre is
unknown. Let us therefore simply demand that $\bar{U}_2^1$ tends smoothly to
some constant at the centre, by imposing the condition
\beq
\td{\bar{U}_2^1}{r}(r=0)=0.
\eeq
So far $X_1^1$ is unfixed; it cannot be written in terms of the other
quantities at the centre, and its actual value there is unknown. We
also cannot evade the problem as done above for $\bar{U}_2^1$, by imposing
$X_1^1{}'(0)=0$, since our system of equations is only first-order in
$X$. Instead we will impose an outer boundary condition on $X^1_1$;
see below.

For $l_\textrm{max}=4$ and $m=2$, the only $X$-function is $X_3^2$. The
method of subsection
\ref{DAE-centrecond} gives $X_3^2(0)$ in terms of the other two radial
functions at the centre, $U_2^2(0)$ and $U_4^2(0)$:
\beq
X_3^2(0) = -\frac{2}{15\sqrt{7}}U^2_2(0)+\frac{207}{550}\sqrt{\frac{3}{7}}U^2_4(0).
\eeq
Again, we ensure that the $\bar{U}$ functions of this expression approach
some constant value at the centre by imposing
\beq
\td{U_2^2}{r}(r=0)=\td{U_4^2}{r}(r=0)=0.
\eeq

We now move to the conditions at the outer boundary, corresponding to
the crust-core boundary for a neutron star; here, we would ideally like to
ensure that the interior perturbed magnetic field matches smoothly to its exterior
counterpart. Mathematically speaking this matching is not necessary, as steps in the non-radial field
components across the boundary may be matched consistently using a
current sheet (steps in the radial component would, however, violate
$\div\bB=0$). Physically, however, such a current sheet is undesirable,
as it must be balanced by additional forces beyond those included
in our problem, or otherwise would result in a net acceleration
acting tangential to the boundary and oscillating on the precession
timescale.

For the continuity of the poloidal component, we need to match the
interior $U_l$ and $V_l$ functions smoothly to the exterior, but by
equation \eqref{elimV} $V_l\sim U'_l$, meaning that we need the
continuity of $U_l$ \emph{and} its radial derivative across the
boundary $R_\textrm{out}$. This means we can safely differentiate the
expression \eqref{Uext_orig} at the boundary, as we were able to do outside the
star, and extend the result \eqref{Uext_diff} to the outer boundary:
\beq \label{U_surface}
R_\textrm{out} U'_l(R_\textrm{out})+(l+2)U_l(R_\textrm{out})=0.
\eeq
We have two boundary conditions to specify for each $U_l$, and since
we have only specified one at the centre for each of them, we may
indeed impose equation \eqref{U_surface}, ensuring the continuity
of the poloidal field.

For the continuity of the toroidal component, we see from
\eqref{deltaB_VSH} that it suffices to match the interior and exterior
$W_l$ functions (or equivalentally the $X_l$ functions) at
$r=R_\textrm{out}=0.9R_*$. But the exterior condition is $X_l=0$, so these
functions must also be zero at the outer boundary:
\beq \label{X_surface}
X_l(R_\textrm{out})=0.
\eeq
For the $X_l$ functions we only have the
freedom to impose a single boundary condition, however, and for
$X_3^1$ and $X_3^2$ this \emph{must} be the one at the centre, so that
we pick out the correct DAE solution. Only for $X_1^1$, which does not
enter the centre conditions, are we able to enforce \eqref{X_surface}
at the outer boundary. In our
results we will see that the $l=3$ $X$ functions do indeed not vanish at the
outer boundary, although their values there are numerically small in
comparison with the typical value of the $X$ functions in the
interior. If one insisted on smooth interior-exterior matching for the
toroidal field, it should be possible to solve the equations with
the additional boundary condition \eqref{X_surface} for \emph{all}
$X$, but at the expense of making the system into an eigenvalue problem.

The boundary conditions used in our solution of the ODE
system \eqref{DAE1},\eqref{diff_DAE2} are summarised in table \ref{BCs}.

\begin{table*}
\begin{center}
\caption{\label{BCs}
               Summary of boundary conditions for our problem,
               truncated at $l=l_\textrm{max}=4$. There is one fewer
               boundary condition for $m=2$, since $X_1$ does not
               exist for this case.}
\begin{tabular}{ccc}
\hline
$m$ & centre & crust-core boundary \\
\hline
   & $U'_2(0)=0$ & $R_\textrm{out}U'_2(R_\textrm{out})+4U_2(R_\textrm{out})=0$ \\
1 & $U_4(0) = -\frac{55}{6\sqrt{6}}U_2(0)$ & $R_\textrm{out}U'_4(R_\textrm{out})+6U_4(R_\textrm{out})=0$\\
   & $X_3(0) = -\frac{7}{12}\sqrt{\frac{35}{2}}U_2(0)$ & $X_1(R_\textrm{out})=0$\\
\hline
   & $U'_2(0)=0$ & $R_\textrm{out}U'_2(R_\textrm{out})+4U_2(R_\textrm{out})=0$ \\
2 & $U'_4(0) =0$ & $R_\textrm{out}U'_4(R_\textrm{out})+6U_4(R_\textrm{out})=0$\\
   & $X_3(0) = -\frac{2}{15\sqrt{7}}U_2(0)+\frac{207}{550}\sqrt{\frac{3}{7}}U_4(0)$ & --\\
\hline
\end{tabular}\\
\end{center}
\end{table*}

\section{The $\bxi$-motions}
\label{xi-motions}

We have derived a set of differential equations whose solution gives
us $\delta\bB$. Together with the simple solution for
$\delta\rho_\alpha$ (see equation \eqref{delta_rho_alpha}), this gives us enough
information to solve for the perturbed velocity field $\dot\bxi$,
using the perturbed continuity and induction equations \eqref{continuity-xi},
\eqref{induction-xi}. In this section we show that $\dot\bxi$ may be found 
explicitly in terms of the magnetic functions $U_l^m$ and $W_l^m$,
through relatively simple algebraic manipulation of equations \eqref{continuity-xi}
and \eqref{induction-xi}.

The first task is to unwrap the curl operator in the induction
equation, for which we use the fact that the magnetic field must be
expressible as a Mie representation \eqref{Mie_rep}:
\beq
\curl(\dot\bxi\times\bB_0)=\delta\dot\bB=\curl(\nabla\dot{\mathcal{P}}\times\br+\dot{\mathcal{T}}\br)
 \implies \dot\bxi\times\bB_0=\nabla\dot{\mathcal{P}}\times\br+\dot{\mathcal{T}}\br+\nabla\dot{\mathcal{G}},
\eeq
where $\mathcal{P},\mathcal{T}$ are poloidal and toroidal scalar
functions, respectively, and $\mathcal{G}$ is a third scalar function
representing the freedom that -- upon equating terms under a curl
operator -- the resulting equation is only defined up to the addition
of curl-free terms $\nabla\mathcal{G}$.

We first need, therefore, to rewrite the general expression for
$\delta\bB$ from \eqref{deltaB_VSH} in its Mie representation. Note
that we could, in fact, commence with $\delta\bB$ in the simpler form
given in \eqref{deltaB-positivem}, which is specific to our
problem. We prefer, however, to consider the general case, so that
parts of the analysis here may be directly applied to other problems
(for example, precession of a magnetic star where $\bB_0$ is a
poloidal or mixed poloidal-toroidal field).

To find the toroidal scalar $\mathcal{T}$, decomposed in spherical harmonics as
$\mathcal{T}=\sum_{l,m}\mathcal{T}_l^m Y_l^m$, we need only rearrange
the toroidal component $\delta\bB_\textrm{tor}$ from equation \eqref{deltaB_VSH}:
\beq
\delta\bB_\textrm{tor}
 = \sum\limits_{l,m} W_l^m\be_r\times\nabla Y_l^m
 = -\sum\limits_{l,m} \curl\brac{\frac{W_l^m Y_l^m}{r}\br}
 = \sum\limits_{l,m} \curl\brac{\mathcal{T}_l^m Y_l^m \br},
\label{dBtor_Mie}
\eeq
using the product rule. In the above we have, as before, defined
\beq
\sum\limits_{l,m}\equiv
\sum\limits_{l=1}^\infty\sum\limits_{m=-l,m\neq 0}^{l}
\eeq
for brevity.

For the poloidal component $\delta\bB_\textrm{pol}$ we begin by
writing $\mathcal{P}=\sum_{l,m} \mathcal{P}_l^m Y_l^m$ and look at
one arbitrary $(l,m)$-component in the sum for
$\curl(\nabla\mathcal{P}\times\br)$:
\begin{align}
\curl[\nabla(\mathcal{P}_l^m Y_l^m)\times\br]
 &= \nabla\mathcal{P}_l^m\times(\nabla Y_l^m\times\br) + \mathcal{P}_l^m\curl(r\nabla Y_l^m\times\be_r)\nn\\
 &= r\mathcal{P}_l^m{}'\nabla Y_l^m + \mathcal{P}_l^m\nabla Y_l^m + r\mathcal{P}_l^m\curl(\nabla Y_l^m\times\be_r)\nn\\
 &= (r\mathcal{P}_l^m{}'+\mathcal{P}_l^m)\nabla Y_l^m + \frac{l(l+1)\mathcal{P}_l^m}{r}Y_l^m\be_r,
\end{align}
using similar algebraic steps to those in equation \eqref{curldeltaB}.
Now comparing the above result with the expression obtained by
substituting \eqref{elimV} into \eqref{deltaB_VSH}:
\beq
\delta\bB_\textrm{pol}
 = \sum\limits_{l,m} \left[ \frac{(r^2 U_l^m{}'+2rU_l^m)}{l(l+1)}\nabla Y_l^m + U_l^m Y_l^m\be_r \right],
\eeq
we see that the poloidal radial functions in the Mie representation
are given by
\beq
\mathcal{P}_l^m = \frac{rU_l^m}{l(l+1)}.
\label{dBpol_Mie}
\eeq
We are now able to write $\dot\bxi\times\bB_0$ in vector spherical
harmonic form:
\begin{align}
\dot\bxi\times\bB_0
 &= \pd{ }{t}\brac{\nabla\mathcal{P}\times\br+\mathcal{T}\br+\nabla\mathcal{G}}\nn\\
 & = \pd{ }{t}\sum\limits_{l,m}\left[r\mathcal{P}_l^m\nabla Y_l^m\times\be_r
                        +r\mathcal{T}_l^m Y_l^m\be_r +\nabla(\mathcal{G}_l^m Y_l^m) \right]\nn\\
 &= \sum\limits_{l,m}\left[ (\mathcal{G}_l^m{}'-W_l^m) \pd{Y_l^m}{t}\be_r +\mathcal{G}_l^m\nabla\brac{\pd{Y_l^m}{t}}
                                         + \frac{r^2 U_l^m}{l(l+1)}\nabla \brac{\pd{Y_l^m}{t}}\times\be_r \right] \nn\\
 &= \omega\sum\limits_{l,m}\left[im\brac{
           (\mathcal{G}_l^m{}'-W_l^m) Y_l^m\be_r +\mathcal{G}_l^m\nabla Y_l^m + \frac{r^2 U_l^m}{l(l+1)}\nabla Y_l^m\times\be_r } \right],
\label{xiB_VSH}
\end{align}
where we have used the result
\beq \label{dYdt}
\pd{Y_l^m(\theta,\lambda)}{t}=\pd{Y_l^m}{\lambda}\pd{\lambda}{t}=\pd{Y_l^m}{\lambda}\pd{}{t}(\omega t+\phi)
 = im\omega Y_l^m
\eeq
and have, predictably, defined $\mathcal{G}_l^m$ as the radial terms
in a spherical-harmonic expansion of $\mathcal{G}$,
i.e. $\mathcal{G}=\sum_{l,m}\mathcal{G}_l^m Y_l^m$.
Now, since we have a toroidal background field $\bB_0=B_0\be_\lambda$,
\beq \label{xiB_prod}
\dot\bxi\times\bB_0 = B_0(\dot\xi_\theta\be_r-\dot\xi_r\be_\theta).
\eeq
We now use equations \eqref{xiB_VSH} and \eqref{xiB_prod} to equate the $r$, $\theta$
and $\lambda$ components of $\dot\bxi\times\bB_0$:
\begin{align}
[\dot\bxi\times\bB_0]_r &=\ B_0\dot\xi_\theta\ 
 = \omega\sum\limits_{l,m} [imY_l^m(\mathcal{G}_l^m{}'-W_l^m)],\label{xiB_r}\\
[\dot\bxi\times\bB_0]_\theta &= -B_0\dot\xi_r
 = \omega\sum\limits_{l,m} \left[ \frac{1}{r\sin\theta} \brac{
         -\frac{m^2 r^2 U_l^m Y_l^m}{l(l+1)} + im\mathcal{G}_l^m\sin\theta Y_{l,\theta}^m} \right], \label{xiB_theta}\\
[\dot\bxi\times\bB_0]_\lambda &=\ \ \ 0\ \ \ \ 
 = \omega\sum\limits_{l,m} \left[ -\frac{1}{r\sin\theta} \brac{
          \frac{imr^2U_l^m\sin\theta Y_{l,\theta}^m}{l(l+1)} + m^2\mathcal{G}_l^m Y_l^m} \right]. \label{xiB_lambda}
\end{align}
Firstly, by the orthogonality of the $e^{im\lambda}$ terms within
spherical harmonics, we know that equation \eqref{xiB_lambda} must
hold for each $m$ separately, so that this equation corresponds to a set of equalities
involving sums over $l$ alone.  In each of these equalities it is permissible to
take $m$ outside the sum, and in particular to divide by $m^2$, so
that
\beq \label{elimG}
\sum\limits_{l\geq |m|}\mathcal{G}_l^m Y_l^m
 = -\sum\limits_{l\geq |m|}\frac{ir^2 U_l^m\sin\theta Y_l^m{}_{,\theta}}{l(l+1)m}
 = -\sum\limits_{l\geq |m|}\frac{ir^2 U_l^m}{l(l+1)m} [lQ_{l+1}Y_{l+1}^m-(l+1)Q_l Y_{l-1}^m],
\eeq
using equation \eqref{sinYlm}. Differentiating this relation with
respect to $r$ gives $(\mathcal{G}_l^m Y_l^m)_{,r}=\mathcal{G}_l^m{}'
Y_l^m$; we substitute this result back into equation \eqref{xiB_r} to
eliminate $\mathcal{G}_l^m$ and get an explicit expression for the
$\theta$-component of $\dot\bxi$:
\beq\label{xi_theta}
\dot\xi_\theta
 = -\frac{\omega}{B_0}\sum\limits_{l,m}\bigg\{
         imW_l^m Y_l^m - \frac{(r^2 U_l^m{}'+2rU_l^m)}{l(l+1)}[lQ_{l+1}Y_{l+1}^m-(l+1)Q_l Y_{l-1}^m] \bigg\}.
\eeq
Next, with a view to eliminating $\mathcal{G}$ from equation
\eqref{xiB_theta}, we use \eqref{elimG} to show that
\beq
\sum\limits_{l\geq |m|}\mathcal{G}_l^m Y_{l,\theta}^m = \sum\limits_{l\geq |m|} (\mathcal{G}_l^m Y_l^m)_{,\theta}
 = -\sum\limits_{l\geq |m|}\frac{ir^2 U_l^m}{l(l+1)m}(\sin\theta Y_{l,\theta}^m)_{,\theta}
 = -\sum\limits_{l\geq |m|}\frac{ir^2 U_l^m}{l(l+1)m} [ lQ_{l+1}Y_{l+1}^m{}_{,\theta}-(l+1)Q_l Y_{l-1}^m{}_{,\theta} ].
\eeq
Now applying \eqref{sinYlm} to the spherical harmonic derivatives on
the right-hand side of the above equation, we can show that
\beq
\sum\limits_{l\geq |m|}\mathcal{G}_l^m\sin\theta Y_l^m{}_{,\theta}
 = -\sum\limits_{l\geq |m|}\frac{ir^2 U_l^m}{l(l+1)m}\Big\{
          l(l+1)Q_{l+1}Q_{l+2}Y_{l+2}^m - [ l(l+2)Q_{l+1}^2+(l-1)(l+1)Q_l^2 ]Y_l^m 
         + l(l+1)Q_{l-1}Q_l Y_{l-2}^m \Big\}. 
\eeq
Substituting this result into \eqref{xiB_theta} yields an explicit
expression for the $r$-component of $\dot\bxi$:
\beq \label{xi_r}
\dot\xi_r
 = -\frac{\omega}{B_0}\sum\limits_{l,m} \Bigg\{ \frac{rU_l^m}{l(l+1)\sin\theta} \Big\{
         l(l+1)Q_{l+1}Q_{l+2}Y_{l+2}^m - [m^2+l(l+2)Q_{l+1}^2+(l-1)(l+1)Q_l^2 ]Y_l^m 
         + l(l+1)Q_{l-1}Q_l Y_{l-2}^m \Big\} \Bigg\}.
\eeq
The induction equation has given us expressions for $\dot\xi_r$ and
$\dot\xi_\theta$. To find the azimuthal component $\dot\xi_\lambda$ we
now use the continuity equation:
\beq
-\delta\dot\rho_\alpha = \div(\rho_0\dot\bxi)
 = \rho'_0\dot\xi_r + \rho_0 \left[ \frac{(r^2\dot\xi_r)_{,r}}{r^2}
  + \frac{(\sin\theta\dot\xi_\theta)_{,\theta}}{r\sin\theta} + \frac{\dot\xi_{\lambda,\lambda}}{r\sin\theta} \right],
\eeq
which we rearrange to give
\beq \label{xi_lamlam}
\dot\xi_{\lambda,\lambda}
 = -\frac{r\sin\theta\delta\dot\rho_\alpha}{\rho_0} - \frac{r\sin\theta\rho'_0\dot\xi_r}{\rho_0}
    -\frac{(r^2\dot\xi_r)_{,r}\sin\theta}{r} - (\sin\theta\dot\xi_\theta)_{,\theta}.
\eeq
The middle two terms on the right-hand side of this equation may be
written down immediately, whilst the other two need minor algebraic manipulation. Firstly, differentiating
\eqref{delta_rho_alpha} with respect to time yields:
\beq
\delta\dot\rho_\alpha
 = \frac{5\pi\alpha^2\omega}{8G} j_2(\Mr)
        \sin\theta [ \sin(2\chi)\cos\theta\sin\lambda + \sin^2\chi\sin\theta\sin(2\lambda) ].
\eeq
The final term on the right-hand side of \eqref{xi_lamlam} is
\begin{align}
-(\sin\theta\dot\xi_\theta)_{,\theta}
 =& \omega\pd{ }{\theta}\brac{\frac{\sin\theta}{B_0}}\sum\limits_{l,m}\bigg\{
         imW_l^m Y_l^m - \frac{(r^2 U_l^m{}'+2rU_l^m)}{l(l+1)}[lQ_{l+1}Y_{l+1}^m-(l+1)Q_l Y_{l-1}^m] \bigg\}\nn\\
 &+ \frac{\omega}{B_0}\sum\limits_{l,m}\bigg\{
         imW_l^m\sin\theta Y_l^m{}_{,\theta} - \frac{(r^2 U_l^m{}'+2rU_l^m)}{l(l+1)}[lQ_{l+1}\sin\theta Y_{l+1}^m{}_{,\theta}-(l+1)Q_l\sin\theta Y_{l-1}^m{}_{,\theta}] \bigg\}.
\end{align}
However, the first right-hand-side term of this vanishes, since
\beq
\pd{ }{\theta}\brac{\frac{\sin\theta}{B_0}} = \pd{ }{\theta}\brac{\frac{1}{\Lambda r\rho_0}} = 0
\eeq
for our background field, given by equation \eqref{B0}. The result,
upon a further application of \eqref{sinYlm}, is
\begin{align}
-(\sin\theta\dot\xi_\theta)_{,\theta}
 = \frac{\omega}{B_0}\sum\limits_{l,m}\bigg\{
    & imW_l^m [ lQ_{l+1}Y_{l+1}^m-(l+1)Q_l Y_{l-1}^m ]
       - \frac{(r^2 U_l^m{}'+2rU_l^m)}{l(l+1)}\Big\{
              l(l+1)Q_{l+1}Q_{l+2}Y_{l+2}^m \nn\\
    & - [ l(l+2)Q_{l+1}^2+(l-1)(l+1)Q_l^2 ]Y_l^m + l(l+1)Q_{l-1}Q_l Y_{l-2}^m \Big\} \bigg\}.
\end{align}
Now evaluating the right-hand side of \eqref{xi_lamlam}, we find that
a large number of the terms involving $U_l^m$ functions cancel, and we are left with
\begin{align}
\dot\xi_{\lambda,\lambda}
 =& -\frac{5\pi\alpha^2\omega}{8G}\frac{r j_2(\Mr)}{\rho_0}
        \sin^2\theta [ \sin(2\chi)\cos\theta\sin\lambda + \sin^2\chi\sin\theta\sin(2\lambda) ]\nn\\
   &\ + \frac{\omega}{B_0}\sum\limits_{l,m} \Big\{
              imW_l^m [lQ_{l+1}Y_{l+1}^m-(l+1)Q_l Y_{l-1}^m]
              -\frac{m^2}{l(l+1)}(r^2 U_l^m{}'+2rU_l^m)Y_l^m \Big\}.
\end{align}
We now integrate this expression with respect to $\lambda$ to
give
\begin{align}
\dot\xi_\lambda
 =& \frac{5\pi\alpha^2\omega}{16G}\frac{r j_2(\Mr)}{\rho_0}
        \sin^2\theta [ 2\sin(2\chi)\cos\theta\cos\lambda + \sin^2\chi\sin\theta\cos(2\lambda) ]\nn\\
    &+ \frac{\omega}{B_0}\sum\limits_{l,m} \Big\{
             W_l^m [lQ_{l+1}Y_{l+1}^m-(l+1)Q_l Y_{l-1}^m]
              +\frac{im}{l(l+1)}(r^2 U_l^m{}'+2rU_l^m)Y_l^m \Big\}.
\label{xi_lambda}
\end{align}
Finally, by following the reasoning of section \ref{analysis_of_eqs}, we rewrite
the three components of $\dot\bxi$ (equations \eqref{xi_theta},\eqref{xi_r} and
\eqref{xi_lambda}) in a manifestly real-valued form, with sums over
$l$ alone:
\begin{align}
\dot\xi_r =&
  -\!\frac{\omega r}{B_0\sin\theta}\sum\limits_{l=1}^\infty \Bigg\{\!
      \frac{U_{2l}^1}{2l(2l+1)} \Big\{
         2l(2l+1)Q_{2l+1}Q_{2l+2}(Y_{2l+2}^1\!+\!Y_{2l+2}^{-1})
         - [1+2l(2l+2)Q_{2l+1}^2+(2l-1)(2l+1)Q_{2l}^2 ](Y_{2l}^1\!+\!Y_{2l}^{-1}) \nn\\
 & \hspace{2.2cm} 
        + 2l(2l+1)Q_{2l-1}Q_{2l} (Y_{2l-2}^1+Y_{2l-2}^{-1}) \Big\}
        + \frac{U_{2l}^2}{2l(2l+1)} \Big\{
          2l(2l+1)Q_{2l+1}Q_{2l+2}(Y_{2l+2}^2-Y_{2l+2}^{-2}) \nn\\
 & \hspace{2.2cm} 
        - [4+2l(2l+2)Q_{2l+1}^2+(2l-1)(2l+1)Q_{2l}^2 ](Y_{2l}^2-Y_{2l}^{-2})
        + 2l(2l+1)Q_{2l-1}Q_{2l} (Y_{2l-2}^2-Y_{2l-2}^{-2}) \Big\} \Bigg\},\label{final_xi_r}\\
\dot\xi_\theta =&
 -\frac{\omega}{B_0}\sum\limits_{l=1}^\infty\bigg\{
      i W_{2l-1}^1 (Y_{2l-1}^1+Y_{2l-1}^{-1})
      - \frac{(r^2 U_{2l}^1{}'+2rU_{2l}^1)}{2l(2l+1)}[2lQ_{2l+1}(Y_{2l+1}^1+Y_{2l+1}^{-1})-(2l+1)Q_{2l}(Y_{2l-1}^1+Y_{2l-1}^{-1})]\nn\\
 & \hspace{1.7cm} 
         +2i W_{2l-1}^2 (Y_{2l-1}^2-Y_{2l-1}^{-2})
         - \frac{(r^2 U_{2l}^2{}'+2rU_{2l}^2)}{2l(2l+1)}[2lQ_{2l+1}(Y_{2l+1}^2-Y_{2l+1}^{-2})-(2l+1)Q_{2l}(Y_{2l-1}^2-Y_{2l-1}^{-2})]\bigg\},\label{final_xi_theta}\\
\dot\xi_\lambda =&
 \ \frac{5\pi\alpha^2\omega}{16G}\frac{r j_2(\Mr)}{\rho_0}
                 \sin^2\theta [ 2\sin(2\chi)\cos\theta\cos\lambda + \sin^2\chi\sin\theta\cos(2\lambda) ]\nn\\
 &\ + \frac{\omega}{B_0}\sum\limits_{l=1}^\infty
           \bigg\{ W_{2l-1}^1 [ (2l-1)Q_{2l}(Y_{2l}^1-Y_{2l}^{-1})-2lQ_{2l-1}(Y_{2l-2}^1-Y_{2l-2}^{-1})]
                  + \frac{i(r^2 U_{2l}^1{}'+2rU_{2l}^1)}{2l(2l+1)}(Y_{2l}^1-Y_{2l}^{-1}) \nn\\
 & \hspace{1.7cm} 
                  + W_{2l-1}^2 [ (2l-1)Q_{2l}(Y_{2l}^2+Y_{2l}^{-2}) - 2lQ_{2l-1}(Y_{2l-2}^2+Y_{2l-2}^{-2})]
                  + \frac{i(r^2 U_{2l}^2{}'+2rU_{2l}^2)}{l(2l+1)}(Y_{2l}^2+Y_{2l}^{-2}) \bigg\}.
\label{final_xi_lambda}
\end{align}

\section{Results}
\label{results}

Finally, we come to the results of solving equations \eqref{DAE1} and
\eqref{diff_DAE2}, with the boundary conditions given in table
\ref{BCs}. Having recast the original system of DAEs into a simpler system of ODEs, the
numerical solution then becomes fairly straightforward, subject to a few
numerical subtleties which we detail in appendix
\ref{numerical-details}. In what follows, the radial functions are of
course time-independent, but the results for $\delta\bB$ and
$\dot\bxi$ are presented for $t=0$. Over a period of $2\pi/\omega$,
these 3D figures would rotate around the $z$-axis.

\subsection{The poloidal and toroidal radial functions}

\begin{figure}
\begin{center}
\begin{minipage}[c]{0.8\linewidth}
\psfrag{W11}{$W_1^1$}
\psfrag{W13}{$W_3^1$}
\psfrag{W23}{$W_3^2$}
\psfrag{U12}{$\bar{U}_2^1$}
\psfrag{U22}{$\bar{U}_2^2$}
\psfrag{U14}{$\bar{U}_4^1$}
\psfrag{U24}{$\bar{U}_4^2$}
\psfrag{r}{$\hat{r}$}
\includegraphics[width=\linewidth]{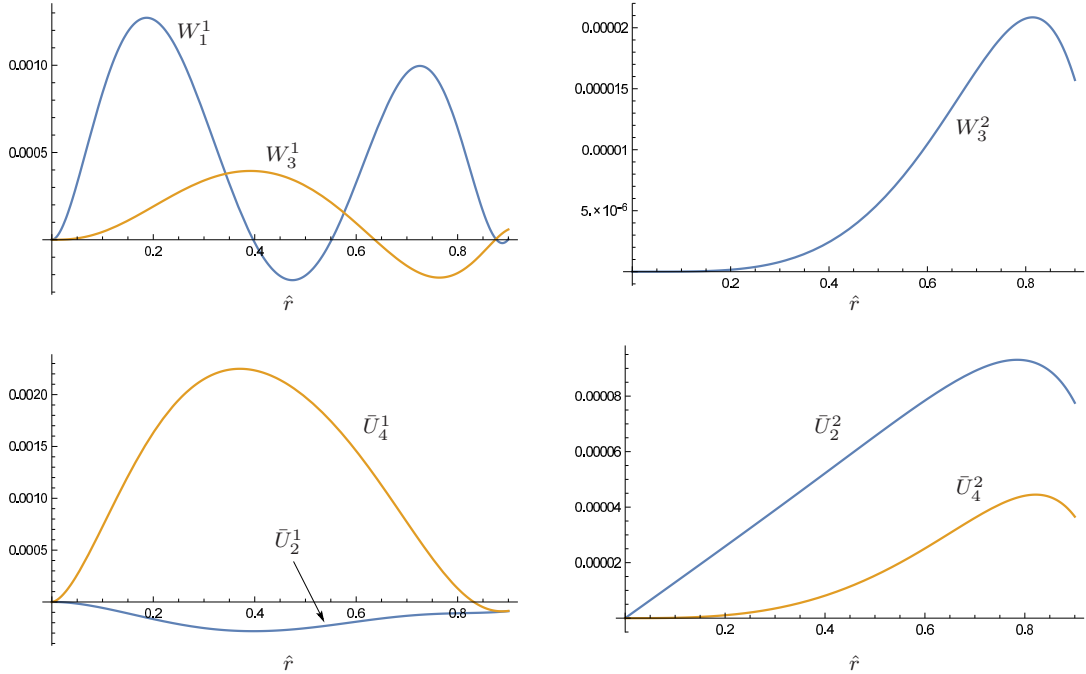}
\end{minipage}
\caption{\label{UW_pi4}
  Plots of the radial functions of $\delta\bB$ against dimensionless
  radius $\hat{r}$, for $l_\textrm{max}=4$, inclination angle
  $\chi=\pi/4$, dimensionless primary rotation
  frequency $\hat\alpha=0.1$ and dimensionless background magnetic field strength
  $\hat\Lambda=0.1$ (see equations \eqref{dim_alpha}-\eqref{dim_r} for formulae to
  redimensionalise these to physical stellar parameters). Varying $\chi$
  only changes the amplitude of the results, and the
  relative size of the $m=1$ and $m=2$ components; in the limit of an
  orthogonal rotator only the $m=2$ functions are non-zero.}
\end{center}
\end{figure} 

We recall that the fundamental quantities to solve
for were the toroidal radial functions $X_l^m\equiv W_l^m/r$ and the
poloidal radial functions $\bar{U}_l^m\equiv iU_l^m$. From these one can
find the perturbed magnetic field $\delta\bB$, by using equation
\eqref{deltaB-positivem}, and the perturbed velocity field -- the $\bxi$-motions -- by
using equations \eqref{final_xi_r}-\eqref{final_xi_lambda}. We begin
by plotting these poloidal and toroidal radial functions, in figure
\ref{UW_pi4}. All of these functions appear to tend to zero at the
origin, suggesting that the central boundary conditions described in
section \ref{DAE-centrecond} were actually unnecessary and that simply
imposing the vanishing of all variables would have been sufficient. Having
tried this, however, we found that the resulting
functions were \emph{not} solutions of the original system of DAEs,
\eqref{DAE1} and \eqref{DAE2}. Numerically, the $U$ and $X$ functions
have tiny -- but non-zero -- values at the centre, and fixing the correct
ratios between them (those of table \ref{BCs}) is in fact crucial
for getting a solution to the DAEs.

The $m=1$ functions in figure \ref{UW_pi4} are numerically larger then
their $m=2$ counterparts for the
chosen inclination angle of $\chi=\pi/4$; only very close to the
orthogonal-rotator case ($\chi=\pi/2$) do the $m=2$ perturbations
become larger than those for $m=1$. Perhaps the most significant other feature to
note is that the $\bar{U}_2^1$ function is a factor of $8$
smaller than the $\bar{U}_4^1$, i.e. the
quadrupolar perturbed magnetic field is significantly smaller than the
$l=4$ component. This contradicts one's experience from simpler
calculations -- in axisymmetric magnetic equilibria, for example, the
solutions feature odd $l$ and are dominantly dipolar, with the power
in the multipoles falling off monotonically \citep{ciolfi09,LAG}.

The solutions plotted in figure \ref{UW_pi4} are numerical, with no
closed-form expressions, and so are of limited applicability. For this
reason, we next explore the scaling of the solutions with rotation
rate, field strength and inclination angle, and present polynomial
approximations to the radial functions $\bar{U}_l^m$ and $W_l^m$.

From equations \eqref{DAE1} and \eqref{DAE2}, the source terms for the
radial functions $U^m_l,X^m_l$ are of the form
$\tilde{\Psi}_l^m/\Lambda$ and $\tilde{\Upsilon}_l^m/\Lambda$;
comparing with equations \eqref{Ups_lm} and \eqref{Psi_lm}, we see
that the $m=1$ source terms are proportional to
$\alpha^2\Lambda\sin(2\chi)$ and that the $m=2$ source terms are
proportional to $\alpha^2\Lambda\sin^2\chi$. We therefore expect the
same scaling for the radial functions $U^m_l,X^m_l$, which we have
confirmed by solving the equations for different values of
$\alpha,\Lambda$ and $\chi$ (plots omitted for brevity).

To avoid very lengthy approximations, we will use four
significant figures in the prefactor for each power of $r$. For each radial
function, we will employ the minimum number of powers of $r$ required so
that the deviation from the actual solution -- defined as the
difference between the approximate and actual solution, divided by the
maximum value attained by the actual solution -- never exceeds $2\%$ (it is
typically below $1\%$, in fact). With these demands, and using the
`Fit' function within Mathematica, we find the following
polynomial approximations to $\bar{U}_l^m$ and $W_l^m$, where we have
terminated the angular sum at $l_\textrm{max}=4$ as usual:
\begin{align}
\bar{U}_2^1(\hat{r}) &= \hat{\alpha}^2\hat{\Lambda}\sin(2\chi) \hat{r}^2 (-7.616 + 20.24 \hat{r} - 13.84 \hat{r}^2 + 0.9159 \hat{r}^3 - 5.827 \hat{r}^4 + 6.186 \hat{r}^5),\\
\bar{U}_4^1(\hat{r}) &= \hat{\alpha}^2\hat{\Lambda}\sin(2\chi) \hat{r}^2 (121.4 - 639.5 \hat{r} + 1548 \hat{r}^2 - 2062 \hat{r}^3 + 1430 \hat{r}^4 - 397.7 \hat{r}^5),\\
W_1^1(\hat{r}) &= \hat{\alpha}^2\hat{\Lambda}\sin(2\chi) \hat{r}^2 (188.6 - 1410 \hat{r} + 4340 \hat{r}^2 - 7531 \hat{r}^3 + 8342 \hat{r}^4 - 5643 \hat{r}^5 + 1720 \hat{r}^6),\\
W_3^1(\hat{r}) &= \hat{\alpha}^2\hat{\Lambda}\sin(2\chi) \hat{r}^3 (81.58 - 474.4 \hat{r} + 1245 \hat{r}^2 - 1819 \hat{r}^3 + 1409 \hat{r}^4 - 442.6 \hat{r}^5),\\
\bar{U}_2^2(\hat{r}) &= \hat{\alpha}^2\hat{\Lambda}\sin^2\chi \hat{r} (0.2287 + 0.3832 \hat{r} - 1.611 \hat{r}^2 + 2.844 \hat{r}^3 - 1.779 \hat{r}^4),\\
\bar{U}_4^2(\hat{r}) &= \hat{\alpha}^2\hat{\Lambda}\sin^2\chi \hat{r}^2 (-0.08230 + 1.079 \hat{r} - 2.798 \hat{r}^2 + 4.073 \hat{r}^3 - 2.286 \hat{r}^4),\\
W_3^2(\hat{r}) &= \hat{\alpha}^2\hat{\Lambda}\sin^2\chi \hat{r}^3 (0.0595 - 0.2124 \hat{r} + 0.9521 \hat{r}^2 - 0.8182 \hat{r}^3).
\end{align}
Note that the number of terms required for an accurate approximation
of each function differ, and that the minimum and maximum powers
featured in the above set of relations are $r$ and $r^8$ respectively. Constant (i.e. $r^0$) terms are
unnecessary, since the $U$ and $X$ functions are very small at the
origin, and the $W$ functions are exactly zero (since $W_l^m=rX_l^m$,
and the $X_l^m$ functions are finite at $r=0$); see figure \ref{UW_pi4}.

These functions may be redimensionalised by multiplying by the
appropriate combinations of $G,\rho_\textrm{c}$ and $R_*$, to give
values for a particular star. For a neutron star with mass
$1.4M_\odot$ and radius $R_*=10$ km, one can redimensionalise the
above functions to chosen values of rotation rate and field strength
using the replacements
\begin{align} 
\hat{\alpha} &= \frac{\alpha}{\sqrt{G\rho_\textrm{c}}}
 = \frac{\alpha}{12100\textrm{ rad s}^{-1}} = \frac{\alpha/2\pi}{1920\textrm{ Hz}},\label{dim_alpha}\\
\hat{\Lambda} &= \frac{\Lambda}{\sqrt{G}}
 = \frac{\langle B\rangle}{7.94\times 10^{16}\textrm{ G}},\label{dim_Lambda}\\
\hat{r} &= \frac{r}{R_*}=\frac{r}{10^6\textrm{ cm}},\label{dim_r}
\end{align}
where we have used the central density 
$\rho_\textrm{c}=2.19\times 10^{15}$ g cm${}^{-3}$ of a $\gamma=2$ polytrope with
the above physical mass and radius \citep{LJ09}, and have
rewritten $\Lambda$ in terms of the average field strength 
$\langle B\rangle$ by combining equation \eqref{epsB} with the ellipticity
formula (38) from \citet{LJ09}.

\subsection{Behaviour of higher multipoles at the centre}
\label{possible_multipoles}

As discussed in section \ref{DAE-centrecond}, we are limited to
truncating our solutions at a maximum angular index
$l=l_\textrm{max}=4$, and we anticipate that higher multipoles will
gradually become negligible. However, we have already seen from
figure \ref{UW_pi4} that the magnitude of the multipolar functions
does not decrease monotonically, and so we would like some idea of the
relative strength of higher multipoles.

Returning to our system of DAEs, \eqref{DAE1} and \eqref{DAE2}, we
note a few features. They couple together $X$ functions over an
$l$-range of 4, from $X_{l-2}$ to $X_{l+2}$, and $\bar{U}$ functions
over an $l$-range of 6, so it is natural to expect solutions to have
significant contributions from many multipoles. The highest multipoles
in the source terms, on the other hand, are an $l=4$ piece of
$\tilde{\Psi}$ and an $l=3$ piece of $\tilde{\Upsilon}$ (see section
\ref{Ylm_decomp_Ups_Psi}). From the way the source terms enter the DAEs, this means that
the $l=5$ system of equations is the last to contain any explicit
source, through the $\tilde{\Upsilon}_{l-2}$ term of equation \eqref{DAE2}. For all higher
values of $l$, the DAEs merely relate different $\bar{U}$ and $X$
functions to one another, and the higher-$l$ functions are only
non-zero by virtue of coupling to their lower-$l$ counterparts.

In lieu of solving the full system of DAEs for high $l_\textrm{max}$,
we may evaluate the equations at the centre of the star -- as done for
the second DAE, \eqref{DAE2}, in section \ref{DAE-centrecond}. For
$r=0$ all source terms vanish (as in the full DAEs at high $l$), and
we simply have a system of algebraic simultaneous equations relating
the values of different functions at the centre, $\bar{U}_l(0)$
and $X_l(0)$. When truncating for odd values of $l_\textrm{max}$, the
equations may be `solved' to give each quantity as a fraction of
$X_1(0)$, thus giving some idea of the way different multipoles couple
together, and their behaviour at high $l$. For lower values of
$l_\textrm{max}$ the results show some variation with the exact
truncation value, so we terminate at
$l_\textrm{max}=101$, well after the point at which any variation can
be seen.

\begin{figure}
\begin{center}
\begin{minipage}[c]{\linewidth}
\psfrag{l}{$l$}
\psfrag{Xcent}{$\displaystyle{\frac{X_l(0)}{X_1(0)}}$}
\psfrag{Ucent}{$\displaystyle{\frac{U_l(0)}{X_1(0)}}$}
\includegraphics[width=\linewidth]{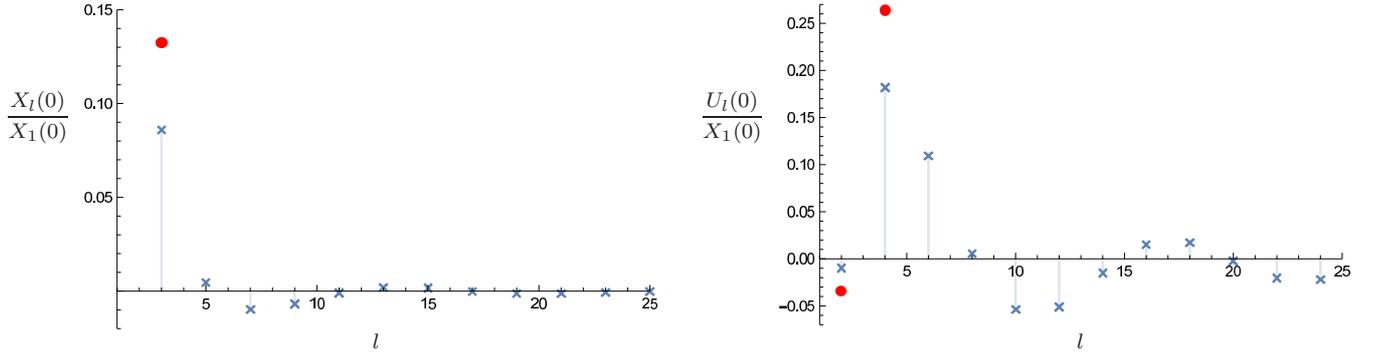}
\end{minipage}
\caption{\label{multipoles}
  Setting the radius to zero in the original system of DAEs,
  \eqref{DAE1} and \eqref{DAE2}, results in
  a system of simultaneous algebraic equations. We truncate these at
  $l_\textrm{max}=101$ to solve them, plotting the results for $l\leq
  25$ here. As solutions, we find the relative magnitudes of different multipoles
  $\bar{U}_l(0),X_l(0)$ at the centre, as a fraction of the
  value of the dipole piece $X_1(0)$ there (blue crosses). This suggests that the
  $U$ functions fall off more slowly with increasing $l$ than the $X$
  functions. For reference, the maximum values of $U_2,U_4$ and $X_3$ from
  our full solution are plotted as ratios of the maximum value of
  $X_1$ (filled red circles).}
\end{center}
\end{figure}

We plot the results of this experiment in figure \ref{multipoles},
where we see that the quantities $X_l(0)/X_1(0)$ and
$\bar{U}_l(0)/X_1(0)$ decay in an oscillatory manner with increasing
$l$, alternating between negative and positive values, rather than in
a more predictable monotonic manner. We also see that the $U_l(0)$
quantities decay far more slowly, with even the $l=24$ multipole being
more than 1\% of the value of $X_1(0)$; by contrast, every $X_l(0)$
quantity with $l>3$ is below 1\% of $X_1(0)$.

The only rigorous statement that one can make about figure
\ref{multipoles} is, of course, that it shows the solutions to the DAEs with $r$
set to zero. Nonetheless, we believe it also captures the essence of how
higher multipoles couple together in our problem. Evidence in favour
of this is the fact that it predicts the magnitudes of $\bar{U}_2^\textrm{max}/X_1^\textrm{max}$,
$\bar{U}_4^\textrm{max}/X_1^\textrm{max}$ and $X_3^\textrm{max}/X_1^\textrm{max}$, 
from our $l_\textrm{max}=4$ solutions, in the correct order, and in
particular the fact that $\bar{U}^1_2$ is far smaller than $\bar{U}_4^1$ and
has the opposite sign.

\subsection{Magnetic-field perturbations}

\begin{figure}
\begin{center}
\begin{minipage}[c]{\linewidth}
\psfrag{chi=1/16}{$\chi=\displaystyle{\frac{\pi}{16}}$}
\psfrag{chi=7/16}{$\chi=\displaystyle{\frac{7\pi}{16}}$}
\psfrag{chi=1/2}{$\chi=\displaystyle{\frac{\pi}{2}}$}
\psfrag{Bpol}{$|\delta\bB_\textrm{pol}|$}
\psfrag{Btor}{$|\delta\bB_\textrm{tor}|$}
\includegraphics[width=\linewidth]{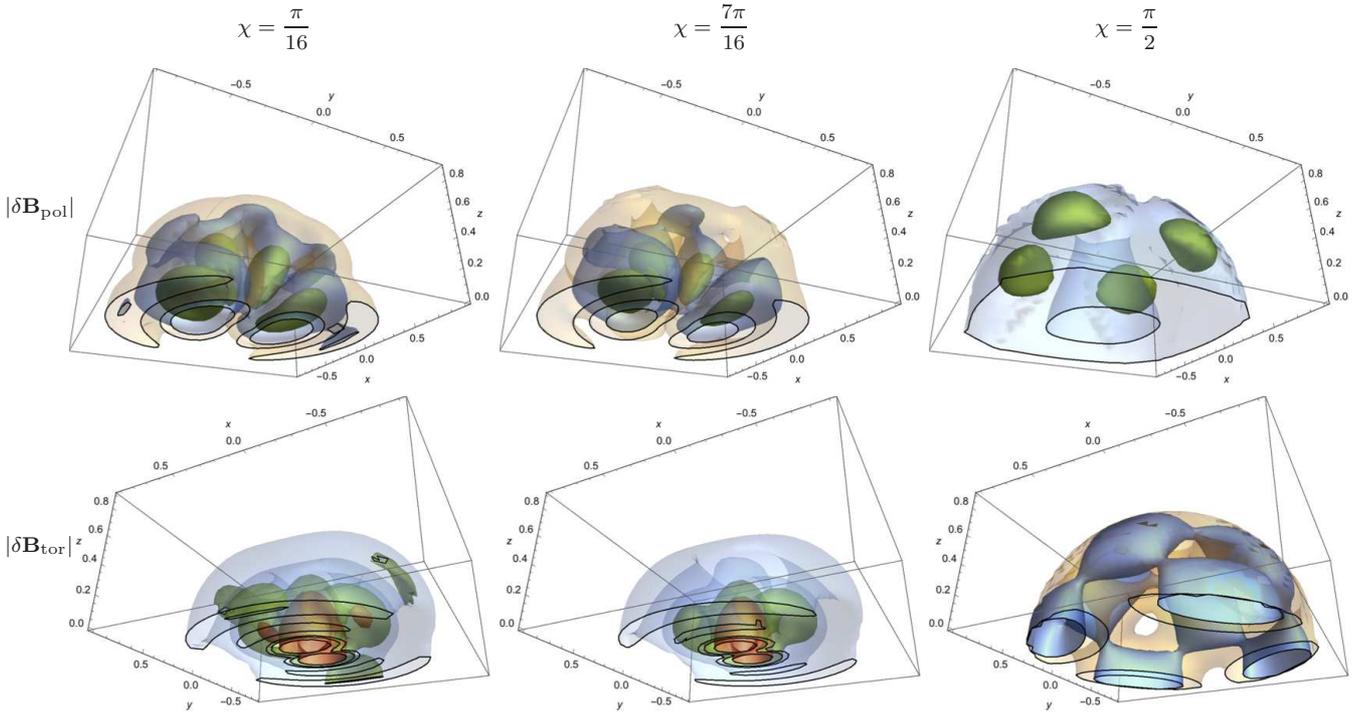}
\end{minipage}
\caption{\label{3D_Bmag}
  3-dimensional contour plots showing the magnitude of the perturbed
  magnetic field, separated into its poloidal (top row) and toroidal
  (bottom row) components. Note that viewing angles differ by $\pi/2$
  for the poloidal and toroidal plots. The results are symmetric about the $z=0$ plane,
  so we show only the northern hemisphere. Results are shown for three
  different values of inclination angle $\chi$. Contours are colour coded so that the blue contour
  represents a field strength double that of the light brown contour,
  green double the value of blue, and red double the
  value of green. For a fixed inclination angle the contour colouring is
  fixed, so that poloidal and toroidal components may be compared
  directly, but different inclination angles use different keys. The $m=1$
  field vanishes for the case of an orthogonal rotator ($\chi=\pi/2$),
  leaving only the $m=2$ component, and resulting in a very
  different-looking field (which is actually
  time-independent). However, the $m=2$ component is generally weak,
  so that even the near-orthogonal configuration ($\chi=7\pi/16$) is
  similar to the near-aligned model ($\chi=\pi/16$).}
\end{center}
\end{figure}

\begin{figure}
\begin{center}
\begin{minipage}[c]{\linewidth}
\includegraphics[width=\linewidth]{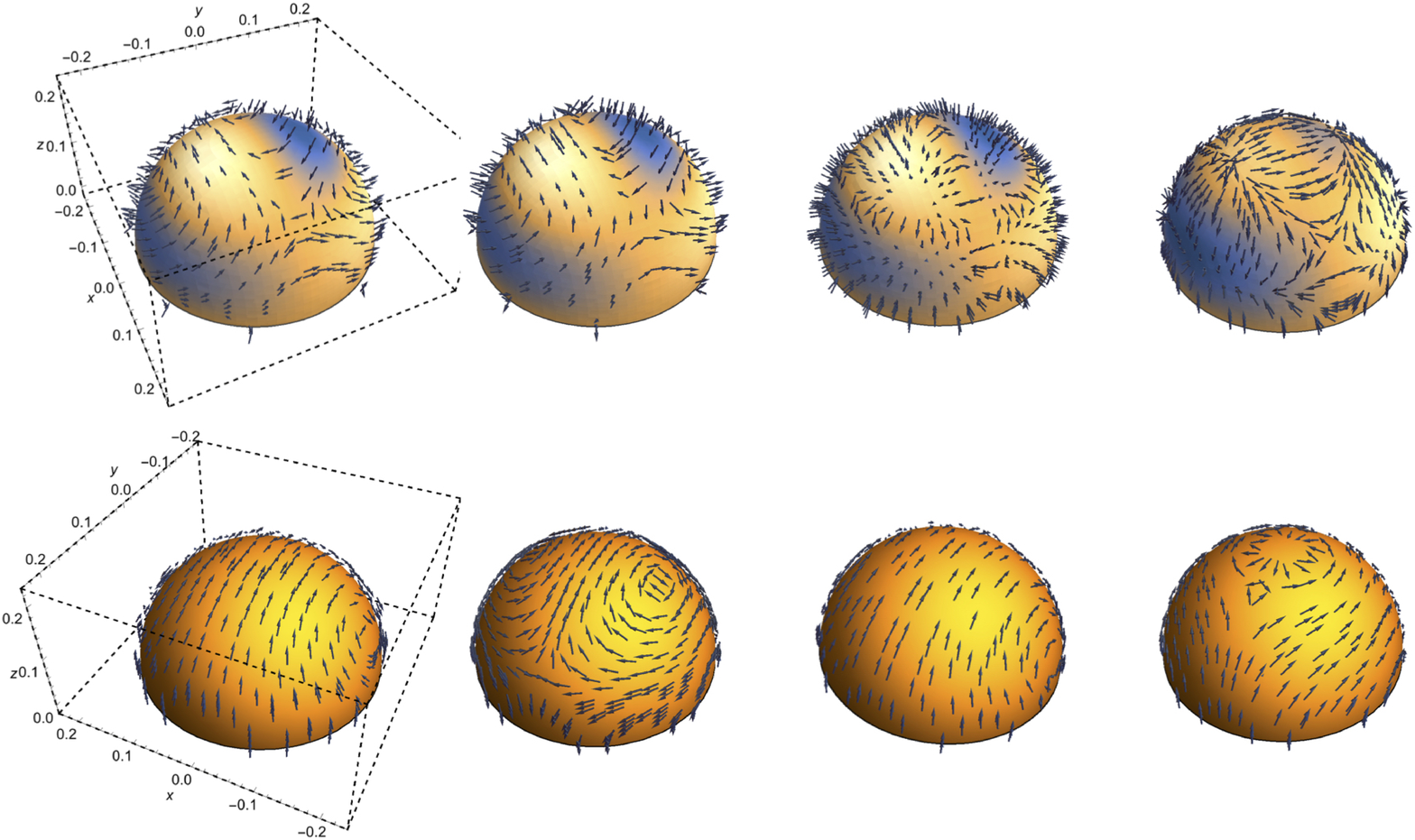}
\end{minipage}
\caption{\label{3D_Bvec}
  The direction of the perturbed magnetic field, plotted along a
  series of hemispherical shells beginning near the centre and ending
  near the surface. From left to right, the shells represent
  $r/R_*=0.2,0.4,0.6,0.8$ respectively. The upper row of plots shows
  the unit poloidal vector and the lower row the unit toroidal vector.
  Since arrows into and out of contours are hard to distinguish, we
  use colour coding for the strength of the radial field component:
  blue represents a region of field lines pointing towards the centre
  of the star, whilst white indicates outward-pointing field
  lines. This only applies to the poloidal-field plots, as toroidal
  fields are by definition non-radial, and so their arrows are all
  tangential to the hemispherical shells. Each poloidal/toroidal plot
  is viewed from the same $(x,y)$ position as in figure \ref{3D_Bmag},
  but from above -- i.e. for a $z>0$ position, unlike the $z<0$
  vantage points of figure \ref{3D_Bmag}. To save space the axes and bounding box are
  only drawn for the left-hand plots. Finally, we take $\chi=\pi/16$ for
  all plots, though there is little variation of the field's direction with inclination
  angle.}
\end{center}
\end{figure}

Having found solutions for $U_l$ and $W_l$, we may now insert these
into equation \eqref{deltaB-positivem} to find $\delta\bB$. The resulting field
configuration is complicated, and we have found it clearest to
separate it into its poloidal and toroidal components. First, in
figure \ref{3D_Bmag} we plot the magnitudes of these
components. Broadly speaking, the perturbed magnetic field is strongest towards
the centre, and the toroidal component is about double the strength of
the poloidal one. For most values of the inclination angle the $m=1$ component dominates
over $m=2$, so the near-aligned ($\chi=\pi/16$) and near-orthogonal
($\chi=7\pi/16$) results are visibly similar. For $\chi=\pi/2$, when
the magnetic and rotation axes are orthogonal, there is only an $m=2$
component. However, this situation is actually stationary (and
therefore does not correspond to precession); from
equation \eqref{omega} we see that $\omega=0$ for $\chi=\pi/2$, and
consequently (by equation \eqref{dYdt}),
$\partial (Y_l^m(\theta,\lambda))/\partial t = 0$.

Next, we show the direction of the magnetic field, in figure
\ref{3D_Bvec}. The clearest visualisation we could find was to plot
the poloidal and toroidal unit vectors over a nested sequence of
constant-radius shells with $r/R_*=0.2,0.4,0.6$ and $0.8$. Comparing
the field across consecutive shells should, hopefully, allow the reader to
visualise the full 3-dimensional field-line structure. The toroidal
component is clearer, since (by definition) its field lines run along
constant-$r$ shells. For the poloidal component, note that the blue
and white patches are in roughly the same position on each contour,
meaning that a given field line moves towards the centre of the star
through the blue regions, then bends back and comes out through a
neighbouring white region to its `starting' location, thus forming a
closed loop (as dictated by the requirement that $\div\delta\bB=0$).

\subsection{$\bxi$-motions}

\begin{figure}
\begin{center}
\begin{minipage}[c]{0.8\linewidth}
\psfrag{chi=1/16}{$\chi=\displaystyle{\frac{\pi}{16}}$}
\psfrag{chi=7/16}{$\chi=\displaystyle{\frac{7\pi}{16}}$}
\psfrag{chi=1/2}{$\chi=\displaystyle{\frac{\pi}{2}}$}
\psfrag{Bpol}{$|\delta\bB_\textrm{pol}|$}
\psfrag{Btor}{$|\delta\bB_\textrm{tor}|$}
\includegraphics[width=\linewidth]{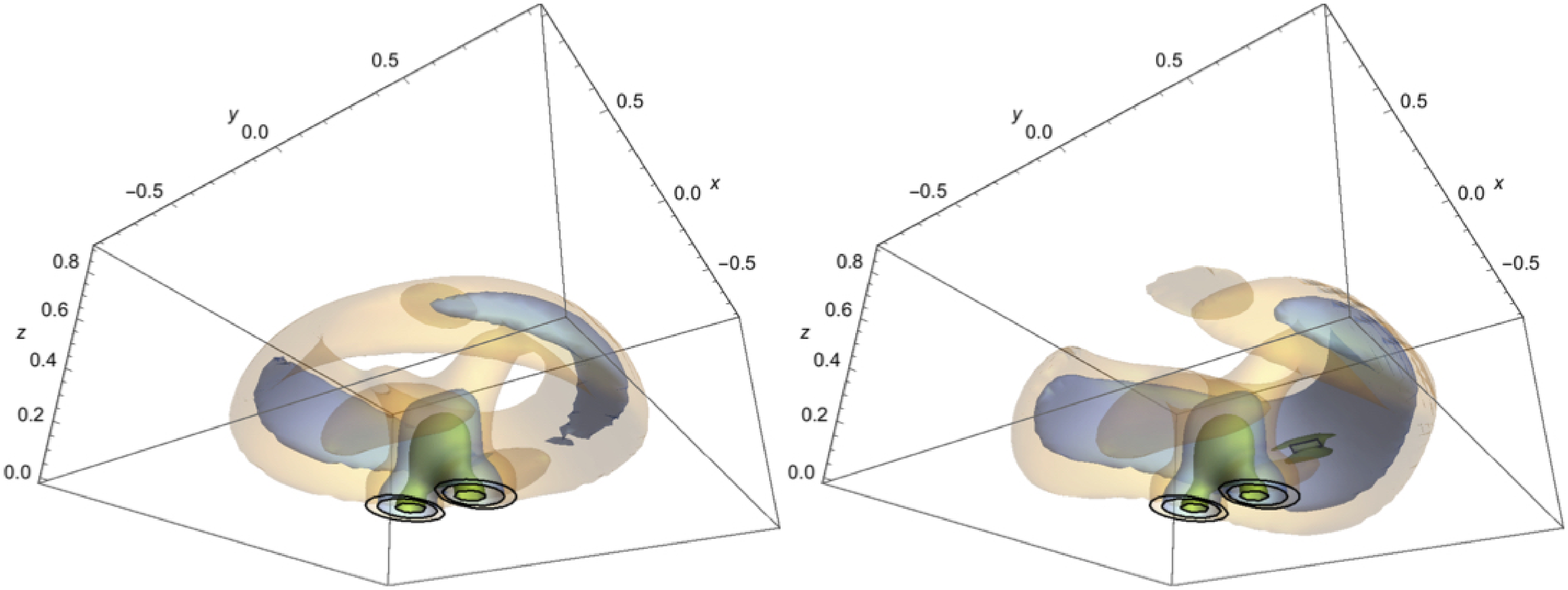}
\end{minipage}
\caption{\label{3D_ximag}
  3-dimensional contour plots showing the magnitude of $\dot\bxi$, for
  $\chi=\pi/16$ (left) and $\chi=7\pi/16$ (right). The details are the
  same as for figure \ref{3D_Bmag}, except that each contour
  represents a value $\sqrt{2}$ times that of the last, meaning that
  the innermost contour has twice the value of the outermost.}
\end{center}
\end{figure}

At last we come to the $\bxi$-motions: the non-rigid velocity field
present in a precessing fluid star, and the original goal of this
paper. We recall that these may now be calculated from our solutions
$\bar{U}_l,X_l$ using equations \eqref{final_xi_r}-\eqref{final_xi_lambda}.
As for $\delta\bB$, we separate the information about
$\dot\bxi$ into two figures, one showing the magnitude of $\dot\bxi$
and the other its direction. The contours of $\dot\bxi$, shown in
figure \ref{3D_ximag}, represent smaller differences than those used above for
$\delta\bB$ -- the speed $|\dot\bxi|$ only differs by a factor of $\sqrt{2}$ between
neighbouring contours, and therefore only varies by a factor of two across most of the stellar
interior. Like the perturbed toroidal field from the lower panels in
figure \ref{3D_Bmag}, the speed is greatest
in a torus centred on the origin and aligned along the $x=0$
plane. For the near-aligned case ($\chi=\pi/16$) $\dot\bxi$ is
approximately reflection symmetric about the $x=0$ plane, but as the
inclination angle is increased the speed lowers more quickly in the
$x<0$ region ($\chi=7\pi/16$ plot). Finally, as mentioned in the
previous subsection, $\dot\bxi={\bf 0}$ for the limiting case of
$\chi=\pi/2$.

\begin{figure}
\begin{center}
\begin{minipage}[c]{\linewidth}
\includegraphics[width=\linewidth]{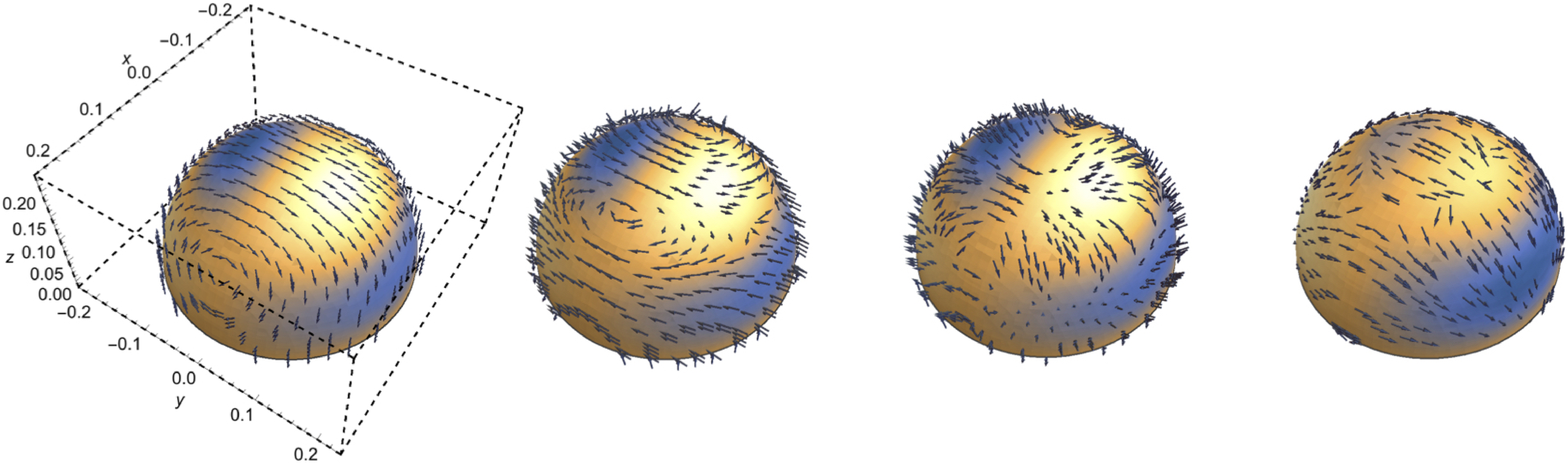}
\end{minipage}
\caption{\label{3D_xivec}
  The direction of the $\dot\bxi$ vector field on constant-radius shells
  of $r/R_*=0.2,0.4,0.6$ and $0.8$ (left to right), for
  $\chi=\pi/16$. See the caption of figure \ref{3D_Bvec} for details.}
\end{center}
\end{figure}

The direction of $\dot\bxi$ is shown in figure \ref{3D_xivec}. Close
to the centre and to the outer boundary, the velocity is roughly
tangential to constant-$r$ shells, but in between it has a large
radial component. As for the poloidal-field direction plots, the blue
and white patches are in the same location for each of the four shells
in figure \ref{3D_xivec}, meaning that there is a meridional
circulation of fluid, with streamlines moving towards the centre of
the star through the blue regions and back out through the white
regions. Unlike the magnetic-field case, however, these streamlines do
not need to close, since $\div\dot\bxi\neq 0$ (i.e. the flow is
compressible). Not restricting to incompressible flow was, in fact,
the place at which our approach first diverged from that of
\citet{mestel1}; recall the discussion of section \ref{fluidprec}. For this
reason, it is logical to close this section by looking at the
behaviour of $\div\dot\bxi$ itself.

\begin{figure}
\begin{center}
\begin{minipage}[c]{0.55\linewidth}
\psfrag{divxi}{$\displaystyle{\frac{r\div\dot\bxi}{|\dot\bxi|}}$}
\psfrag{r}{$\hat{r}$}
\psfrag{th=pi8}{$\theta=\pi/8$}
\psfrag{th=pi4}{$\theta=\pi/4$}
\psfrag{th=3pi8}{$\theta=3\pi/8$}
\includegraphics[width=\linewidth]{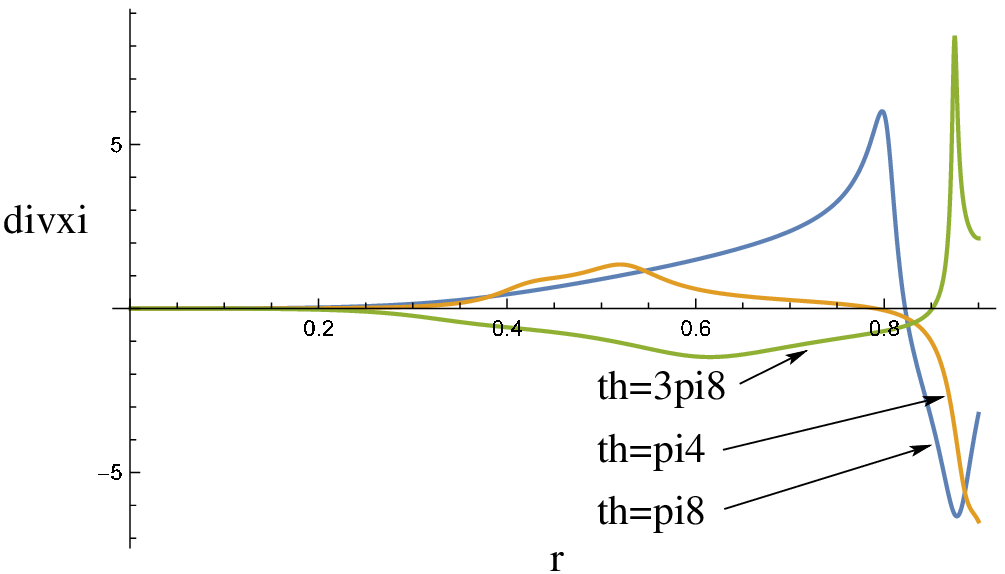}
\end{minipage}
\caption{\label{divxi}
  Divergence of $\dot\bxi$, normalised by the speed
  $|\dot\bxi|$ divided by the radius, to make a dimensionless
  quantity. We show the behaviour of this quantity for $\lambda=\pi/2$
  and three values of $\theta$, as labelled. This plot demonstates
  that for barotropic stellar models, the higher-order motions in a
  precessing fluid star are in fact highly compressible. It also gives
  an idea of how strong buoyancy forces would need to be in order to
  produce the incompressible motions discussed in \citet{mestel1}.}
\end{center}
\end{figure}

In figure \ref{divxi} we plot the divergence of $\div\dot\bxi$,
normalised by $|\dot\bxi|/r$ to give a dimensionless quantity. In a
barotropic star one cannot assume incompressible motions, and indeed
we see that $\div\dot\bxi$ deviates greatly from zero, especially in
the outer regions. Thermal- or composition-gradient stratification in
real stars would provide some additional buoyancy force to suppress
compressible motions, but for a given model one still needs to account
for such a buoyancy force consistently, and to verify it is
sufficiently strong to impose $\div\dot\bxi=0$. Figure \ref{divxi}
gives an idea of how and where such forces would have to act to ensure
$\div\dot\bxi=0$.

\section{Discussion}
\label{conclusions}

This paper has been concerned with a detailed description of the
dynamics of a rotating magnetised star, in the general `oblique
rotator' case when the magnetic and rotational axes are misaligned by
some angle $0<\chi<\pi/2$. The star's bulk motion is that of free
precession, but sustaining this in a fluid star requires additional
dynamics in the interior: time-dependent velocity and magnetic
fields. The concept of non-rigid precession was first described by \citet{mestel1}, who
produced simplified models of the 
perturbed field $\dot\bxi$ using first-order perturbation
equations. A new calculation of $\dot\bxi$ was presented in \citet{mestel2},
who also used this, in turn, to calculate magnetic field
perturbations, given a poloidal background
field. \citet{nittmann-wood} performed the same analysis for a
toroidal background field.

In this paper we argue that a consistent solution for both $\dot\bxi$
and $\delta\bB$ can only come from solving a higher-order set of
equations than those previously considered. A large part of the work
presented here involves a detailed formalism to describe
these perturbations (section \ref{pert_scheme}), algebraic manipulation to reduce them
to a set of differential equations in radial functions (section \ref{solution_second}),
and then solving these (section \ref{results} and appendix \ref{numerical-details}).

In order to make the problem tractable, we have had to make a number
of assumptions, and now need to consider how these affect the validity
of our model when applied to different classes of star. Nonetheless,
we emphasise that our model contains the minimum essential physics for
non-rigid precession, and so everything that follows concerns
\emph{additional} effects. Even when our assumptions are
unjustifiable, then, the work presented here can be regarded as a
`background' model upon which to build.

The most major simplification we have made is assuming a toroidal background
magnetic field. \skl{Purely toroidal fields, like purely poloidal fields, suffer rapidly-growing
  dynamical instabilities -- and so cannot be realistic models of
  stellar magnetic fields \citep{tayler73}, which must instead feature both poloidal and toroidal
  components. However, in the absence of an additional stabilising
  mechanism, there are indications that even mixed poloidal-toroidal
  fields are generically unstable \citep{LJ12,mitchell15}. On the other
  hand, toroidal-dominated stable equilibria have been found for
  stellar models with an additional thermal-buoyancy force
  \citep{braith09}. The issue of stable magnetic equilibria is not
  fully settled, but if the toroidal component is generically the
  dominant one in stars, our results may indeed be a reasonable first
  approximation to their precessional dynamics.

For a stellar model with a purely poloidal field, we believe an
analogous analysis to ours could be carried out. However, whilst a
toroidal field has a component only in the direction of one spherical-polar coordinate
(the azimuthal one, $\lambda$), a poloidal field's depends on the other two ($r$ and
$\theta$), suggesting that spherical-polar coordinates will not be
appropriate for studying this problem. Instead, one can work in a coordinate system
fitted to poloidal-field lines \citep{bernstein}, in which case all
the steps of our toroidal-field calculation should be tractable for a
poloidal field too. There is, however, no coordinate system in which a
mixed poloidal-toroidal field becomes similarly simple (see, e.g.,
\citet{tayler80}); for a stellar model with such a background field,
it seems inevitable that a fully numerical approach would be required.}

\emph{Neutron stars}: most of the assumptions made in this paper are
typical ones in the context of neutron-star modelling. This is
certainly not equivalent to saying that they are all justifiable on
physical grounds, however, so each needs further attention. The
$\gamma=2$ polytrope we adopt is quite suitable for a \skl{single-fluid model
  of a neutron star, but the cores of mature neutron stars are
  better described by a two-fluid equation of state in which the
  neutrons are superfluid; simple magnetic equilibrium models for
  mature neutron-star cores may be found in which
  each fluid component obeys its own polytropic relation
  \citep{LAG,GAL}. The neutron superfluid will decouple from the
  precessional dynamics we consider in this paper, and so only the
  charged component is relevant. This component is, however, better
  described by a polytropic index $\gamma\approx 3/2$.

  It is reasonable to assume that no additional buoyancy forces enter
  the precessional dynamics of a mature neutron star. A neutron star
  in equilibrium has a density-dependent fraction of neutrons to
  protons. When the neutrons are \emph{not} superfluid, displacing a fluid element
  results in chemical reactions which convert between neutrons and
  protons to restore the equilibrium fraction of protons and neutrons, and this represents a buoyancy force associated with the
  composition-gradient stratification \citep{reis_gold}. However, these
  reactions are strongly suppressed (along with the associated
  buoyancy) when the neutrons are superfluid. In
  addition, neutron stars become isothermal very early on in their
  life, so there is negligible thermal-gradient stratification.} Finally, we
have already argued (in section \ref{cc-boundary}) that crustal
elasticity \skl{and the matching between crust and core} could seriously modify our solutions in the outermost
region, $0.9R_*\leq r\leq R_*$, and so we considered it wisest to exclude
this region altogether, and place the outer boundary for our
solutions at a radius of $0.9R_*$.

The most significant neutron-star physics we have failed to account
for is the superconducting nature of the stellar core. Since most of this region is
believed to form a type-II superconductor, however, the magnetic field
will still penetrate the core and produce a macroscopically uniform
magnetic force \citep{baym_pp}. The conceptual picture should thus remain intact: the
magnetic field lends rigidity to the fluid, inducing bulk precession
but at the expense of requiring additional internal motions and
magnetic-field perturbations. Quantitatively, however, the nature of
these perturbations will undoubtedly be different -- and qualitatively
we would expect them to scale with the superconducting magnetic
force\footnote{$H_{c1}$ is the lower critical field for a type-II 
superconductor and is associated with the microscopic field along fluxtubes.}
$\propto H_{c1}B\sim 10^{15} B$ 
instead of the normal Lorentz force $\propto
B^2$ \citep{easson_peth,wass03}.

The velocity field $\dot\bxi$ represents a `macroscopic' coupling between the star's
rotation and magnetic field. In a neutron-star core with type-II
superconducting protons and superfluid neutrons, however, there is a
competing `microscopic' coupling mechanism: the interaction of rotational neutron
vortices with magnetic proton fluxtubes \citep{rud98,link03}. Ascertaining the
relative importance of these two mechanisms would be an important step
towards a quantitative picture of non-rigid neutron-star precession.

\emph{White dwarfs:} unlike neutron stars, white dwarfs form a broad
class of stars, with differences in chemical composition and the
physics governing their different regions. The major subdivision of
white dwarfs of relevance to this paper is between the magnetic and
non-magnetic white dwarfs. Like neutron stars, magnetic white dwarfs
have field strengths spanning
several orders of magnitude, but with an upper limit of $\lesssim 10^9$
G. Many of these strongly magnetised objects have, like magnetars,
long rotation periods, although there are some rapidly-rotating
\emph{and} highly-magnetised white dwarfs \citep{ferrario15} -- the ideal combination for
relatively short precession periods.

The equation of state for a white dwarf is a degenerate Fermi gas. In
the high-density limit this reduces to a $\gamma=4/3$ polytrope, and
in the low-density limit to a $\gamma=5/3$ polytrope
\citep{nauenberg}. For lower densities then, the $\gamma=2$
relation adopted here is probably not an egregious approximation to
white-dwarf matter. Imposing an outer
boundary below the surface is also justifiable, since the degenerate core
of white dwarfs gives way abruptly to a low-mass nondegenerate
envelope region -- although the envelope is only $\sim 30$km \citep{althaus},
meaning that a more appropriate radius to impose the outer boundary at
would be $\sim 0.997R_*$ (taking care that the perturbative scheme is
still applicable at this radius; see section \ref{surface_issues}).

In this work we have assumed a fluid region which extends to
the stellar centre, which is not always the case for white dwarfs: as
they become older and colder, white
dwarfs begin to form solid cores \citep{vanhorn}, with the
crystallisation proceeding from the centre outwards and solidifying
the whole star over a timescale of $\sim 10^9$ yr
\citep{iben}. Elasticity will then affect the induced velocity and
magnetic-field perturbations in the solid core; for a
sufficiently rigid core lattice these perturbations would become
negligible, and the core's motion would then be well described by standard
rigid-body free precession. Finally, more detailed modelling of white-dwarf precession should account for the
fact that these stars are non-barotropic, with stratification due to
both composition and thermal gradients.

\emph{Nondegenerate (main-sequence) stars}: with some caveats, our
analysis applies to objects along the upper part of
the main sequence of nondegenerate stars. Upper-main-sequence stars
have higher masses and more simple field topologies than other
nondegenerate stars, and since their envelopes appear to be in
radiative equilibrium (rather than convective), dynamo action seems
unlikely. The conclusion is that the magnetic fields in these
envelopes are likely to be in dynamically stable equilibrium
states. Of these stars, the most highly magnetised are the chemically
peculiar A and B-type stars, with field strengths up to $\sim 10^4$
G. Upper-main-sequence stars with strong magnetic fields tend to rotate far more slowly than their
weakly-magnetised counterparts, such as the normal A and B stars \citep{don-land}.

In the radiative envelopes of main-sequence stars our analysis is
broadly applicable, but as for white dwarfs should be modified by
the inclusion of a thermal buoyancy force. This will constrain the form of $\dot\bxi$ and
$\delta\bB$, but in a way which can only be confirmed by
including thermal effects in a self-consistent matter in the whole
analysis. As for partially-crystallised white dwarfs, our analysis
will not be valid all the way to the centre of main-sequence stars. In
this case the reason is the presence of an 
inner core region where vigorous convection disrupts and regenerates
the magnetic field -- and so our assumption of a large-scale stable
background magnetic field becomes invalid here. Moving to the outer
boundary, there is no distinct outer region of the radiative
zone where one would expect the $\bxi$-motions to stop. Our approach
requires us to cut off the solution before reaching the stellar
surface $R_*$ (see section \ref{surface_issues}), but as long as we choose an outer boundary close to
$R_*$ we will only be neglecting a small, low-density region of the star.

For all classes of star, the internal velocity and magnetic fields
solved for in this paper may have significant effects on their
evolution and dynamics. In particular, since they couple the star's
rotation and magnetic fields, the dissipation of these motions could
cause the evolution of the inclination angle \citep{spitzer}.
For toroidal fields, the case considered here,
the dissipation will tend to cause the star to become an orthogonal
rotator, $\chi\to\pi/2$; for poloidal fields $\chi\to 0$ \citep{mestel1}. For the more
realistic case of mixed poloidal-toroidal fields the evolution of
$\chi$ will be dictated by the dominant field component\footnote{The
possibility of no evolution of $\chi$ can probably be dismissed, since
it would require very fine tuning of the field components.}, and so
observations of this evolution could allow one to infer details of the
star's internal magnetic field. Although
we do not intend to tackle a fully self-consistent 
evolution of dissipative processes, a qualitative discussion of this
and applications to astrophysics will be presented in a follow-up paper.

\section{Summary}
\label{summary}

We have presented a formalism and solution method for models of
rotating magnetic stars with misaligned rotation and magnetic axes --
sometimes known as oblique rotators. We began with arguably the simplest
necessary pieces of physics to describe this problem: a rigid primary
rotation of the star, and a large-scale $m=0$ dipolar toroidal magnetic field. Despite this
simplicity, we have found that the dynamics of such stars
involve internal velocity and magnetic-field perturbations with highly
complex geometries. The perturbed velocity and magnetic field are both stronger towards the
centre of the star, and the toroidal perturbed field is a factor of
two more intense than the poloidal. There are indications that whilst
the perturbed toroidal field is dominated by its dipole ($l=1$) piece, the
poloidal component of the magnetic-field perturbation may have significant contributions from many higher
multipoles -- and in fact its $l=4$ component is stronger than its
lowest-order multipole, a quadrupole ($l=2$) piece. The velocity field has significant
azimuthal and meridional components, and deviates strongly from being
solenoidal -- i.e., the internal motions are highly compressible. The
secular dissipation of these perturbations will cause the star to
evolve over time into either an aligned or orthogonal rotator, and so
the distribution of inclination angles in stars is likely to be
connected with their internal magnetic fields.

\section*{Acknowledgements}

We acknowledge support from STFC via grant number ST/M000931/1, and from
NewCompStar (a COST-funded Research Networking Programme). \skl{We
  thank the anonymous referee for a number of helpful suggestions for
  improving this paper.}


\appendix

\section{Spherical-harmonic decompositions of source terms}
\label{Ylm_decomp_Ups_Psi}

Since we solve equations for a given $m$, we need the source terms
decomposed into $Y_l^m$ pieces too.  
We use the following relations between trigonometric functions and
spherical harmonics $Y_l^m(\theta,\lambda)$:
\begin{align}
\sin(\theta)\sin\lambda   
    &= i\sqrt{\frac{2\pi}{3}}(Y_1^1+Y_1^{-1}),\\
\sin^2\theta\sin(2\theta)\cos\lambda
    &= \frac{8}{21}\sqrt{\frac{\pi}{5}}(Y_4^1-Y_4^{-1}) - \frac{8}{7}\sqrt{\frac{2\pi}{15}}(Y_2^1-Y_2^{-1}),\\
\sin^4\theta\cos(2\lambda)
    &= -\frac{4}{21}\sqrt{\frac{2\pi}{5}}(Y_4^2+Y_4^{-2}) + \frac{4}{7}\sqrt{\frac{6\pi}{5}}(Y_2^2+Y_2^{-2}),
\end{align}
together with those given in equations \eqref{Y21}, \eqref{Y22}, \eqref{Y1131}
and \eqref{Y32}. These allow us to decompose the scalar functions $\Upsilon_{\delta\clL}$
and $\Psi_{\delta\clL}$ from equations \eqref{Upsilon_dL} and
\eqref{Psi_dL}, which enter as source terms in our final system of
equations:
\beq \label{Ups_lm}
\Upsilon_{\delta\clL} =
   \frac{i\alpha^2\Lambda^2 r}{16G}\sqrt{\frac{\pi}{6}}\left\{
       \sin(2\chi)\left[j_2(\Mr)-16k_B\right](Y_1^1+Y_1^{-1})
      + \frac{1}{3\sqrt{7}} j_2(\Mr) \left[ \sqrt{2}\sin(2\chi)(Y_3^1+Y_3^{-1})
                                                            - \sqrt{5}\sin^2\chi(Y_3^2-Y_3^{-2}) \right] \right\}
\eeq
and
\begin{align}
\Psi_{\delta\clL}
  = \frac{\alpha^2\Lambda^2}{336G}\sqrt{\frac{5\pi}{6}}
        & \bigg\{ -\sin(2\chi)\left[ \Mr j_1(\Mr) +11 j_2(\Mr) +12\Mr\cot(\Mr)j_2(\Mr) \right](Y_2^1-Y_2^{-1})\nn\\
        &\ \ \ \   + \sin^2\chi\left[ -2\Mr j_1(\Mr) +20 j_2(\Mr) +18\Mr\cot(\Mr)j_2(\Mr)\right](Y_2^2+Y_2^{-2})\nn\\
        &\ \ \ \   + \sqrt{6}\sin(2\chi)\left[-\Mr j_1(\Mr)+3j_2(\Mr)+2\Mr\cot(\Mr)j_2\right](Y_4^1-Y_4^{-1})\nn\\
        &\ \ \ \  - \sqrt{3}\sin^2\chi\left[-\Mr j_1(\Mr)+3j_2(\Mr)+2\Mr\cot(\Mr)j_2\right](Y_4^2+Y_4^{-2})\bigg\}.
 \label{Psi_lm}
\end{align}

\section{Numerical solution and error estimation}
\label{numerical-details}

We use the NDSolve command in the software package Mathematica in
order to solve our system of equations. NDSolve is not capable of
solving differential-algebraic equations as boundary-value problems,
as we would need, so we instead convert our original equations into a system of
ODEs by working with equation \eqref{DAE1} and equation \eqref{diff_DAE2}. We implement the boundary conditions as
summarised in table \ref{BCs}. The full solution to our precession problem
should be an infinite sum of harmonics, but we terminate at a value of
$l_\textrm{max}=4$, as we have failed to find a successful method of
solving the equations for arbitrary truncation values; implementation
of the techniques of section \ref{DAE-centrecond} to
$l_\textrm{max}>4$ typically results in a solution which does not
satisfy the the highest-$l$ equation to acceptable accuracy. Finally,
we non-dimensionalise the variables in our equations by suitable
combinations of $G,\rho_c$ and $R_*$ (these three quantities contain
instances of mass, length and time, so together they may be used to
non-dimensionalise any other quantity). This allows us to work with
quantities whose magnitude does not deviate excessively from unity,
unlike the unwieldy physical parameters for stellar models. It
also allows one to scale results easily to parameters appropriate
for models of main-sequence stars, white dwarfs and neutron stars.

The problems discussed up to this point have been mathematical in
origin, but the numerical solution has also required some care. We
have found that the numerical algorithm typically requires 25-digit
precision or higher for success. Finally, Mathematica returns
zero-divided-by-zero errors with the equations in their original
form. To avoid these, certain quantities need to be evaluated at a point
slightly displaced from the centre. To minimise the introduced error, and to prevent
it from affecting conditions at the outer boundary (the crust-core
boundary), we define the function
\beq \label{frakd}
\mathcal{D}(r)\equiv \mathfrak{d} (r-R_\textrm{out}),
\eeq
where $\mathfrak{d}$ is some small constant, and in the equations in
the Mathematica notebook we make the replacement
$r\mapsto r+\mathcal{D}(r)$ in the quantities
$rU'_l(r),\Upsilon'_l(r),(r\rho'_0/\rho_0)'$. At the end of this
section, we verify that the
introduced error is small, and converges to zero as $\mathfrak{d}\to 0$.

\subsection{Verification of DAE solution method}

In section \ref{DAE-centrecond}, we presented a method to solve the system of DAEs
\eqref{DAE1} and \eqref{DAE2}. We differentiated the second of these DAEs
in order to get a system of ODEs, \eqref{DAE1} and \eqref{diff_DAE2};
however, the set of solutions to these ODEs is larger than the set of
solutions to the original DAEs, and so a typical solution to the ODEs
would differ from a solution to the DAEs by some `integration
constant' factor. The method of section \ref{DAE-centrecond} was designed to pick
out the correct integration constant, so let us now check that our
solutions do indeed satisfy the original DAEs.

\begin{table*}
\begin{center}
\caption{\label{wrongBCs}
               An example set of superficially sensible boundary
               conditions on the ODE system \eqref{DAE1} and \eqref{diff_DAE2}, which
               are nonetheless not acceptable, as they do not ensure
               the resulting solution is also a solution of the
               original DAEs \eqref{DAE1} and \eqref{DAE2}.}
\begin{tabular}{ccc}
\hline
$m$ & centre & crust-core boundary \\
\hline
   & $\bar{U}'_2(0)=0$ & $R_\textrm{out}\bar{U}'_2(R_\textrm{out})+4\bar{U}_2(R_\textrm{out})=0$ \\
1 & $\bar{U}_4'(0)=0$ & $R_\textrm{out}\bar{U}'_4(R_\textrm{out})+6\bar{U}_4(R_\textrm{out})=0$\\
   &  & $X_1(R_\textrm{out})=0$\\
   &  & $X_3(R_\textrm{out})=0$\\
\hline
   & $\bar{U}'_2(0)=0$ & $R_\textrm{out}\bar{U}'_2(R_\textrm{out})+4\bar{U}_2(R_\textrm{out})=0$ \\
2 & $\bar{U}_4'(0)=0$ & $R_\textrm{out}\bar{U}'_4(R_\textrm{out})+6\bar{U}_4(R_\textrm{out})=0$\\
   &  & -- \\
   &  & $X_3(R_\textrm{out})=0$\\
\hline
\end{tabular}\\
\end{center}
\end{table*}

\begin{figure}
\begin{center}
\begin{minipage}[c]{\linewidth}
\psfrag{Mathematica error}{Inherent Mathematica error}
\psfrag{integration-const error}{\ \ \ Boundary conditions from table \ref{wrongBCs}}
\psfrag{error removed}{\ \ \ Boundary conditions from table \ref{BCs}}
\psfrag{e1}{$\mathfrak{e}_1$}
\psfrag{e2}{$\mathfrak{e}_2$}
\psfrag{e3}{$\mathfrak{e}_3$}
\psfrag{e4}{$\mathfrak{e}_4$}
\psfrag{r}{$\hat{r}$}
\includegraphics[width=\linewidth]{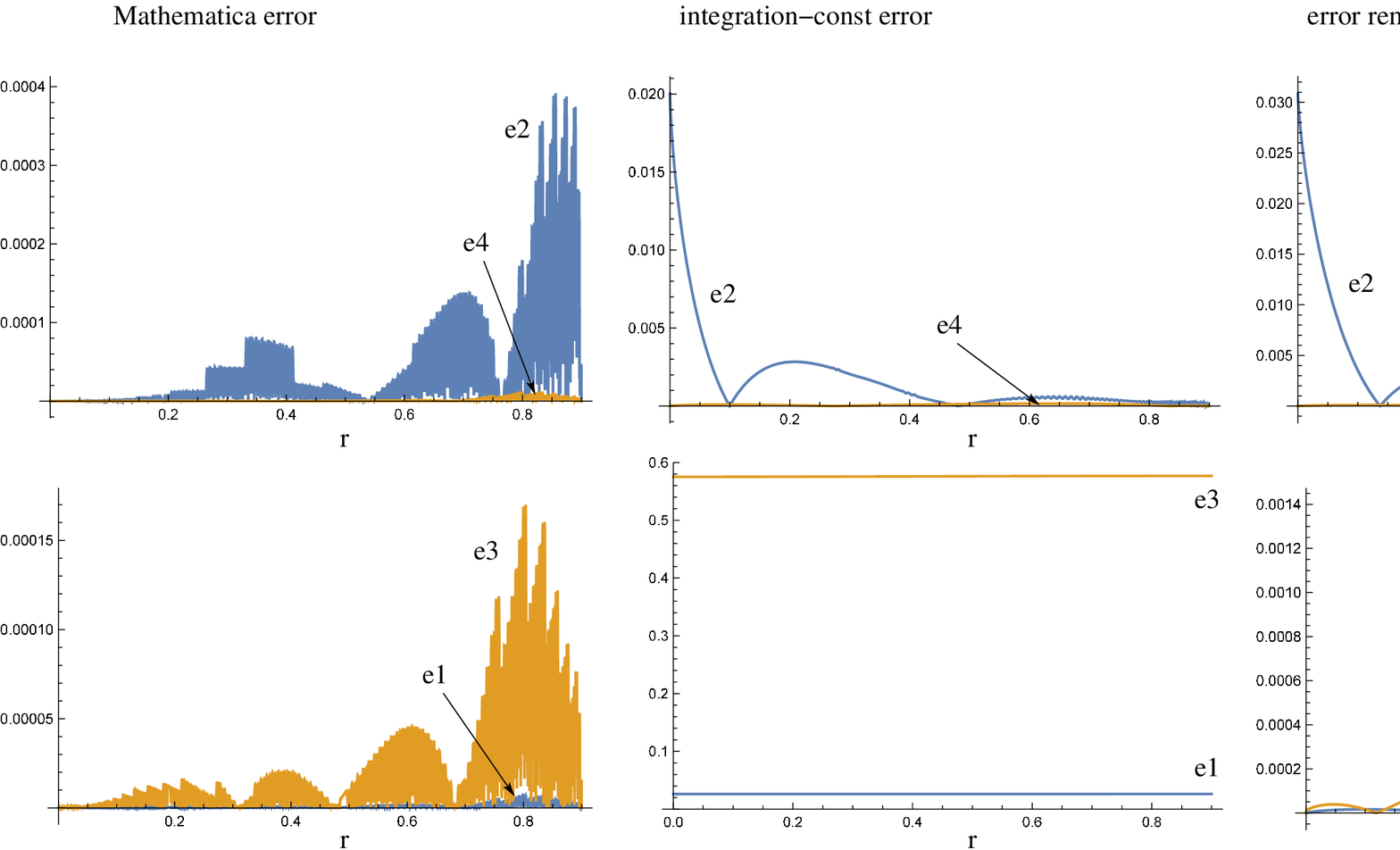}
\end{minipage}
\caption{\label{error-plots}
  Comparison of errors $\mathfrak{e}_l$ (see equation \eqref{frak-e})
  for different boundary conditions. We convert our original
  differential-algebraic equations \eqref{DAE1} and \eqref{DAE2} into
  ODEs which Mathematica is capable of solving by differentiating the
  latter equation, giving equation \eqref{diff_DAE2}. The left-hand plots show the result of
  subsituting the obtained solution back into \eqref{DAE1} and
  \eqref{diff_DAE2}; this is merely a consistency check
  of Mathematica's algorithm, and the resulting error is predictably
  small (regardless of which set of boundary conditions we use). We then solve \eqref{DAE1} and
  \eqref{diff_DAE2} together with a na\"ive set of boundary conditions,
  table \ref{wrongBCs}, which do not contain information about the
  original DAEs. We substitute the solution obtained with these
  boundary conditions back into the original DAEs, \eqref{DAE1} and
  \eqref{DAE2}, and show the resulting error in the central plots. For the lower
  of these plots, corresponding to the equation which was differentiated, a
  large, constant error can be seen. This reflects the fact that a
  solution to \eqref{diff_DAE2} will typically differ by
  some integration constant from a solution of the original equation
  \eqref{DAE2}. To fix this constant at the correct value (of zero),
  we need to use the more careful boundary conditions of table
  \ref{BCs}. With these, as shown in the right-hand plots, the error
  becomes small.
}
\end{center}
\end{figure}

We substitute our numerical solution
$\mathcal{S}\equiv\{X_l,\bar{U}_l\}$
back into equations \eqref{DAE1} and \eqref{DAE2}. To produce a normalised error estimate, we
recall that the $X_l$ functions are only non-zero
for odd $l$, and the $\bar{U}_l$ functions are non-zero for even
$l$. Since we solve \eqref{DAE2} (in differentiated form) for odd $l$ and
\eqref{DAE1} for even $l$, this means that for a particular value of
$l$, we can associate one equation with one particular function,
e.g. equation \eqref{DAE2} for $l=1$ may be associated with $X_1$.
We produce normalised errors $\mathfrak{e}_l(r)$ using this
observation:
\beq \label{frak-e}
\mathfrak{e}_l(r)\equiv
 \begin{cases}
    \displaystyle{\frac{\textrm{$\mathcal{S}$ substituted into equation \eqref{DAE1}}}{\max{(\bar{U}_l)}}\textrm{ (for even $l$)}},\\
    \displaystyle{\frac{\textrm{$\mathcal{S}$ substituted into equation \eqref{DAE2}}}{\max{(X_l)}}\textrm{ (for odd $l$)}}.
 \end{cases}
\eeq
In figure \ref{error-plots} we show the resulting errors, whose small size
indicate we have successfully obtained a solution to the DAE system. For comparison, we also
perform a consistency check by substituting the solution back into
\emph{the ODEs} \eqref{DAE1} and \eqref{diff_DAE2}; since these were the
equations actually solved by Mathematica, the resulting error is very
small.

If we had not worried about fixing the integration constant correctly,
we might instead have adopted a set of boundary conditions like those
of table \ref{wrongBCs}, which are natural-looking ones to use to ensure that
solutions are well-behaved at the centre and do not give rise to
a current sheet at the outer boundary. Figure \ref{error-plots} clearly shows
that the resulting solutions fail to satisfy the original DAE system.

\subsection{Convergence}

\begin{figure}
\begin{center}
\begin{minipage}[c]{\linewidth}
\psfrag{frakd-3}{$\mathfrak{d}=10^{-3}$}
\psfrag{frakd-4}{$\mathfrak{d}=10^{-4}$}
\psfrag{frakd-5}{$\mathfrak{d}=10^{-5}$}
\psfrag{e1}{$\mathfrak{e}_1$}
\psfrag{e2}{$\mathfrak{e}_2$}
\psfrag{e3}{$\mathfrak{e}_3$}
\psfrag{e4}{$\mathfrak{e}_4$}
\psfrag{r}{$\hat{r}$}
\includegraphics[width=\linewidth]{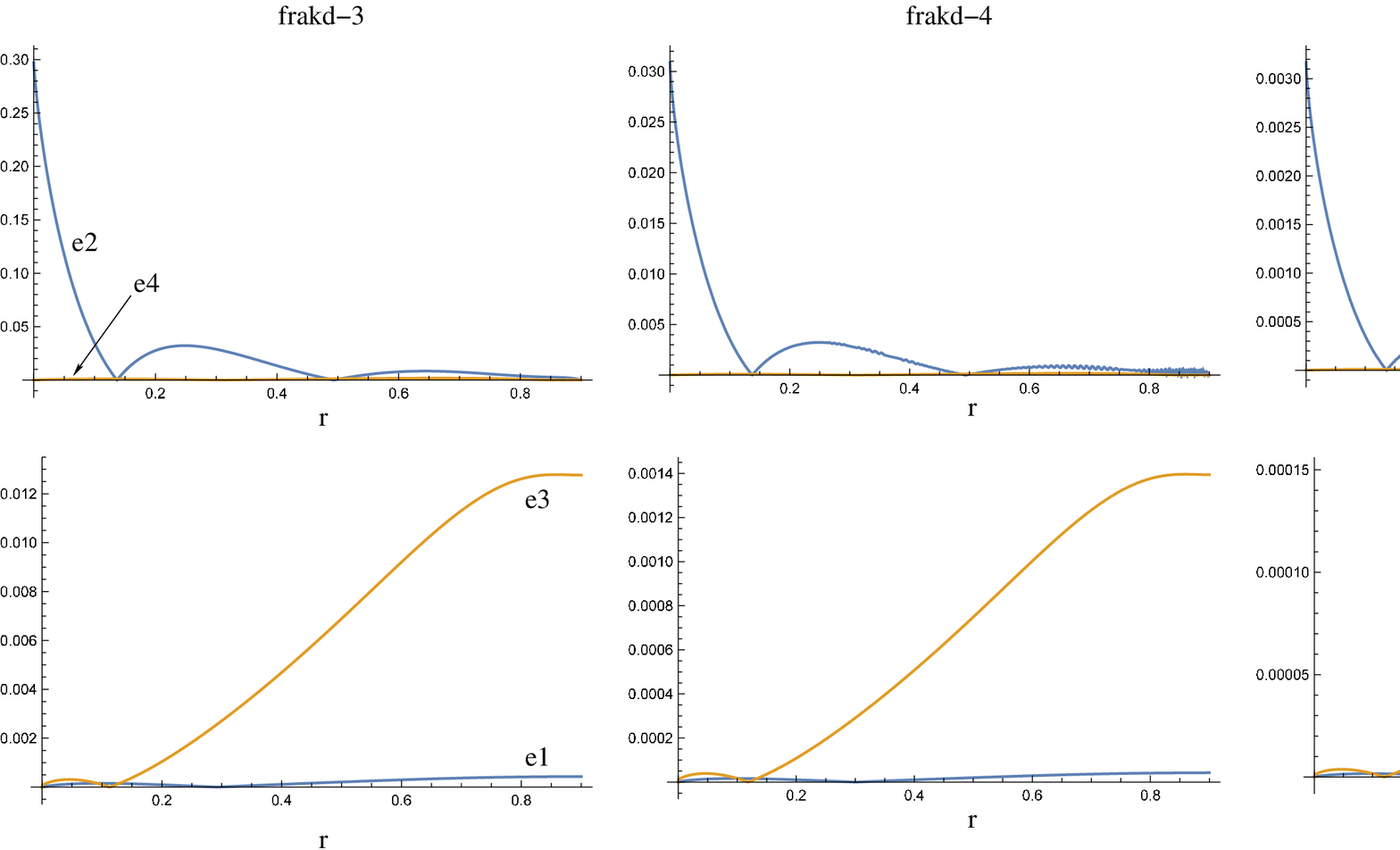}
\end{minipage}
\caption{\label{frakd-conv}
  Convergence of the error in solutions as $\mathfrak{d}\to 0$. For
  $\mathfrak{d}=10^{-5}$, the noisy inherent error present from
  Mathematica's solution becomes visible; i.e. the error introduced
  from evaluating quantities slightly away from the origin reduces to
  the level of intrinsic error from Mathematica's solution method. Note
  that the actual physical results for $\bar{U}_l,X_l$ for these three
  values of $\mathfrak{d}$ are indistinguishable.}
\end{center}
\end{figure}

To prevent numerical errors at the centre, we evaluated
some quantities at a small radial distance --  given by equation
\eqref{frakd} -- from their correct location. Our final check is to make
sure the introduced error is small, and converges to zero in the limit
$\mathfrak{d}\to 0$; we do this in figure \ref{frakd-conv}. Here we
have used the correct boundary conditions, from table \ref{BCs}; note
that the large `integration constant' error resulting from using the
boundary conditions of table \ref{wrongBCs} does \emph{not} converge
away as $\mathfrak{d}\to 0$.

\label{lastpage}

\end{document}